\documentclass[sigconf,review=false,anonymous=false]{acmart}

\usepackage{multirow}
\usepackage{array}
\newtheorem{problem}{Problem}[section]
\usepackage{graphicx}
\usepackage{caption}
\usepackage{subcaption}
\usepackage{makecell}
\usepackage{balance}
\usepackage{xspace}
\usepackage{booktabs}

\newcommand{\wrt}{\emph{w.r.t.}\xspace}
\newcommand{\ie}{\emph{i.e.,}\xspace}
\newcommand{\eg}{\emph{e.g.,}\xspace}

\newcommand{\eat}[1]{}

\AtBeginDocument{%
  }

\begin{document}

\title{How Cohesive Are Community Search Results on Online Social Networks?: An Experimental Evaluation}
\subtitle{[Technical Report]}

\setcopyright{none} 
\copyrightyear{}
\acmConference[Conference acronym 'XX]{xx}{xx}{xx}
\renewcommand\footnotetextcopyrightpermission[1]{} 
\renewcommand{\authornote}[1]{}

\author{Yining Zhao}
\email{yining002@e.ntu.edu.sg}
\orcid{0009-0000-0897-8431}
\affiliation{
  \institution{Nanyang Technological University}
  \country{Singapore}
}

\author{Sourav S Bhowmick}
\email{assourav@ntu.edu.sg}
\orcid{0000-0003-1957-8016}
\affiliation{
  \institution{Nanyang Technological University}
  \country{Singapore}
}

\author{Nastassja L. Fischer}
\email{nastassja.fischer@nie.edu.sg}
\orcid{0000-0002-1383-1917}
\affiliation{
  \institution{National Institute of Education, Singapore}
  \country{Singapore}
}

\author{SH Annabel Chen}
\email{annabelchen@ntu.edu.sg}
\orcid{0000-0002-1540-5516}
\affiliation{
  \institution{Nanyang Technological University}
  \country{Singapore}
}

\begin{abstract}
Recently, numerous \textit{community search} methods for large graphs have been proposed, at the core of which is defining and measuring \textit{cohesion}. This paper experimentally evaluates the effectiveness of these community search algorithms \wrt cohesiveness in the context of online social networks. Social communities are formed and developed under the influence of group cohesion theory, which has been extensively studied in social psychology. However, current \emph{generic} methods typically measure cohesiveness using \textit{structural} or \textit{attribute-based} approaches and overlook domain-specific concepts such as group cohesion. We introduce five novel \textit{psychology-informed cohesiveness measures}, based on the concept of \textit{group cohesion} from social psychology, and propose a novel framework called \textsf{CHASE} for evaluating eight representative community search algorithms \wrt these measures on online social networks. Our analysis reveals that there is no clear correlation between structural and psychological cohesiveness, and no algorithm effectively identifies psychologically cohesive communities in online social networks. This study provides new insights that could guide the development of future community search methods.
\end{abstract}

\maketitle

\vspace{-1ex}
\section{Introduction} \label{sec: intro}

Real-world networks frequently exhibit community structures~\cite{girvan2002community}, making \textit{community search (CS)} a key area of focus in graph search~\cite{fang2020survey}. It seeks to retrieve \textit{cohesive} subgraphs from an input graph based on query nodes. Communities extracted by CS algorithms have potential applications in event organization, recommendations, protein complex identification, and targeted advertising~ \cite{fang2020survey, gao2021ics}. In particular, human beings thrive in communities, forming connections through shared experiences and values, which promote cooperation, resilience, and a sense of belonging~\cite{baumeister2017need, wellman2001physical}. Thus, community search is particularly important for applications involving human interaction, such as those in online social networks.

Extensive research on the community search problem has primarily focused on enhancing the efficiency, scalability, and quality of results~\cite{fang2020survey}. Central to this issue is the concept of \textit{cohesion} within communities. Most CS algorithms define communities through \textit{subgraph cohesiveness metrics} derived from structural network information, such as \textit{k-core}, \textit{k-truss}, and \textit{k-ECC}, along with attribute data~\cite{fang2020survey}. These measures are \textit{generic}, meaning they are not tied to any specific application domain. Recently, several studies have investigated \textit{learning-based} methods for identifying cohesive subgraphs, which integrate both structural information and vertex attributes into neural network models, thereby removing the need for predefined rules for communities~\cite{gao2021ics, li2023coclep, hashemi2023cs, fang2024inductive, wang2024scalable, wang2024efficient}. \emph{In this paper, we evaluate the cohesiveness of communities produced by these existing techniques within online social networks}.

In online social networks, communities are fundamentally characterized as groups of individuals connected through ``invisible bonds'', a phenomenon that has been widely studied in social psychology as \textit{cohesion} or \textit{group cohesion} since the 1950s \cite{festinger1950informal,zimbardo2003psychology}. Most psychology studies recognize cohesion as \textit{multi-dimensional}~\cite{salas2015measuring, mcleod2013towards, carron2000cohesion} and have demonstrated its validity in various contexts, such as sports \cite{carron1982cohesiveness, pescosolido2012cohesion}, work \cite{lu2015building, paul2016global}, and technology-mediated learning environments \cite{zamecnik2022cohesion}, as well as in both offline and online settings \cite{galyon2016comparison, lyles2018longitudinal, cai2023strengthening}. Consequently, algorithmically identified communities can be hypothesized to manifest psychologically grounded cohesion properties. Notably, structural cohesiveness metrics used in the CS problem have their origins in social psychology \cite{wasserman1994social, fortunato2010community}. 

Existing CS methods evaluate their solutions in three main ways~\cite{das2021attribute, li2017most, miao2022reliable, lin2024qtcs, wang2024efficient, wang2024scalable, zhang2023size}: (a) by comparing retrieved communities to known ground-truth communities using various performance metrics, though the interpretation and definition of ground truth vary \cite{yang2012defining, yin2017local, leskovec2007graph, madani2022dataset, klymko2014using}; (b) by assessing communities against those produced by other algorithms through ``effectiveness metrics'' or ``goodness metrics'' that measure community characteristics \cite{behrouz2022firmtruss, lin2024qtcs}, especially when the ground truth is unavailable; and (c) by conducting case studies to demonstrate the interpretability and significance of the communities identified by CS algorithms.

None of the existing evaluation approaches systematically assesses the effectiveness of cohesiveness measures in networks involving interacting individuals. Consequently, they fail to address key questions specific to online social networks, such as: (1) Do the generic cohesiveness measures from different CS techniques accurately capture group cohesion in online social networks? (2) How cohesive are the communities identified by these techniques when evaluated using measures based on group cohesion in social psychology?

In this paper, we address these key questions about the cohesiveness of communities identified by community search (CS) algorithms in online social networks. We begin by discussing how research on group cohesion can inform the assessment of cohesion in online social networks. We critique the limitations of existing generic cohesiveness measures used by various CS techniques in capturing the multi-dimensional nature of group cohesion and introduce a set of \textit{psychology-informed cohesiveness measures} grounded in social psychology’s definition of cohesion to address them. Additionally, we present a framework called \textit{\textsf{CHASE}} to conduct extensive experiments comparing the cohesiveness of results from eight representative CS techniques using real-world social network data.

Our experimental evaluation reveals three key findings. First, existing algorithms, when using the same query node, often identify communities that differ significantly in terms of structural and psychology-based cohesiveness, with varying query hit rates. Second, recent learning-based algorithms tend to produce communities with low structural cohesiveness or fail to identify valid communities. Third, there is no clear relationship between structural cohesiveness and psychology-based cohesion. More importantly, no current algorithm identifies cohesive communities in online social networks that align with the concept of cohesion in social psychology.

In summary, this work makes the following novel contributions. (a) It is the first study to assess the cohesiveness of social network communities generated by various community search methods, analyzed through the lens of cohesion theories in social psychology. (b) It introduces a set of psychology-informed cohesiveness measures for evaluation, which is unavailable in existing community search evaluation toolkits. (c) The experimental findings presented in this work are novel and have not been reported in previous studies.

\vspace{-1ex}
\section{Related Work} \label{sec: related_work} 
\citet{fang2020survey} evaluated CS results from traditional algorithms on simple and attributed graphs. For simple graphs, it experimentally assesses the average running time and reports community quality using four generic structure-based measures: diameter, degree, density, and clustering coefficient. For attributed graphs, the evaluation focuses on keyword-based CS solutions for efficiency and F1 scores. In social psychology, cohesion measures are developed based on a specific definition and are tailored to a target population. Their quality is assessed through \textit{reliability} and \textit{validity} \cite{bhattacherjee2012social}, focusing on how well these measures align with theoretical constructs \cite{shaver2015principles, blackstone2018principles}. Recently, Bhowmick et al.~\cite{s2024social} introduced the paradigm of \textit{psychology-informed design} and reviewed existing social influence computation techniques that embrace it. None of these efforts undertakes a systematic evaluation of the cohesiveness of CS solutions on online social networks through the lens of social psychology.

\vspace{-1ex}
\section{The Community Search Problem} \label{sec: cs} 
In this section, we review the community search problem and describe a taxonomy of existing CS algorithms. We begin by introducing some terminology to facilitate the exposition of the paper.
    
\vspace{1ex}\noindent\textbf{Terminology.} An \textit{online social network} can be modeled as a directed multigraph $\mathcal{G}=(\mathcal{U}, \mathcal{E})$, where $\mathcal{U}$ indicates the set of users (vertices) and $\mathcal{E}$ includes edges representing individual and interaction activities. \textit{Individual activity} occurs when a user $u_i$ posts information at time $t$ without targeting anyone, represented as a self-loop $e_{ii}^t \in \mathcal{E}$. \textit{Interaction activity} refers to an interaction between user $u_i$ and $u_j$ at time $t$, forming a directed edge $e_{ij}^t \in \mathcal{E}$. An \textit{induced subgraph} $H = (\mathcal{U}_H, \mathcal{E}_H)$ of $\mathcal{G}$ consists of a vertex subset from $\mathcal{G}$ and all edges between them. 
  
\textbf{Problem Definition.} Based on \cite{sozio2010community, fang2020survey}, community search problem can be formalized as follows:

\begin{problem}{\em\textbf{[Community Search (CS) Problem]}\/} \label{CS_Prob}
    Given a graph $G = (V, E)$ and a set of query nodes $Q \subseteq V$, the community search problem seeks to find a community that contains $Q$ and satisfies the following properties:
    \vspace{-1ex}
    \begin{itemize}
        \item[(i)] \textit{Connectivity}: Vertices in the community are connected; 
        \item[(ii)] \textit{Cohesiveness}: Vertices in the community are intensively linked to each other based on a particular goodness metric.
    \end{itemize}
\end{problem}

Note that $\mathcal{G}$ can be reduced to a simple graph $G = (V, E)$ where $V = \mathcal{U}$ and each set of parallel edges in $\mathcal{E}$ between vertices can be reduced to undirected or directed connections.

\textbf{Taxonomy of CS Algorithms.} Following \cite{wang2024scalable}, we classify existing CS works into \textit{traditional} and \textit{learning-based} categories.
    
\textit{\underline{Traditional CS algorithms}.} Generally, traditional CS algorithms formulate a community by specifying its subgraph properties \cite{fang2020survey}. The commonly adopted cohesiveness properties are structural-based, including \textit{k-core} \cite{sozio2010community, cui2014local, barbieri2015efficient, yao2021efficient}, \textit{k-truss} \cite{huang2014querying, akbas2017truss, liu2020truss, zhang2023size}, \textit{k-clique} \cite{yuan2017index}, and \textit{k-ECC} \cite{chang2015index, hu2016querying, hu2017minimal}. These metrics are often referred to as \textit{structural cohesiveness metrics} (\textit{structural metrics} for brevity). Additionally, studies incorporate attribute information to formalize cohesiveness by refining existing structural metrics or proposing new \textit{attribute cohesiveness metrics} alongside the structural ones. Examples of the former approach include \textit{attributed $pkd$-truss community} \cite{xie2021effective}, \textit{$(k, \theta)$-community} \cite{miao2022reliable}, \textit{$(k, p)$-core} \cite{lu2022time}, and \textit{$\beta$-temporal core} \cite{lin2024qtcs}. Examples of the latter include \textit{keyword cohesiveness} \cite{fang2016effective}, \textit{spatial cohesiveness} \cite{fang2017effective, chen2018maximum}, and \textit{query cohesiveness} \cite{das2021attribute}.

\textit{\underline{Learning-based CS algorithms}.} Due to the structural inflexibility of traditional CS algorithms, recently learning-based methods have been proposed to retrieve communities by learning their hidden patterns without relying on predefined structures. Depending on their reliance on ground-truth communities for training, they can be further categorized as \textit{supervised} \cite{jiang2021query, chen2023communityaf, hashemi2023cs, fang2024inductive, wang2024scalable}, \textit{semi-supervised} \cite{li2023coclep}, and \textit{unsupervised} \cite{wang2024efficient}. However, most learning-based solutions weakly correlate with community cohesiveness, struggle with large graphs due to neural network constraints, and risk of model invalidation when the graph topology changes \cite{wang2024scalable}.

\vspace{-1ex}
\section{Cohesiveness Framework for Social Networks} \label{sec: cohesive}
This section outlines the methodology for measuring cohesion based on social psychology and highlights the limitations of current generic cohesiveness measures used in community search techniques. The goal is to determine whether these generic measures align with social psychology's concept of cohesion, which is crucial for effective community retrieval in online social networks.

\vspace{-1ex}
\subsection{Methodology}
\begin{figure}[t]
    \centering
    \includegraphics[scale=0.18]{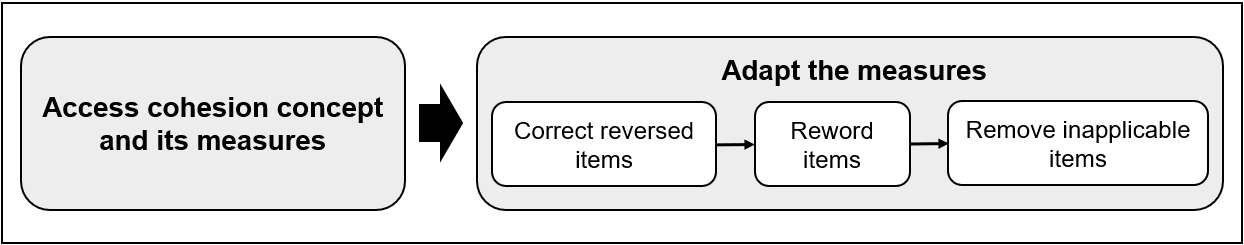}  
    \vspace{-2ex}
    \caption{Workflow for adapting cohesiveness measures.}
    \label{fig: workflow}
    \vspace{-3ex}
\end{figure}

In social psychology, measurements are developed from proposed concepts or adapted from existing ones to fit different study subjects and contexts \cite{shaughnessy2000research, blackstone2018principles}. With well-studied cohesion concepts and measures available, in Figure \ref{fig: workflow}, we summarize a workflow to adapt existing cohesiveness measures in social psychology to the context of online social networks. Here, ``measures'' particularly refer to questionnaires, which are widely used in psychological studies to assess cohesion \cite{santoro2015measuring, forsyth2021recent, zamecnik2022cohesion}.

We start by selecting a cohesion concept from psychological studies and accessing its \textit{measures}. To ensure consistent positive scoring of cohesiveness, \textit{reversed-word items}, introduced alongside \textit{positively-worded items} to reduce response style bias \cite{suarez2018using}, are corrected. Besides, two common adaptation methods used in group cohesion studies—rewording and removing inapplicable items \cite{spink1994group, bollen1990perceived}—are applied to fit the remaining items into our study context.

\textbf{Cohesion Concept and Measures.} In this study, we adopt a widely accepted definition proposed by \citet{carron1982cohesiveness}, which defines cohesion as \textit{``a dynamic process which is reflected in the tendency for a group to stick together and remain united in the pursuit of its goals and objectives''}. It recognizes cohesion as \textit{multidimensional} and \textit{dynamic} that changes its extent and forms throughout group formation, development, maintenance, and dissolution \cite{carron2000cohesion}.

To measure this concept, a conceptual model has been proposed that identifies two key perceptions binding members to their group: a member's perception of the group as a whole (labeled as \textit{group integration (GI)}) and a member's personal attraction to the group (labeled as \textit{individual attractions to group (ATG)}). Each perception is further divided into task and social aspects, resulting in four correlated constructs that influence group cohesion: \textit{group integration-task (GI-T)}, \textit{group integration-social (GI-S)}, \textit{individual attractions to group-task (ATG-T)}, and \textit{individual attractions to group-social (ATG-S)} \cite{carron1985development}. The \textit{Group Environment Questionnaire (GEQ)} derived from this model consists of 18 items and has demonstrated its reliability and validity in various contexts \cite{carron1985development, blanchard2000group, anderson2017social}. The cohesion score for a group is calculated by averaging individual GEQ responses \cite{carron2003individual}.

\textbf{Measure Adaptation.} Regarding our study context, we focus on the social scales of the GEQ, which emphasize social-oriented motivations driving user interactions ~\cite{syn2015social, dindar2018iusetwitterbecause, alsalem2019they, shin2010analysis}. Following our workflow, the GEQ is adapted into Table \ref{tab: GEQ_coded}, which includes four items for the ATG-S scale and two items for the GI-S scale.

\vspace{-1ex}
\subsection{Comparison} \label{subsec: compre}

Table \ref{tab: cohesiveness_comparison} summarizes whether existing measures (introduced in Section~\ref{sec: cs}) capture the cohesion features outlined in the adapted GEQ (Table~\ref{tab: GEQ_coded}). Observe that both ATG-S and GI-S items incorporate underlying structural information, yet in distinct ways. In particular, ATG-S places greater emphasis on personal opinions,  carrying more psychological insights that influence a user's membership. To reflect this distinction, items are labeled \textsf{``S''} for structural information and \textsf{``P''} for psychological information. ``$\checkmark$'', ``$\triangle$'', and ``$\times$'' denote that an item is fully captured, partially captured, or not captured by existing measures, respectively.

Intuitively, all existing structural cohesiveness measures, which focus on the subgraph interconnections, successfully capture Item 13. While strong topological connections may suggest users' enjoyment, friendship formation, and group engagement (Items 1, 5, 7, and 9), these psychological elements are not explicitly characterized. Similarly, it remains unclear to what extent \textit{keywords cohesiveness}, which captures users' attribute similarity, and \textit{spatial cohesiveness}, which reflects spatial relationships, accurately represent ATG-S items. While \textit{query cohesiveness} incorporates network information to characterize the user's temporal activeness, its representation of psychology-based cohesiveness is still implicit and insufficient. 

Notably, none of the existing metrics capture users' willingness to interact within a group (Item 11). Furthermore, they cannot be combined to capture all items in Table~\ref{tab: GEQ_coded}.

\begin{table}[t] \scriptsize
    \caption{Adapted GEQ for online social networks.}
    \vspace{-3ex}
    \label{tab: GEQ_coded}
    \centering
    \begin{tabular}{p{0.7cm}p{7.2cm}} 
        \toprule
        \textbf{Scale} & \textbf{Item} \\
        \midrule

        \multirow{4}{0.7cm}{ATG-S}
            & 1. I enjoy being a part of the social activities of this group. \\
            & 5. Some of my best friends are in this group.\\
            & 7. I enjoy interactions in my group more than in other groups. \\
            & 9. For me, this group is one of the most important social groups to which I belong.\\
        \hline
        
        \multirow{3}{0.7cm}{GI-S}
            & 11. Members of our group would rather interact with each other than post information individually. \\
            & 13. Our group members frequently interact with each other. \\
    \bottomrule
    \end{tabular}
    \vspace{-6ex}
\end{table}

\begin{table*}[t]\scriptsize
    \centering
    \caption{Comparison of existing CS cohesiveness metrics and social psychology-based cohesion.}
    \vspace{-3ex}
    \label{tab: cohesiveness_comparison}
    
    \begin{tabular}
        {|wc{1 cm}|wc{0.5cm}|wc{0.5cm}|wc{0.5cm}|wc{0.5cm}|wc{2cm}|wc{1.5cm}|wc{1cm}|wc{1.3cm}|wc{1cm}|wc{1cm}|wc{1cm}|}
        \toprule
        \textbf{Item (Info)} & \textbf{$k$-core} & \textbf{$k$-truss} & \textbf{$k$-clique} & \textbf{$k$-ECC} & \textbf{\makecell{Attributed\\ $pkd$-truss community}} & 
        \textbf{\makecell{$(k, \theta)$-community}} & 
        \textbf{\makecell{$(k, p)$-core}} & 
        \textbf{\makecell{$\beta$-temporal core}} & 
        \textbf{\makecell{Keyword\\ cohesiveness}} & 
        \textbf{\makecell{Spatial\\ cohesiveness}} & 
        \textbf{\makecell{Query\\ cohesiveness}} \\
         \midrule
             1 (\textsf{S \& P}) & $\triangle$  & $\triangle$  & $\triangle$ & $\triangle$ & $\triangle$ & $\triangle$ & $\triangle$ & $\triangle$ & $\triangle$ & $\triangle$ & $\triangle$ \\
             5 (\textsf{S \& P}) &  $\triangle$ & $\triangle$ & $\triangle$  & $\triangle$ & $\triangle$ & $\triangle$ & $\triangle$ & $\triangle$ & $\triangle$ & $\triangle$ & $\triangle$ \\
             7 (\textsf{S \& P}) &  $\triangle$ & $\triangle$ & $\triangle$ & $\triangle$ & $\triangle$ & $\triangle$ & $\triangle$ & $\triangle$ & $\triangle$ & $\triangle$ & $\triangle$\\
             9 (\textsf{S \& P}) &  $\triangle$ & $\triangle$ & $\triangle$ & $\triangle$ & $\triangle$ & $\triangle$ & $\triangle$ & $\triangle$ & $\triangle$ & $\triangle$ & $\triangle$\\
             11 (\textsf{S}) &  $\times$ & $\times$ & $\times$ & $\times$ & $\times$ & $\times$ & $\times$ & $\times$ & $\times$ & $\times$ & $\times$\\
             13 (\textsf{S}) &  $\checkmark$ & $\checkmark$ & $\checkmark$ & $\checkmark$ & $\checkmark$ & $\checkmark$ & $\checkmark$ & $\checkmark$ & $\times$ & $\times$ & $\times$\\
        \bottomrule
    \end{tabular}
    \vspace{-2ex}
\end{table*}

\vspace{-1ex}
\section{Psychology-informed Cohesiveness Measures} \label{sec: psy_measures}
The preceding section highlighted the limitations of the generic cohesiveness measures in capturing psychological cohesion in social networks. To further assess and compare their effectiveness, in this section, we propose a set of \textit{psychology-informed measures} that mathematically quantifies group cohesion indicators described in items within Table~\ref{tab: GEQ_coded}. 

Intuitively, for ATG-S items, our formulation focuses on the relationship between individual users and the community, yielding individual scores, which are then averaged into a community score. For GI-S items, the community is treated as a single unit to report a group-level score. All measures are positively scored.

\vspace{-1ex}        
\subsection{Measures for ATG-S Items}

\textbf{Enjoyment Index (EI).} Item 1 emphasizes the user's enjoyment level during social interactions. Enjoyment is often measured concerning positive emotions or needs fulfillment during interactions \cite{syn2015social, kawamichi2016increased, lin2009understanding}. Accordingly, we identify the \textit{enjoyment index} as a cohesiveness indicator, quantified by the cumulative sentiments elicited during interactions.

Empirical findings indicate that an individual's sentiment is shaped by the sentiments and frequency of prior interactions \cite{kawamichi2016increased, bareket2011emotional}. Individuals prioritize emotion-congruent stimuli and respond more quickly to emotional stimuli than neutral ones \cite{niedenthal1994emotion, vinson2014does, yang2014emotional, kissler2009emotion, garcia2016dynamics}. Therefore, we focus on the sentiment polarity of each interaction (positive, neutral, or negative, represented as 1, 0, or -1). We model the impact of sentiments as a \textit{sentiment-aware excitation function}, which quantifies the \textit{excitation degree} of previous interactions for the sentiment of the current interaction. Then the \textit{elicited sentiment} can be modeled as the sentiment conveyed in the interaction scaled by the \textit{excitation degree}.

The sentiment-aware excitation function is constructed based on a basic excitation function, represented as a linear combination of the basic decay functions $\Phi(\cdot)$ \cite{rahimi2023sentihawkes} while introducing a sign function to determine whether past activities $\mathcal{E}^{\prec t}$ enhance or diminish the sentiment of the interaction $e^t$ at time $t$, as shown below.

\vspace{-2ex}
\begin{equation*} \label{eq: excitation} 
    \lambda(t, e^t, \mathcal{E}^{\prec t}) = \lambda_0 + \sum_{e^{t^\prime} \in \mathcal{E}^{\prec t}} sgn(Senti(e^{t^\prime}), Senti(e^t)) \cdot \Phi(t-t^\prime)
\end{equation*}
\vspace{-1ex}
            
Wherein, $\lambda_0$ represents the basic sentiment perception level elicited by a single interaction. We set $\lambda_0=1$ to ensure that the elicited sentiment matches the original sentiment in the absence of prior activities. The function $Senti(\cdot) \in [-1, 1]$ extracts the sentiment from activities, and the sign function is defined as $sgn(Senti(e^{t^\prime}), Senti(e^t))  = Senti(e^{t^\prime}) \cdot Senti(e^t)$. The sentiment extraction methods and decay function selection will be introduced in Section \ref{sec: exp_settings}. To ensure non-negative excitation, we set $\lambda(t, e^t, \mathcal{E}^{\prec t}) = \max(\lambda(t, e^t, \mathcal{E}^{\prec t}), 0)$. Formally, the \textit{elicited sentiment} is defined as below.

\begin{definition}\textbf{[Elicited Sentiment]} \label{def: elicited_sentiment}
{\em Given an interaction activity $e^t$ at time $t$, the elicited sentiment is defined as the interaction sentiment scaled by the sentiment excitation degree as follows:} 
    \begin{equation*} \label{eq: elicit}
        ESenti(t, e^t, \mathcal{E}^{\prec t}) = Senti(e^t) \cdot \lambda(t, e^t, \mathcal{E}^{\prec t}).
    \end{equation*}
\end{definition}

Let $\mathcal{E}_{H, i\cdot}^{t_{cur}} = \{e_{xy}^t \mid e_{xy}^t \in \mathcal{E}_H, \{x, y\} = \{i, j\}, j \neq i, t \leq t_{cur}\}$ be activities where user $u_i$ interacts with other group members up to time $t_{cur}$. Assuming that more recent activities better capture user interests \cite{sarker2019recencyminer}, we weight them by a decay function $\Phi(\cdot)$ and define the \textit{enjoyment index} as follows.
             
\begin{definition}\textbf{[Enjoyment Index (EI)]}{\em Given a graph $\mathcal{G}$, community $H$, current time $t_{cur}$, and a user $u_i \in H$, $u_i$'s EI is defined as the cumulative elicited sentiments generated during interactions between $u_i$ and others in $H$ by $t_{cur}$:\/}

    \vspace{-2ex}
    \begin{equation*} \label{eq: le1}
        \mathbb{EI}_i(\mathcal{G}, H, t_{cur}) = \sum_{e^t \in \mathcal{E}_{H, i\cdot}^{t_{cur}}} ESenti(t, e^t, \mathcal{E}_{H, i\cdot}^{\prec t}) \cdot \Phi(t_{cur}-t), \\
    \end{equation*}

    where $\mathcal{E}_{H, i\cdot}^{\prec t} = \{e_{xy}^{t ^ \prime} \mid e_{xy}^{t^\prime} \in \mathcal{E}_H, \{x, y\} = \{i, j\}, j \neq i, t^\prime < t\}$.
\end{definition}

\textbf{Sentimental Interaction Tendency (SIT).} Now consider Item 5 in Table~\ref{tab: GEQ_coded}. Having best friends in the group indicates cohesiveness, prompting the need to measure friendships among group members. Friendship generally develops through the exchange of information \cite{chan2014friendship} and is characterized by reciprocity, closeness, and emotional attachment \cite{amichai2013friendship, paxton2003structure}. Compared to ordinary friends, best friends share greater intimacy, reciprocity, and stability \cite{berndt1982fairness, roberto1989friendships, weisz2005social}. For simplicity, we propose \textit{sentimental interaction tendency} to characterize friendships developed through cumulative information exchange, with higher values denoting closer ties.

Let $\mathcal{M}_{H, i}^{t_{cur}} = \{u_j \mid u_j \in \mathcal{U}_H, j \neq i, \exists t_1, t_2 \leq t_{cur} : e_{ij}^{t_1}, e_{ji}^{t_2} \in \mathcal{E}_H\}$ be the set of subgroup members that mutually interact with $u_i$. Then, for each $u_j \in \mathcal{M}_{H, i}^{t_{cur}}$, define the interactions between $u_i$ and $u_j$ as $\mathcal{E}_{H, \{i, j\}}^{t_{cur}} = \{e_{xy}^t \mid e_{xy}^t \in \mathcal{E}_H, \{x, y\} = \{i, j\}, t \leq t_{cur} \}$. The cumulative, time-decaying sentiments in these interactions are computed as follows.

\vspace{-2ex}
\begin{equation*}\label{eq: accu}
    Accu_{ij}(\mathcal{G}, H, t_{cur}) = \sum_{e^t \in \mathcal{E}_{H, \{i, j\}}^{t_{cur}}} ESenti(t, e^t, \mathcal{E}_{H, \{i, j\}}^{\prec t}) \cdot \Phi(t_{cur}-t)
\end{equation*}
\vspace{-2ex}

Note that $Accu_{ij}(\cdot)$ is symmetric, \ie $Accu_{ij}(\mathcal{G}, H, t_{cur}) = \linebreak Accu_{ji}(\mathcal{G}, H, t_{cur})$, reflecting the reciprocity of friendship.

 \begin{definition}\textbf{[Sentimental Interaction Tendency (SIT)]} 
{\em Given a graph $\mathcal{G}$, community $H$, current time $t_{cur}$, and a user $u_i \in H$, $u_i$'s SIT is defined as the accumulated elicited sentiments between $u_i$ and each user with whom they have mutual interactions up to $t_{cur}$:\/} 

    \vspace{-2ex}
    \begin{equation*} \label{eq: sit}
        \mathbb{SIT}_i(\mathcal{G}, H, t_{cur}) = \sum_{u_j \in \mathcal{M}_{H, i}^{t_{cur}}} Accu_{ij}(\mathcal{G}, H, t_{cur}).
    \end{equation*}
\end{definition}
\vspace{-1ex}

\textbf{Comparative Enjoyment Degree (CED).} Users often engage in multiple groups on online social networks \cite{tan2015all}. They perceive greater cohesiveness in a specific group when it positively contributes to their enjoyment, as indicated by Items 7 and 9. This leads us to propose the \textit{comparative enjoyment degree} to measure how much more enjoyment a user gets from interactions  within a community than from interactions outside that community.

\begin{definition}\textbf{[Comparative Enjoyment Degree (CED)]}
{\em  Given a graph $\mathcal{G}$, community $H$, current time $t_{cur}$, and a user $u_i \in H$, the CED of $u_i$ at time $t_{cur}$ is defined as the difference between the enjoyment derived from interactions within and outside $H$ by $t_{cur}$ :\/

\small{
\vspace{-2ex}
\begin{equation*}
    \begin{split}
        \mathbb{CED}_i(\mathcal{G}, H, t_{cur}) = 
        &\sum_{e^t \in \mathcal{E}_{H, i\cdot}^{t_{cur}}} ESenti(t, e^t, \mathcal{E}_{H, i\cdot}^{\prec t}) \cdot \Phi(t_{cur}-t) \\
        & - \sum_{e^t \in \mathcal{E}_{\mathcal{G} \setminus H, i\cdot}^{t_{cur}}} ESenti(t, e^t, \mathcal{E}_{\mathcal{G}\setminus H, i\cdot}^{\prec t})\cdot \Phi(t_{cur}-t),
    \end{split}
\end{equation*}
}
}

\em{where $\mathcal{E}_{\mathcal{G} \setminus H, i\cdot}^{t_{cur}} = \{e_{xy}^t \mid e_{xy}^t \in \mathcal{E}_\mathcal{G}, \{x, y\} = \{i, j\}, u_j \notin \mathcal{U}_H, t \leq t_{cur}\}$ represents $u_i$'s interactions with users outside $H$.\/}
\end{definition}

\vspace{-1ex}        
\subsection{Measures for GI-S Items}
\textbf{Group Interaction Preference (GIP).} Item 11 captures the group's preference for group interaction over individual posting, which can be formulated as \textit{group interaction preference (GIP)}. Let $\mathcal{E}_{H, inter}^{t_{cur}} = \{e_{ij}^t \mid e_{ij}^t \in \mathcal{E}_H, j \neq i, t \leq t_{cur} \}$ and $\mathcal{E}_{H}^{t_{cur}} = \{e_{ij}^t \mid e_{ij}^t \in \mathcal{E}_H, t \leq t_{cur}\}$ denote the set of interaction activities and all activities in $H$ up to $t_{cur}$, respectively, where $\mathcal{E}_{H, inter}^{t_{cur}} \subseteq \mathcal{E}_{H}^{t_{cur}}$. The GIP ranges from 0 to 1 and is defined as follows.

\begin{definition}\textbf{[Group Interaction Preference (GIP)]}
{\em Given a graph $\mathcal{G}$, community $H$, and current time $t_{cur}$, GIP is defined as the ratio of group members' interactions to their overall activities by $t_{cur}$:\/}
    \begin{equation*}
        \mathbb{GIP}(\mathcal{G}, H, t_{cur}) = \frac{|\mathcal{E}_{H, inter}^{t_{cur}}|}{|\mathcal{E}_{H}^{t_{cur}}|}.
    \end{equation*}
\end{definition}

\textbf{Group Interaction Density (GID).} Lastly, Item 13 emphasizes the interplay between member relationships and interaction frequency. We quantify it by extending the \textit{weighted density} \cite{liu2009complex} with a temporal normalization. Let the observation window be $W_{t_{cur}} = (t_{cur} - t_{0}) / \tau_t$, where $t_0$ is the start of observation and $\tau_t$ is the time-unit (\eg day, week, month). We call the resulting metric the \textit{group interaction density (GID)}.

\begin{definition}\textbf{[Group Interaction Density (GID)]}
{\em Given a graph $\mathcal{G}$, community $H$, and current time $t_{cur}$, GID measures the interaction frequency per possible user pair per unit time within $H$ by $t_{cur}$:}\/
    \begin{equation*}
        \mathbb{GID}(\mathcal{G}, H, t_{cur}) = \frac{|\mathcal{E}_{H, inter}^{t_{cur}}|}{|\mathcal{U}_H| \cdot (|\mathcal{U}_H| - 1) \cdot W_{t_{cur}}}.
    \end{equation*}
\end{definition}

\noindent\textbf{Remark.} Note that these five measures for assessing cohesion can be merged into a single \textit{cohesiveness score} through an aggregated linear combination of weights. However, our work does not aim to calculate such a score. Instead, the evaluation shall focus on analyzing the quality of results produced by existing CS techniques \wrt{ each of these measures}, enabling us to get a fine-grained understanding of how they perform against \emph{each} item in Table~\ref{tab: GEQ_coded}.

\vspace{-1ex}
\section{Experimental Settings} \label{sec: exp_settings} 
In this section, we introduce the datasets, representative community search algorithms, evaluation framework, and parameter settings used to set up the experiments. The codebase is available at~\cite{cohesion_codebase}.
    
\vspace{-1ex}           
\subsection{Dataset}

Due to the lack of sentiment data in most publicly available datasets (\eg SNAP \cite{snapnets}), we constructed four real-world datasets from \textit{X} (formerly \textit{Twitter}) to facilitate evaluation. Table \ref{tab: net_stats} summarizes their topological characteristics, where $|\mathcal{U}|$, $|\mathcal{E}|$, and $|\mathcal{T}|$ denote the numbers of vertices, edges, and edge timestamps, respectively. $Deg_{avg}$ represents the average node degree, with self-loops and parallel edges counted.

\begin{table}[t] \scriptsize
    \caption{Dataset statistics.}
    \vspace{-3ex}
    \label{tab: net_stats}
    \centering
    
    \begin{tabular}{p{1.9cm}wc{0.85cm}wc{0.85cm}wc{0.85cm}wc{0.85cm}wc{1cm}} 
        \toprule
        \textbf{Dataset} & $|\mathcal{U}|$ & $|\mathcal{E}|$ & $|\mathcal{T}|$ & Density & $Deg_{avg}$\\
        
        \midrule
        
        \textbf{BTW}17 & 7,721 & 12,839 & 11,890 & 0.0002 & 3.33\\
        \textbf{C}hicago\_\textbf{C}OVID & 4,917 & 246,633 & 246,001 & 0.0102 & 100.32\\
        \textbf{C}rawled\_Dataset\textbf{144} & 10,388 & 119,836 & 116,639 & 0.0011 & 23.07\\
        \textbf{C}rawled\_Dataset\textbf{26} & 21,263 & 108,538 & 104,575 & 0.0002 & 10.21\\
        
        \bottomrule
   \end{tabular}
   \vspace{-3ex}
\end{table}

\textbf{Data Collection.} The \textsf{BTW17} dataset contains tweets related to the 19th German election campaigns, with over 1.2 million tweets from active users \cite{BTW17_dataset}. The \textsf{Chicago\_COVID} dataset contains tweets related to the COVID-19 pandemic and geolocated near Chicago \cite{chicago_dataset}. Due to its large volume, we sampled 125k tweets for this study. Besides, we collected \textsf{Crawled\_Dataset144} and \textsf{Crawled\_Dataset26} datasets in December 2022 using the official API \cite{twitter_academic_api}. Starting with single users randomly selected from real-time tweets, we employed snowball sampling to gather the users' original tweets and associated replies (conversation utterances). Newly identified users were recorded for future data collection. This process resulted in over 200k original tweets for each dataset. For all four datasets, we extracted metadata for each user, including \textit{user\_id}. The tweet metadata includes \textit{conversation\_id}, \textit{tweet\_id}, \textit{user\_id}, \textit{in\_reply\_to\_user\_id}, \textit{tweet\_text}, and \textit{timestamp}. The number of users and tweet data elements varied across datasets. We retained only English tweets, removed retweets, quoted tweets, duplicates, and empty entries, and cleaned the tweet text by removing newlines and URLs.

\begin{figure}[t]
    \centering
    \includegraphics[scale=0.13]{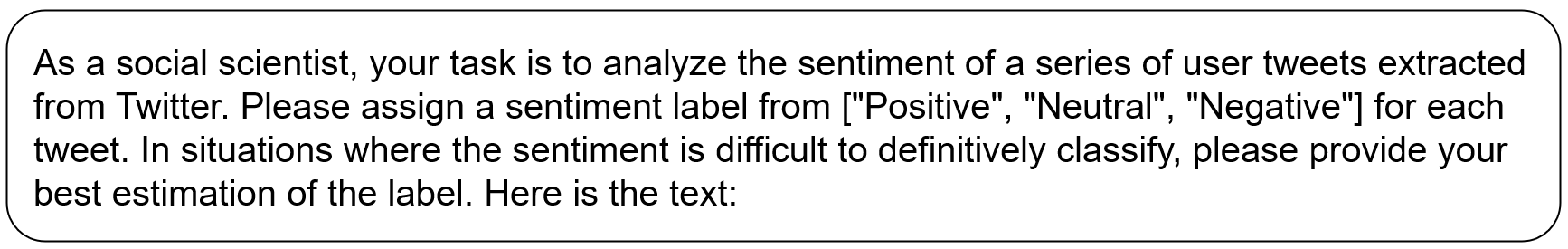}
    \vspace{-2ex}
    \caption{Prompt for sentiment analysis.}
    \vspace{-3ex}
    \label{fig: prompt}
\end{figure}

\textbf{Sentiment Analysis.} The preprocessed tweets were then subjected to a sentiment analysis task to determine their sentiment scores. Following common practice, we treated the task as a three-class classification problem \cite{chang2024survey}. To ensure superior analysis results, we adopted \textit{Llama3-8B}, an open-source Large Language Model (LLM) from Meta \cite{llama3modelcard}, and used a prompt-based approach. The prompt was adapted from \cite{kheiri2023sentimentgpt} to assign sentiment labels (see Figure \ref{fig: prompt}), which were mapped to 1, 0, and -1, respectively. Both \textit{top\_p} and \textit{temperature} were set to 0.01 for more deterministic responses. Only the tweets that were successfully assigned labels were retained.

\textbf{Network Construction.} Networks were constructed based on the \textit{reply} relationships (\ie who replies to whom), including interaction and self-posting tweets (\ie initial posts not directed at specific users). Isolated nodes were removed. Since the constructed networks consist of components with their size distribution highly skewed, only the largest component in each network was considered, resulting in the networks reported in Table \ref{tab: net_stats}.

\vspace{-1ex}           
\subsection{Representative CS Algorithms} \label{subsec: algos}
We selected representative CS algorithms based on the following criteria: (a) their codebases are publicly available; (b) they cover the two categories introduced in Section~\ref{sec: cs}; and (c) they consider various types of information to characterize cohesiveness, potentially capturing different items in the adapted GEQ (Table \ref{tab: cohesiveness_comparison}). As a result, we chose three  $k$-core-based algorithms (\textsf{ALS}, \textsf{WCF-CRC}, \textsf{CSD}), three $k$-truss-based algorithms (\textsf{ST-Exa}, \textsf{Repeeling+}, \textsf{I2ACSM}), and two unsupervised learning algorithms (\textsf{TransZero-LS}, \textsf{TransZero-GS}), as summarized in Table \ref{tab: algo}. Although \cite{das2021attribute} could be a useful technique for our study (based on Table~\ref{tab: cohesiveness_comparison}), unfortunately, its codebase is not available. The last column in Table \ref{tab: algo} indicates whether the algorithms use ground truth for evaluation, with most of them not using it. Note that the ground truth adopted in the three approaches is not based on group cohesion theory and lacks data elements (\eg temporality and sentiments) to calculate our proposed measures. Therefore, we cannot exploit them for our evaluation. We now briefly describe these algorithms.

\textsf{ALS} identifies the \textit{$\beta$-temporal proximity core} with the largest $\beta$ in a temporal graph \cite{lin2024qtcs}, integrating temporal proximity with structural cohesiveness seamlessly. It requires specifying the teleportation probability $\alpha$ to compute temporal proximity. \textsf{WCF-CRC} searches on dynamic networks, represented as a sequence of time-variant weighted graph instances ($\mathbb{G} = {G_{t_1}, G_{t_2}, \ldots, G_{t_T}}$) \cite{tang2022reliable}. It aims to find a \textit{Time Interval based ($\theta, k$)-core Reliable Community (CRC)} with the highest \textit{reliability score}, where a \textit{($\theta, k$)-core} in a $G_t$ is a $k$-core where all edge weights are at least $\theta$. \textsf{CSD} finds a \textit{$(k, l)$-core} in the directed graph, where $k$ and $l$ denote the minimum in-degree and out-degree of subgraph nodes \cite{chen2021efficient}.

\textsf{ST-Exa} identifies a subgraph with the largest \textit{min-support} based on the query node, while satisfying the size constraint $[l, h]$. It is parameter-free, eliminating the need to specify the cohesiveness constraint. \textsf{Repeeling+} is a peeling-based algorithm for finding a maximal \textit{$D$-truss} containing the query node in the snapshot digraph $G^t$ \cite{liao2024truss}. $G^t$ is induced from a streaming directed graph using a sliding window length $\tau$ and a \textit{stride} $\beta$. The algorithm's parameters need to be tuned according to the dataset characteristics to ensure valid results. \textsf{I2ACSM} aims to search an attributed \textit{$pkd$-truss} community that includes the query node in an Attributed and Weighted Graph (AWG) \cite{xie2021effective}, ensuring that any vertex in the $k$-truss subgraph is within distance $d$ from the query node and the maximum path influence from the query node to any subgraph node exceeds $p$~\cite{xie2021effective}.

 \textsf{TransZero-LS} and \textsf{TransZero-GS} are proposed under \textit{TransZero}, a two-phase pre-trained Graph Transformer-based community search framework that aims to find communities without ground-truth data \cite{wang2024efficient}. In the \textit{offline} phase, \textit{CSGphormer} is pre-trained to learn node representations that encode both community information and graph topology. In the \textit{online} phase, the community score is calculated by measuring the representation similarity between the query and the remaining graph nodes, and an \textit{expected score gain function} is introduced to search communities with maximum expected score gain (\textit{ESG}). \textsf{TransZero-LS} and \textsf{TransZero-GS} are two heuristic algorithms based on \textit{ESG} for community identification.

\begin{table}[t] \scriptsize
    \caption{Representative community search algorithms.}
    \vspace{-3ex}
    \label{tab: algo}
    \centering
    
    \begin{tabular}{p{1.4cm}p{1.7cm}p{2.4cm}p{1.4cm}} 
        \toprule
        \multicolumn{1}{m{1.3cm}}{\textbf{Category}} & \textbf{Algorithm} & \textbf{Graph Type} & \textbf{Ground Truth} \\
        \midrule
        \multirow{6}{*}{Traditional} 
        & \multirow{1}{*}{\textsf{ALS} \cite{lin2024qtcs}} & Undirected Temporal & No \\ \cline{2-4} 
        & \multirow{1}{*}{\textsf{WCF-CRC} \cite{tang2022reliable}} & Dynamic & No \\ \cline{2-4} 
        & \multirow{1}{*}{\textsf{CSD} \cite{chen2021efficient}} & Directed Multigraph \footnotemark & No \\ \cline{2-4} 
        & \multirow{1}{*}{\textsf{ST-Exa} \cite{zhang2023size}} & Undirected & No \\ \cline{2-4} 
        & \multirow{1}{*}{\textsf{Repeeling+} \cite{liao2024truss}} & Streaming Directed & No \\ \cline{2-4} 
        & \multirow{1}{*}{\textsf{I2ACSM} \cite{xie2021effective}} & Attributed and Weighted & Yes \\
        \midrule
        \multirow{2}{*}{Learning-based}
        & \multirow{1}{*}{\textsf{TransZero-LS} \cite{wang2024efficient}} &  Undirected & Yes \\ \cline{2-4}
        & \multirow{1}{*}{\textsf{TransZero-GS} \cite{wang2024efficient}} &  Undirected & Yes \\
        \bottomrule
   \end{tabular}
    \vspace{-4ex}
\end{table}

\footnotetext{The original paper referred to the graphs as directed, but upon closer examination, we found they are more accurately characterized as directed multigraphs.}

\begin{table*}[t] \scriptsize
    \caption{Parameter settings in \textsf{Repeeling+} algorithm.}
    \vspace{-3ex}
    \label{tab: streaming_parameters}
    \centering
    
    \begin{tabular}{wc{2cm}|wc{6cm}|wc{3.5cm}|wc{1.3cm}|wc{1.3cm}} 
        \toprule
        \textbf{Dataset} & \textbf{$|Window|$} & \textbf{$|Stride|/|Window|$(\%)} & \textbf{$k_c$} & \textbf{$k_f$}\\
        
        \midrule
        \textsf{BTW17} & \{10k, 50k, 100k, 500k, 1M, 5M, 10M, 50M\} & \{0.01, 0.05, 0.1, 0.5, 1, 5, 10\} & \{0, 1\} & \{0, 1\}\\
        \textsf{Chicago\_COVID} & \{10k, 50k, 100k, 500k, 1M, 5M, 10M, 50M, 100M\} & \{0.01, 0.05, 0.1, 0.5, 1, 5, 10\} & \{0, 1, 2, 3\} & \{0, 1, 2, 3\}\\
        \textsf{Crawled\_Dataset144} & \{100k, 500k, 1M, 1.5M, 2M, 2.5M\} & \{0.001, 0.005, 0.01, 0.05, 0.1\} & \{0, 1, 2, 3\} & \{0, 1, 2, 3\}\\
        \textsf{Crawled\_Dataset26} & \{100k, 500k, 1M, 1.5M, 2M, 2.5M\} & \{0.001, 0.005, 0.01, 0.05, 0.1\} & \{0, 1, 2, 3\} & \{0, 1, 2, 3\}\\
        \bottomrule
    \end{tabular}
    \vspace{-2ex}
\end{table*}

\begin{table*}[t] \scriptsize
    \caption{Evaluation of structural cohesiveness of CS solutions across different datasets.}
    \vspace{-3ex}
    \label{tab: structural}
    \centering
    \resizebox{\textwidth}{!}{
    \begin{tabular}{wc{2cm}| wc{0.5cm}wc{0.5cm}wc{0.8cm}wc{0.5cm}|wc{0.5cm}wc{0.5cm}wc{0.8cm}wc{0.5cm}|wc{0.5cm}wc{0.5cm}wc{0.8cm}wc{0.5cm}|wc{0.5cm}wc{0.5cm}wc{0.8cm}wc{0.5cm}}
    \toprule
    \multirow{2}{*}{\textbf{Algorithm}} & \multicolumn{4}{c}{\textbf{BTW17}} & \multicolumn{4}{c}{\textbf{Chicago\_COVID}} & \multicolumn{4}{c}{\textbf{Crawled144}}  & \multicolumn{4}{c}{\textbf{Crawled26}}\\
       & $\mathbf{d}$  & Size & $Deg_{min}$  & $Q_{hit}$
       & $\mathbf{d}$  & Size & $Deg_{min}$  & $Q_{hit}$
       & $\mathbf{d}$  & Size & $Deg_{min}$  & $Q_{hit}$
       & $\mathbf{d}$  & Size & $Deg_{min}$  & $Q_{hit}$\\
    \midrule
    \textsf{ALS} & 9.24 & 5029.18 & 1.00 & 100 & 16.21 & 3998.33 & 1.18 & 100 & 15.36 & 9385.31 & 1.00 & 100 & 14.87 & 17043.60 & 1.15 & 100\\
    \textsf{WCF-CRC} & 6.88 & 242.80 & 2.19 & 45 & 10.71 & 280.91 & 8.77 & 52 & 5.64 & 451.59 & 11.12 & 93 & 9.86 & 287.89 & 3.93 & 39\\
    \textsf{CSD} & 5.67 & 24.00 & 0.67 & 3 & 4.16 & 19.70 & 59.52 & 10 & 7.70 & 204.22 & 11.14 & 34 & 6.15 & 125.29 & 8.84 & 20 \\
    \textsf{ST-Exa} & 4.49 & 44.96 & 2.20 & 100 & 5.60 & 44.88 & 13.09 & 100 & 3.01 & 44.58 & 26.62 & 100 & 3.60 & 44.74 & 5.59 & 100\\
    \textsf{Repeeling+} & 8.46 & 3228.53 & 1.49 & 100 & 6.80 & 998.94 & 18.20 & 100 & INF & INF & INF & INF & INF & INF & INF & INF\\
    \textsf{I2ACSM} & 0.55 & 1.61 & 1.14 & 100 & 1.22 & 3.48 & 101.28 & 100 & 0.97 & 2.41 & 5.95 & 100 & 0.42 & 1.42 & 4.37 & 100 \\
    \textsf{TransZero-LS} & 10.00 & 3031.42 & 1.00 & 100 & 10.07 & 458.28 & 12.22 & 100 & 11.72 & 4536.41 & 1.11 & 100 & 9.10 & 2367.10 & 1.36 & 100 \\
    \textsf{TransZero-GS} & - & - & - & 0 & - & - & - & 0 & - & - & - & 0 & - & - & - & 0\\
    \bottomrule
    \end{tabular}}
\end{table*}

\vspace{-1ex}           
\subsection{The CHASE Framework}

\begin{figure}[t]
    \centering
    \includegraphics[scale=0.14]{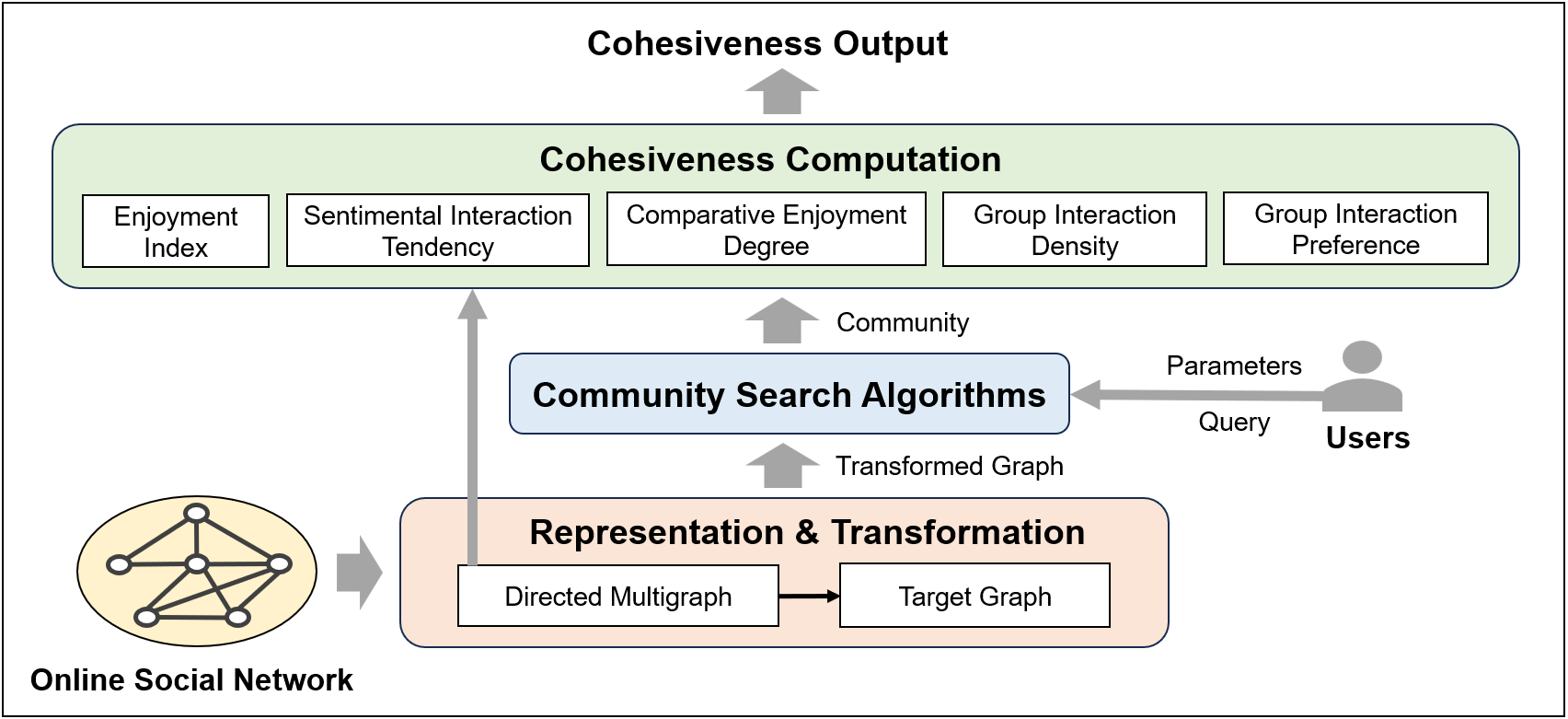}
    \vspace{-2ex}
    \caption{The \textsf{CHASE} framework.}
    \label{fig: framework}
    \vspace{-3ex}
\end{figure}

Figure \ref{fig: framework} presents our framework coined \textsf{CHASE} (\underline{C}o\underline{H}esiveness ev\underline{A}luation for \underline{S}ocial n\underline{E}tworks) for evaluating CS algorithms, which includes query generation, parameter selection, graph transformation, community mapping, and cohesiveness evaluation.

We first generate $Q$ queries ($Q=100$ in our study) per dataset, each with a single query node. Queries are randomly selected from nodes in the top 50\% by degree. Next, for each algorithm, datasets are transformed into algorithm-specific graph types to support its community retrieval. The varying parameter values are chosen from the original paper. For dataset-sensitive parameters, we manually test values to ensure meaningful results. For algorithms with multiple parameters, all value combinations are considered.

Using the query nodes and parameters, we obtain $Q$ communities from the transformed graph by applying the algorithm. To ensure fairness, each community is mapped back to the directed multigraphs for evaluation (\ie cohesiveness computation). In other words, a mapped community is extracted according to nodes in the initial community and includes all edges among them. For algorithms that provide multiple solutions for a single query under various parameter combinations, the cohesiveness of these communities is averaged before being compared with other techniques.

\vspace{-2ex}           
\subsection{Parameter Settings}
The parameter settings for the \textsf{ALS}, \textsf{I2ACSM}, \textsf{TransZero-LS}, and \textsf{TransZero-GS} follow the original papers, with influence values and attributes for \textsf{I2ACSM} generated following the methods mentioned in \cite{xie2021effective}. Dataset-dependent parameters for the remaining algorithms are adjusted for meaningful results. Specifically, for \textsf{WCF-CRC}, the query time intervals for \textsf{BTW} and the rest of the datasets are set to $t \in \{2, 3\}$ and $t \in \{4, 6, 8, 10\}$, respectively, based on the features of transformed datasets. In \textsf{CSD}, $k$ and $l$ are varied from 1 to 10. The size constraint $[l, h]$ in \textsf{ST-Exa} ranges from $[1, 10]$ to $[71, 80]$. The window size, stride size, $k_c$, and $k_f$ in \textsf{Repeeling+} vary across datasets; these parameters for each dataset are manually tested and reported in Table~\ref{tab: streaming_parameters}.

We use an exponential decay function, commonly applied in fields such as biology, physics, and network analysis \cite{hobbie2007exponential, karagiannis2007power, stehle2010dynamical}, to calculate cohesiveness measures. The function is defined as $\Phi(t_{cur} - t) = e^{-\lambda(t_{cur}-t)}$, where $t_{cur}$ is the latest timestamp in each dataset. The decay rate $\lambda$ is selected from $\{0.0001, 0.0005, 0.001, 0.005, 0.01\}$, with $0.0001$ as the default. When calculating GID values, we note that the four datasets span different observation lengths and that no standard choice of $\tau_{t}$ exists. Since $W_{t_{cur}}$ uniformly scales all GID scores of communities within a dataset without affecting their relative ranking, we omit $W_{t_{cur}}$ for clarity and simplicity. As users in the four datasets lack consistent information except their ids, we use only \textit{user\_id} as the node feature during \textit{CSGphormer} training.

\begin{figure*}[t]
    \setlength{\abovecaptionskip}{0.2cm}
    \begin{minipage}[t]{0.24\linewidth}
        \centering
        \includegraphics[width=\linewidth]{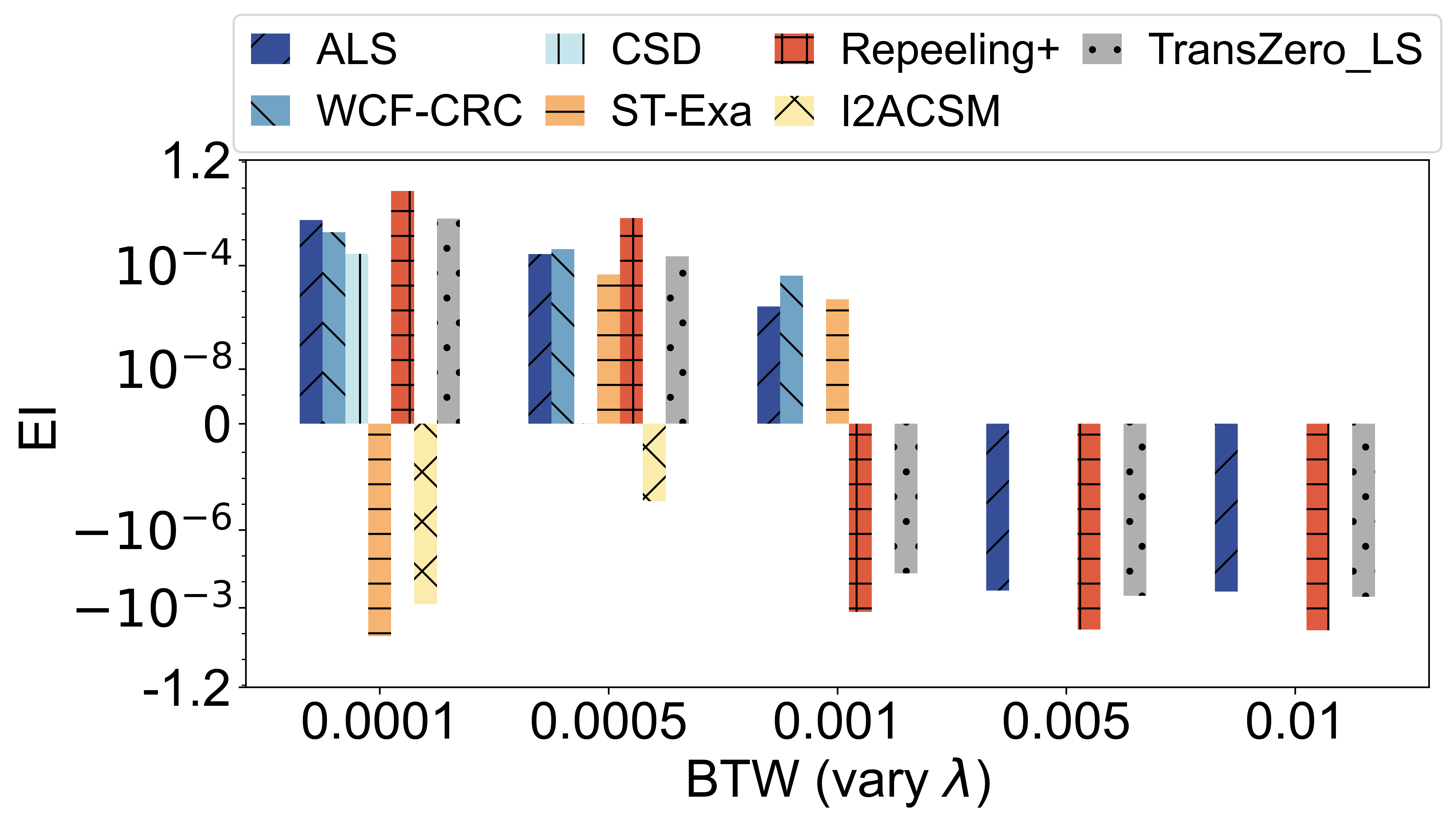}
        \label{fig: EI_btw_lambda}
    \end{minipage} 
    \begin{minipage}[t]{0.24\linewidth}
        \centering
        \includegraphics[width=\linewidth]{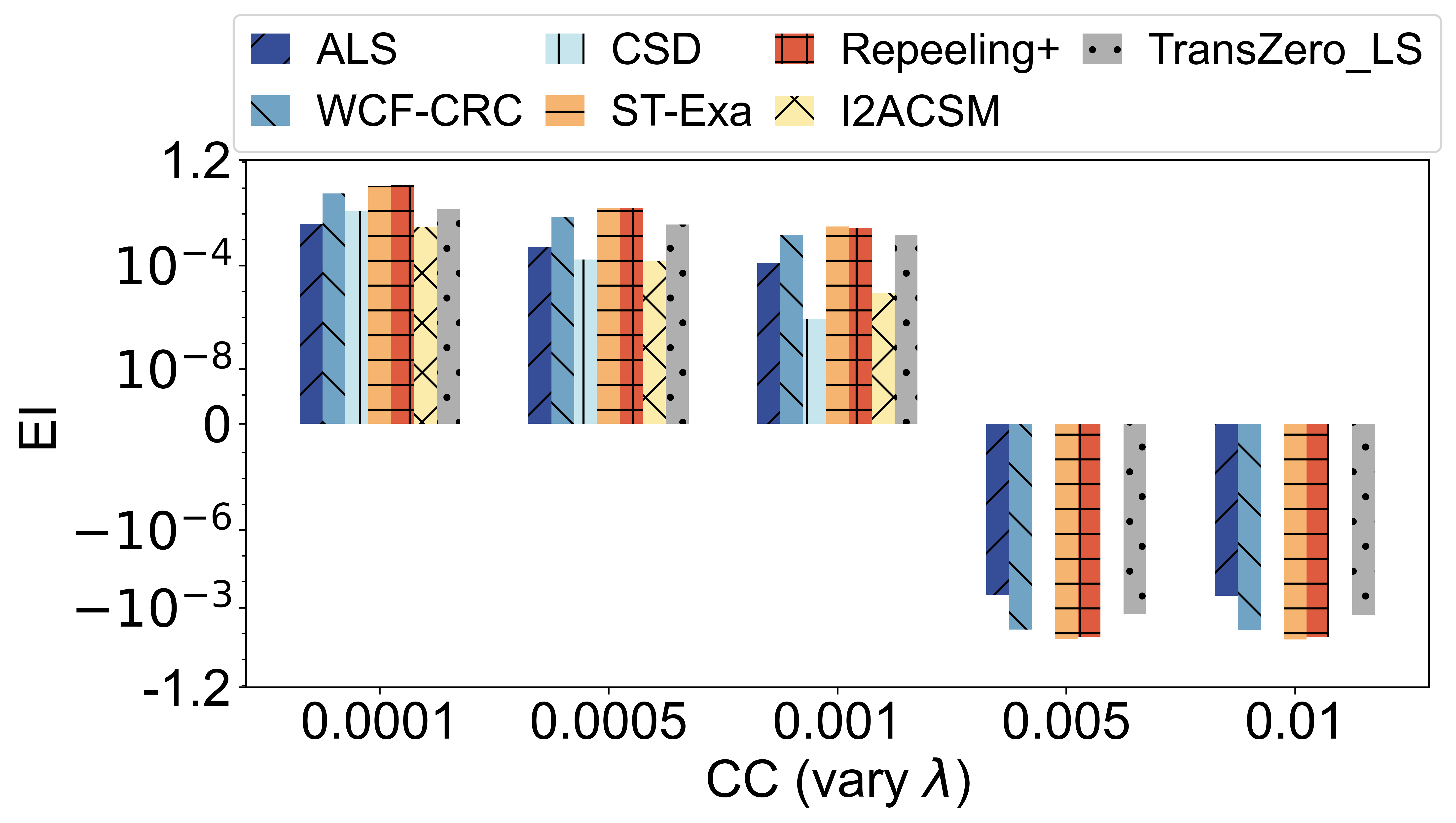}
        \label{fig: EI_cc_lambda}
    \end{minipage}
    \begin{minipage}[t]{0.24\linewidth}
        \centering
        \includegraphics[width=\linewidth]{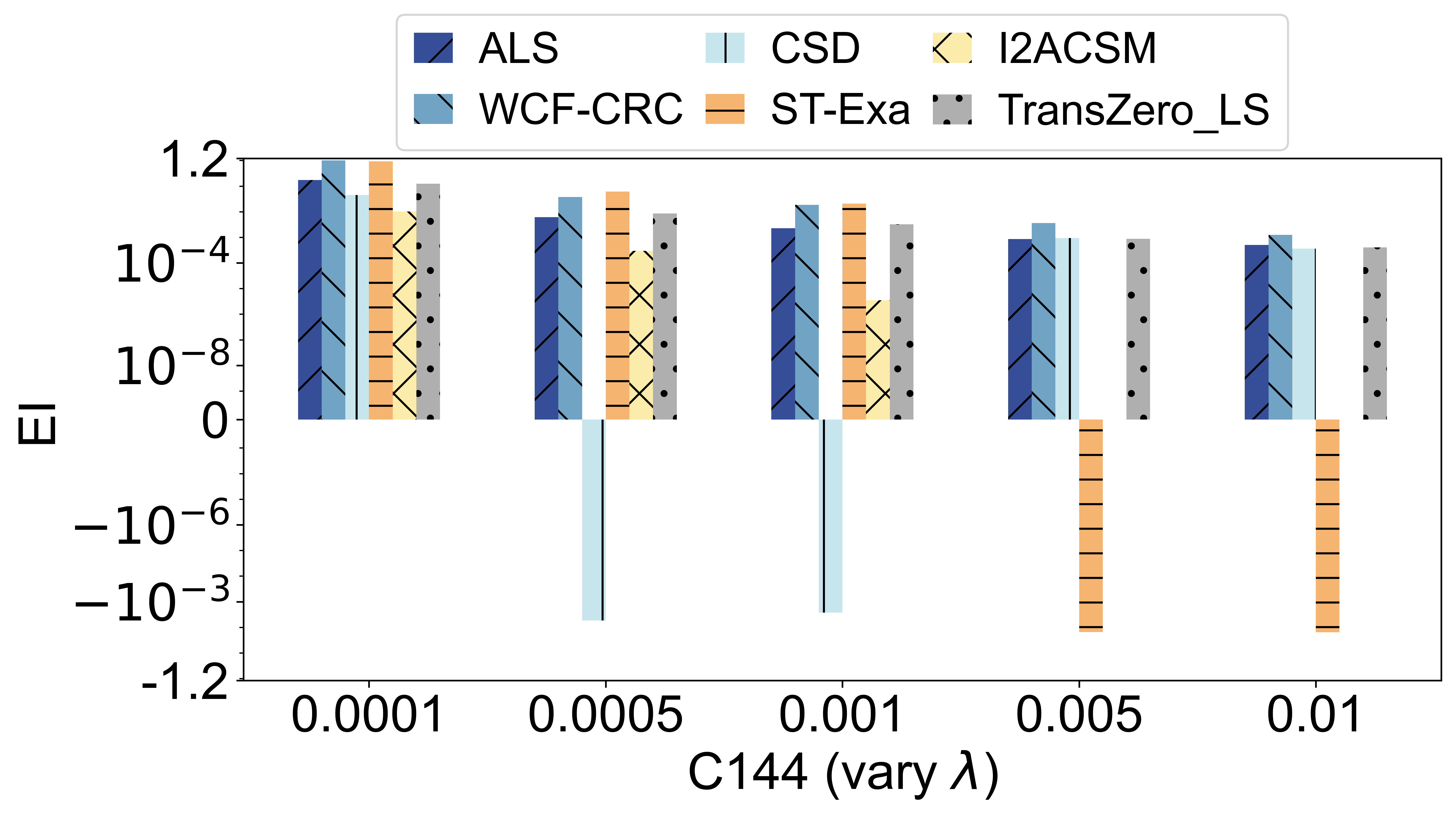}
        \label{fig: EI_c144_lambda}
    \end{minipage}
    \begin{minipage}[t]{0.24\linewidth}
        \centering
        \includegraphics[width=\linewidth]{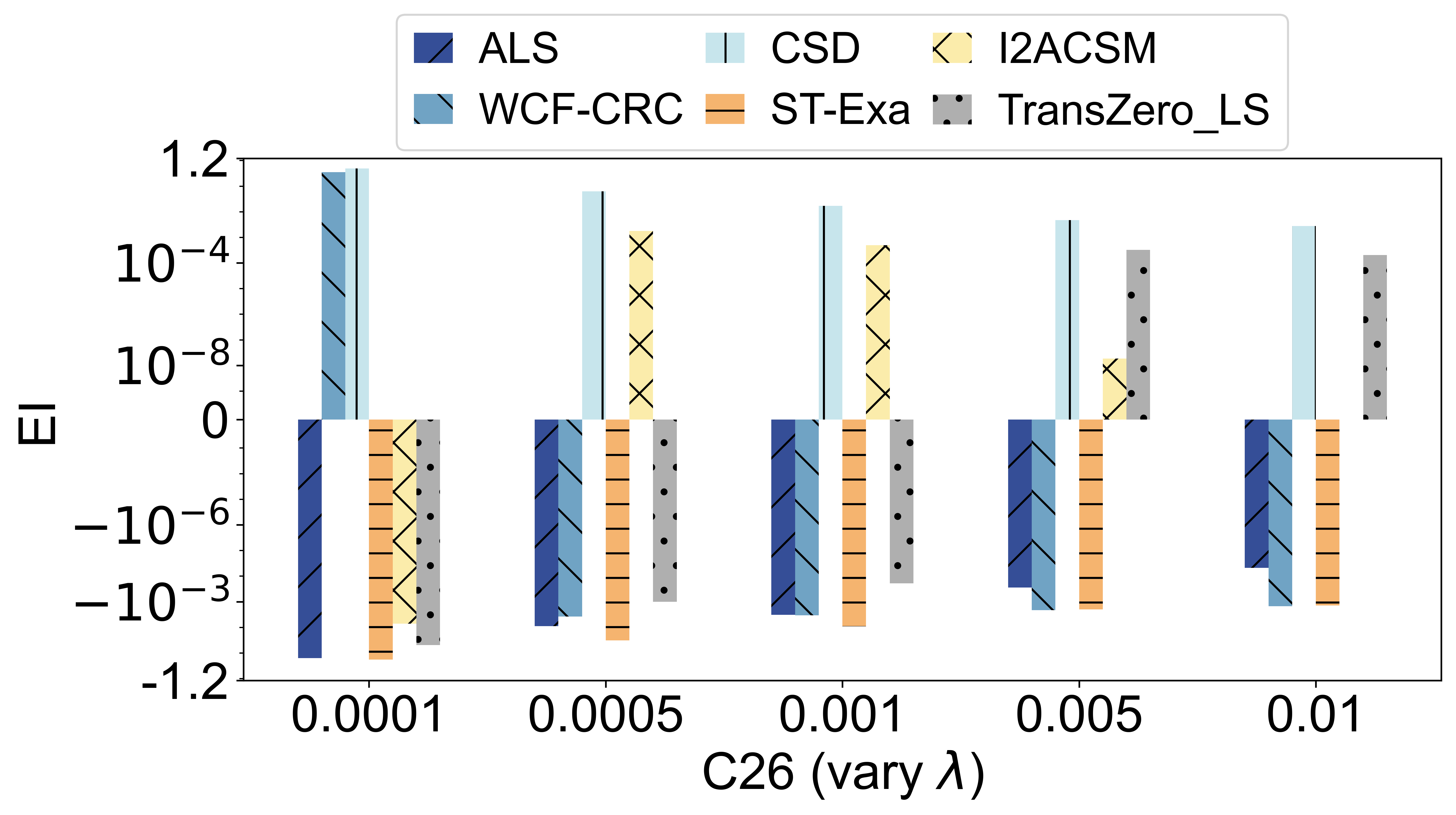}
        \label{fig: EI_c26_lambda}
    \end{minipage}
    
    \begin{minipage}[t]{0.24\linewidth}
        \centering
        \vspace{-3ex}
        \includegraphics[width=\linewidth]{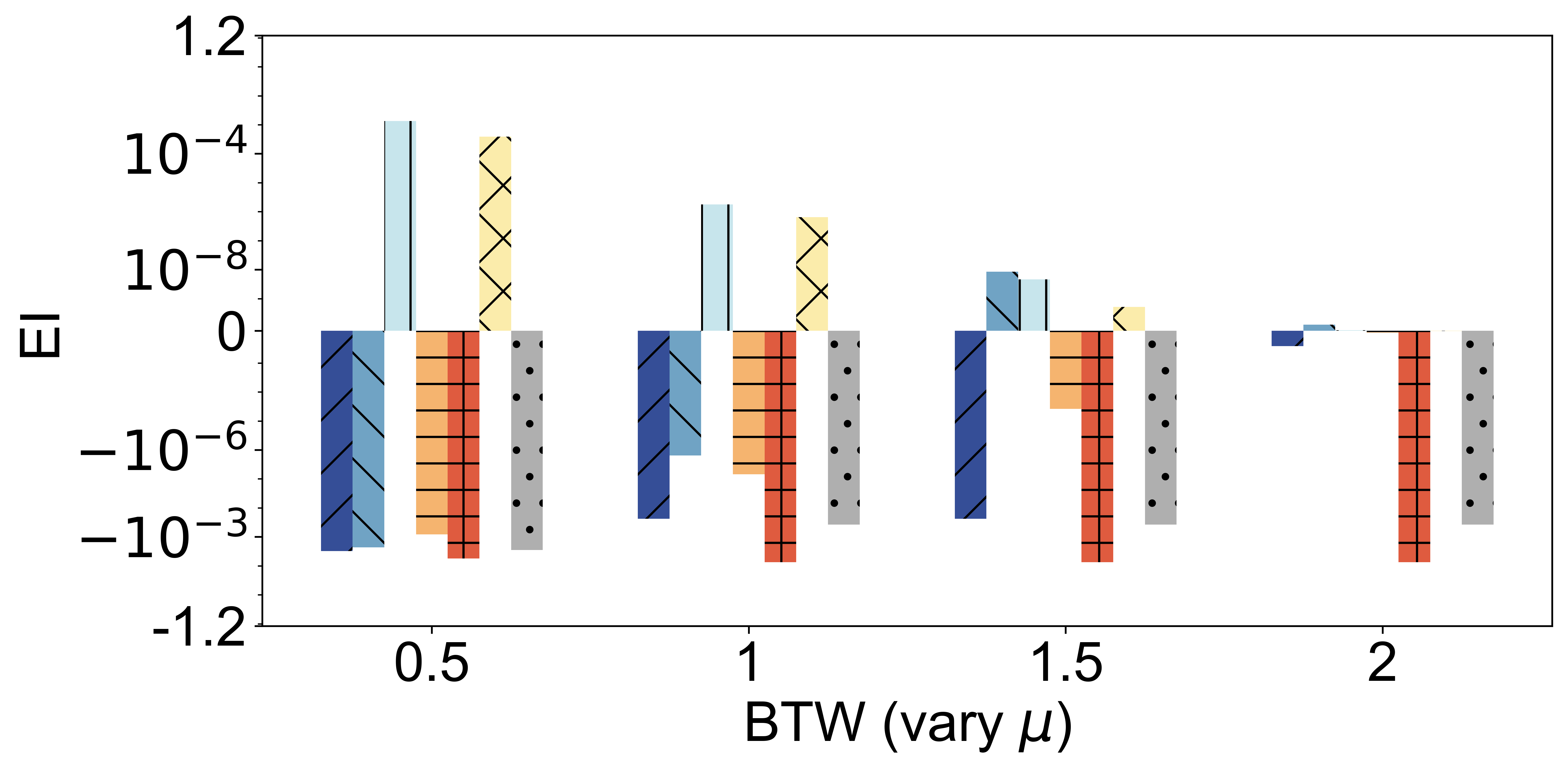}
        \label{fig: EI_btw_mu}
    \end{minipage}
    \begin{minipage}[t]{0.24\linewidth}
        \centering
        \vspace{-3ex}
        \includegraphics[width=\linewidth]{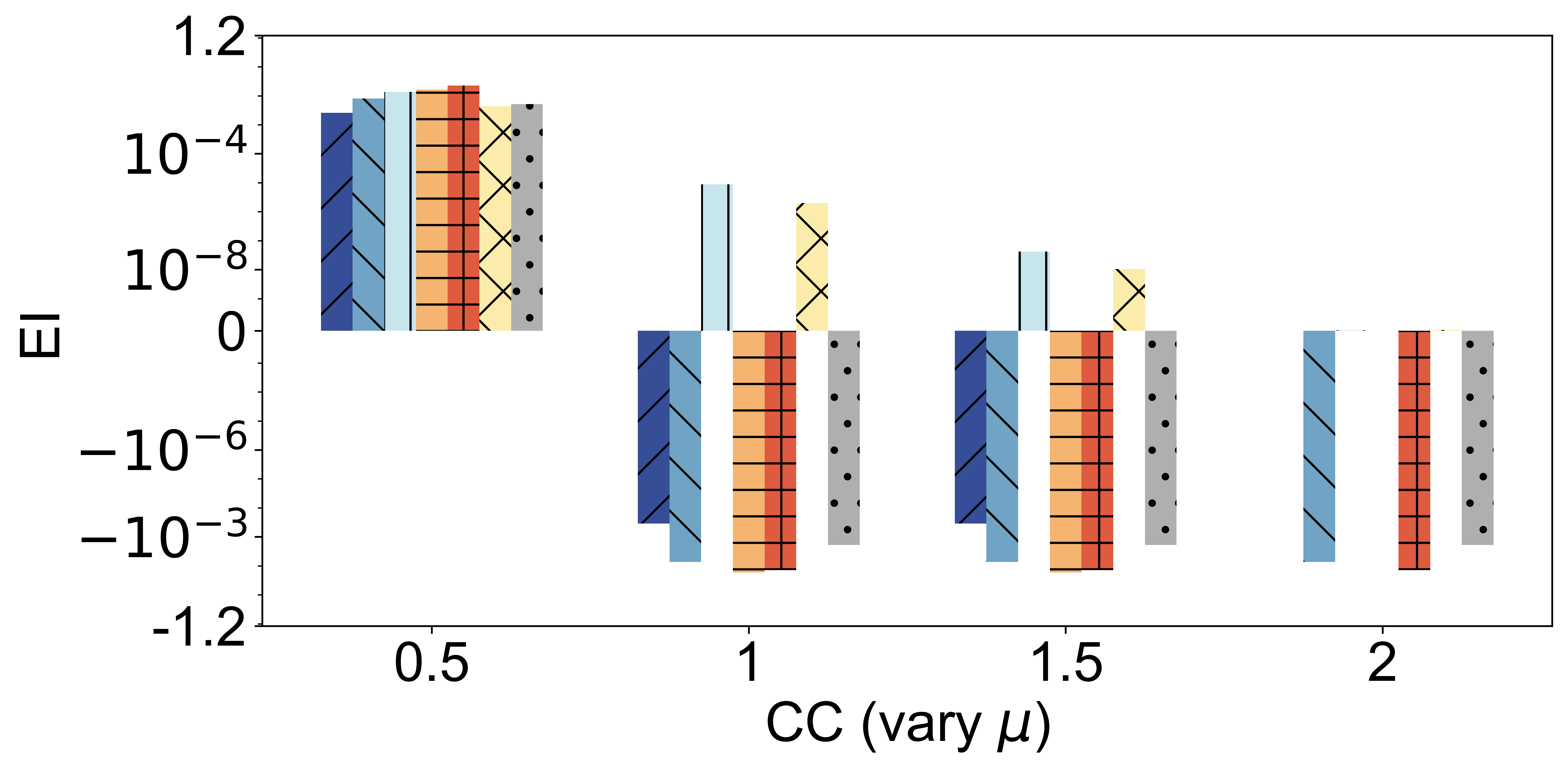}
        \label{fig: EI_cc_mu}
    \end{minipage}
    \begin{minipage}[t]{0.24\linewidth}
        \centering
        \vspace{-3ex}
        \includegraphics[width=\linewidth]{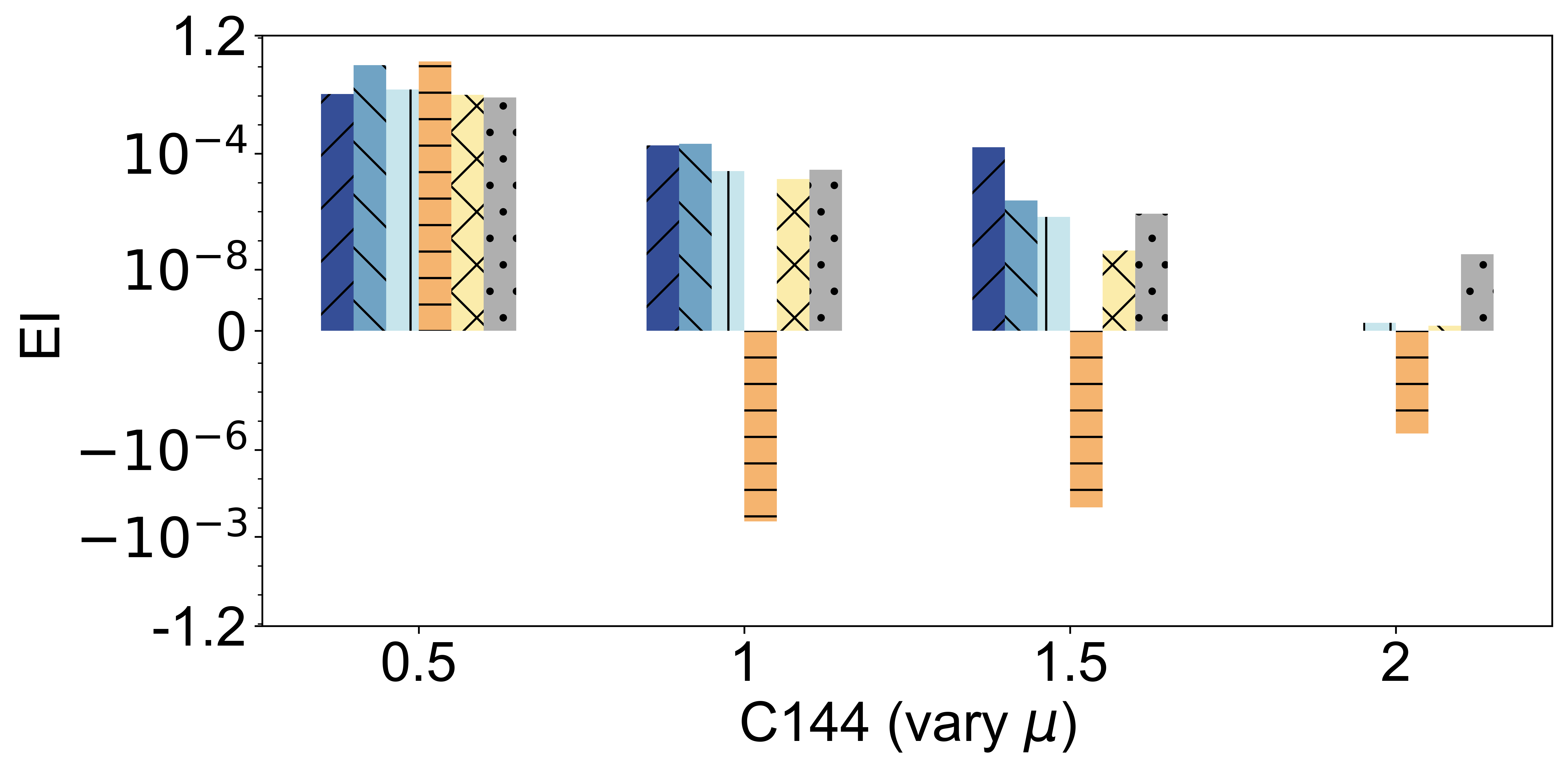}
        \label{fig: EI_c144_mu}
    \end{minipage}
    \begin{minipage}[t]{0.24\linewidth}
        \centering
        \vspace{-3ex}
        \includegraphics[width=\linewidth]{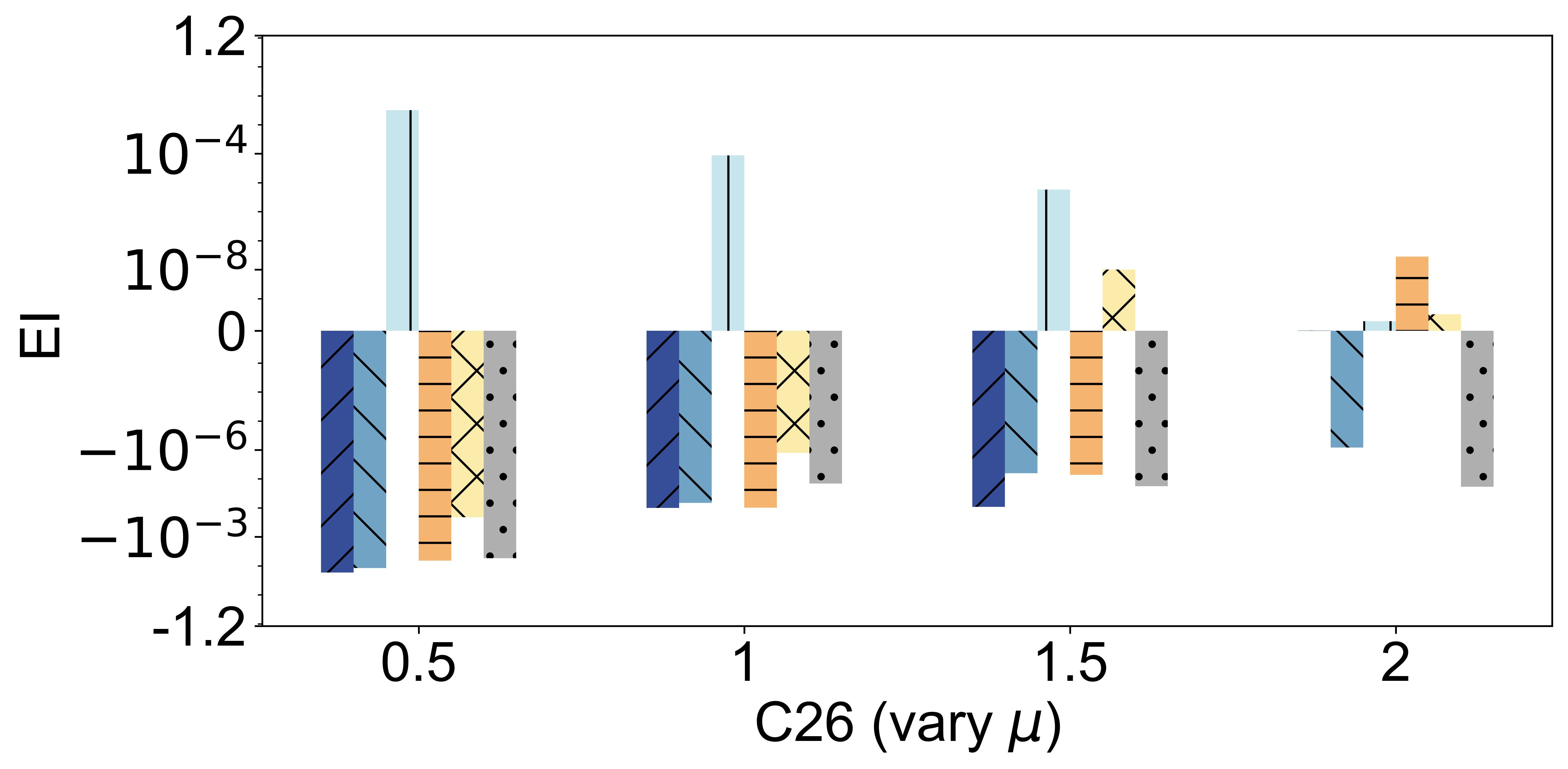}
        \label{fig: EI_c26_mu}
    \end{minipage}
    \vspace{-3ex}
    \caption{EI of communities (top row: community cohesiveness; bottom row: impact of time-decay functions). }
    \label{fig: EI}
    \vspace{-2ex}
\end{figure*}

\begin{figure*}[t]  
    \setlength{\abovecaptionskip}{0.2cm}
    \begin{minipage}[t]{0.24\linewidth}
        \centering
        \includegraphics[width=\linewidth]{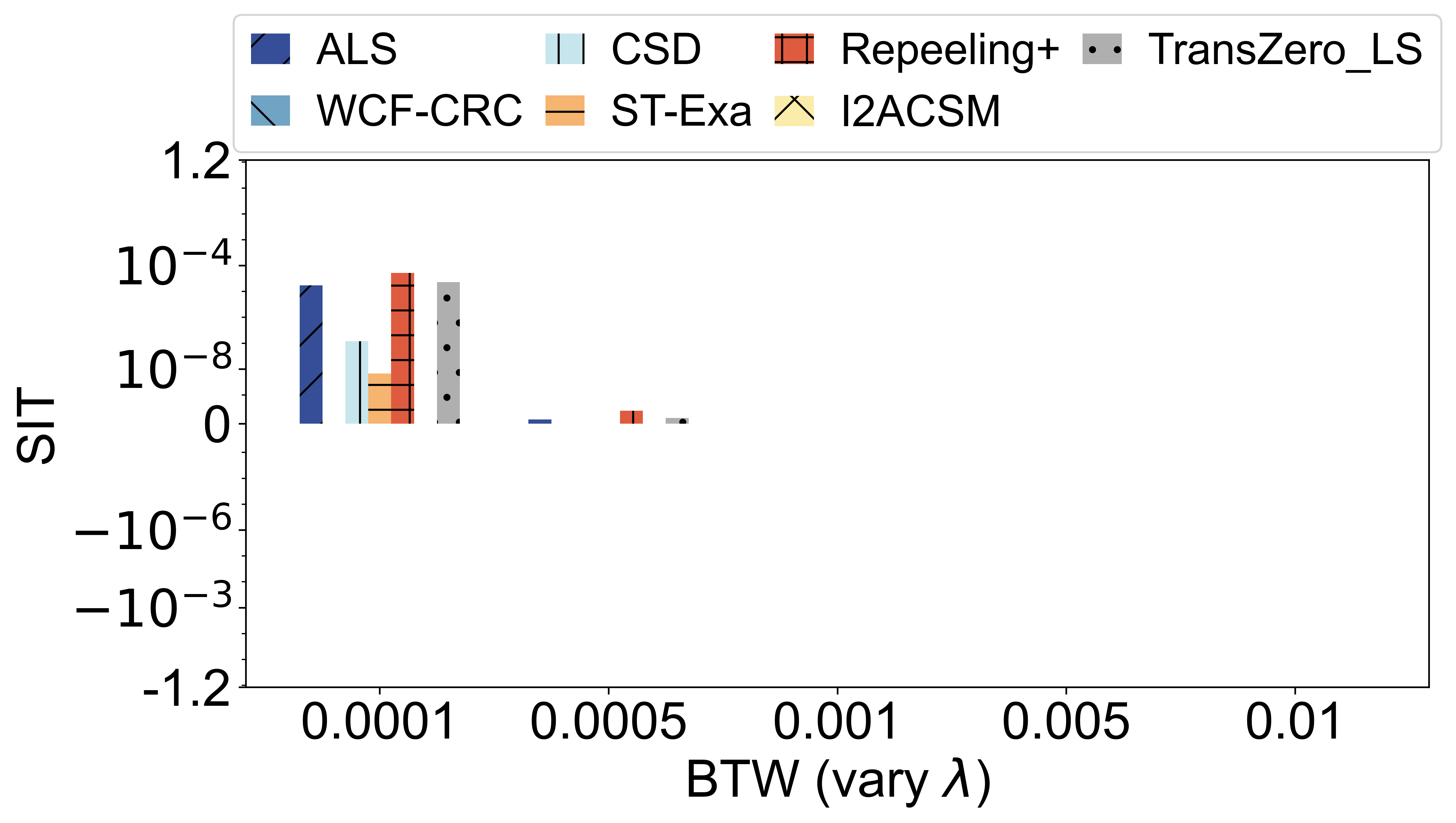}
        \label{fig: SIT_btw_lambda}
    \end{minipage}
    \begin{minipage}[t]{0.24\linewidth}
        \centering
        \includegraphics[width=\linewidth]{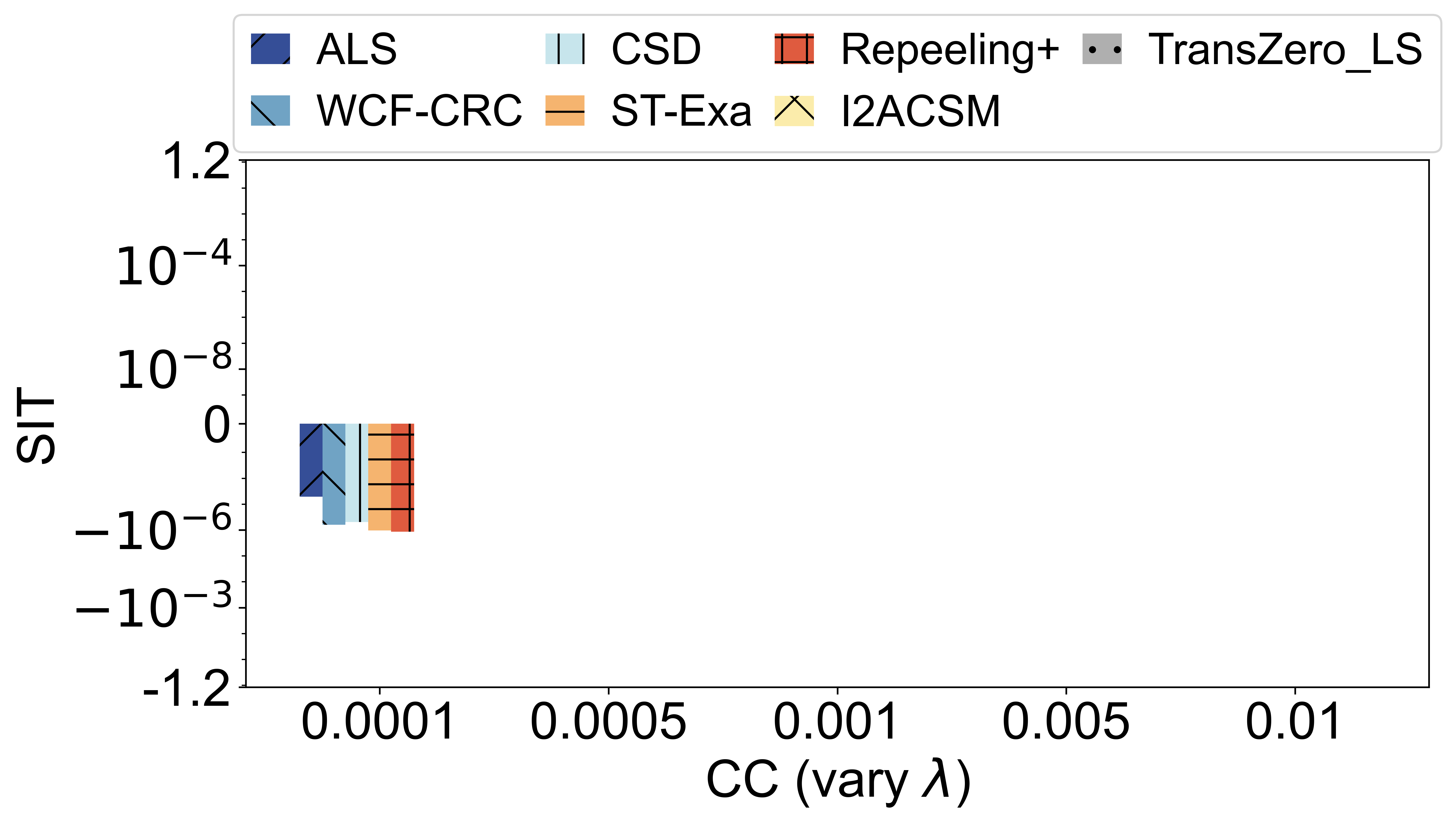}
        \label{fig: SIT_cc_lambda}
    \end{minipage}
    \begin{minipage}[t]{0.24\linewidth}
        \centering
        \includegraphics[width=\linewidth]{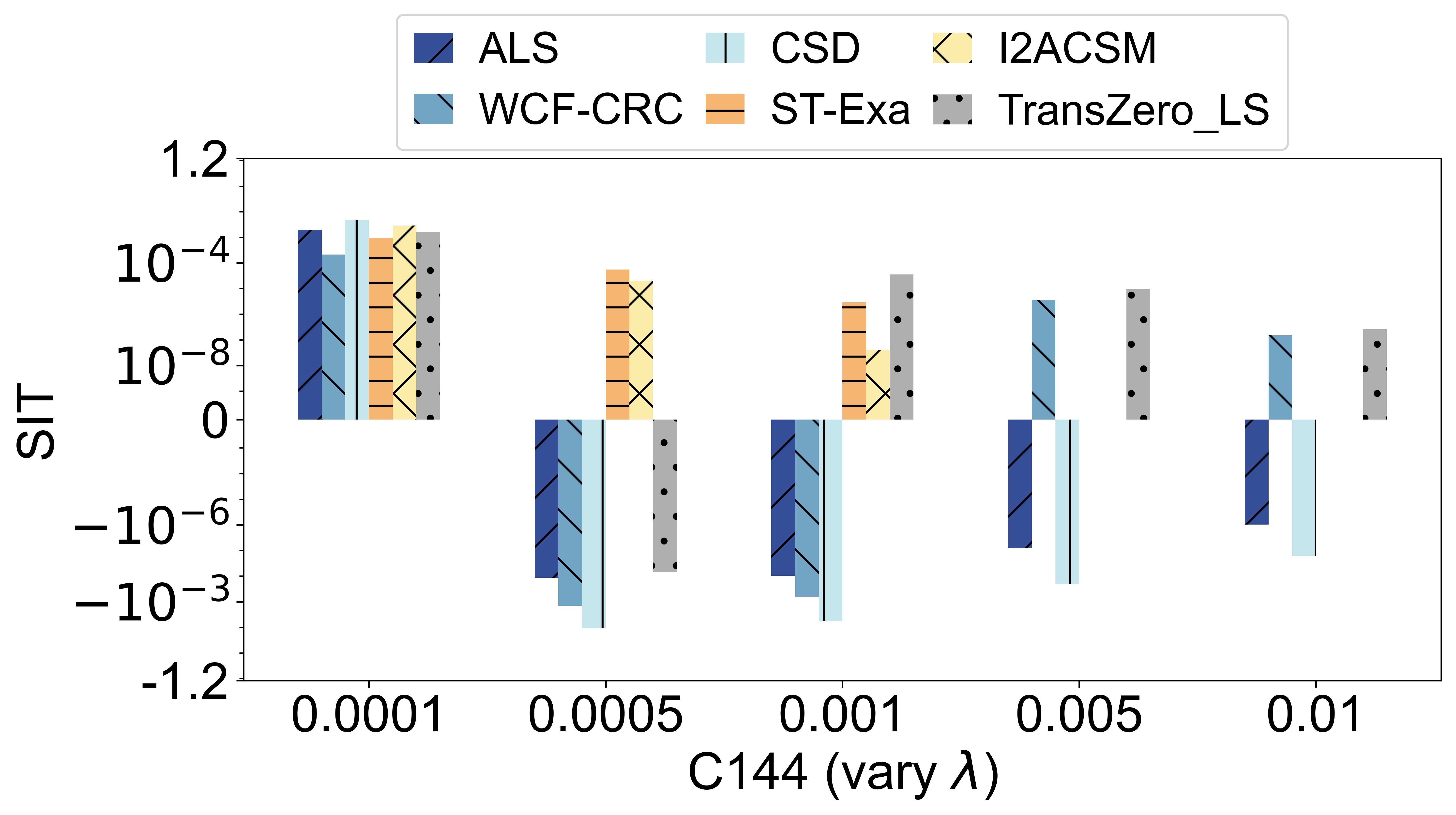}
        \label{fig: SIT_c144_lambda}
    \end{minipage}
    \begin{minipage}[t]{0.24\linewidth}
        \centering
        \includegraphics[width=\linewidth]{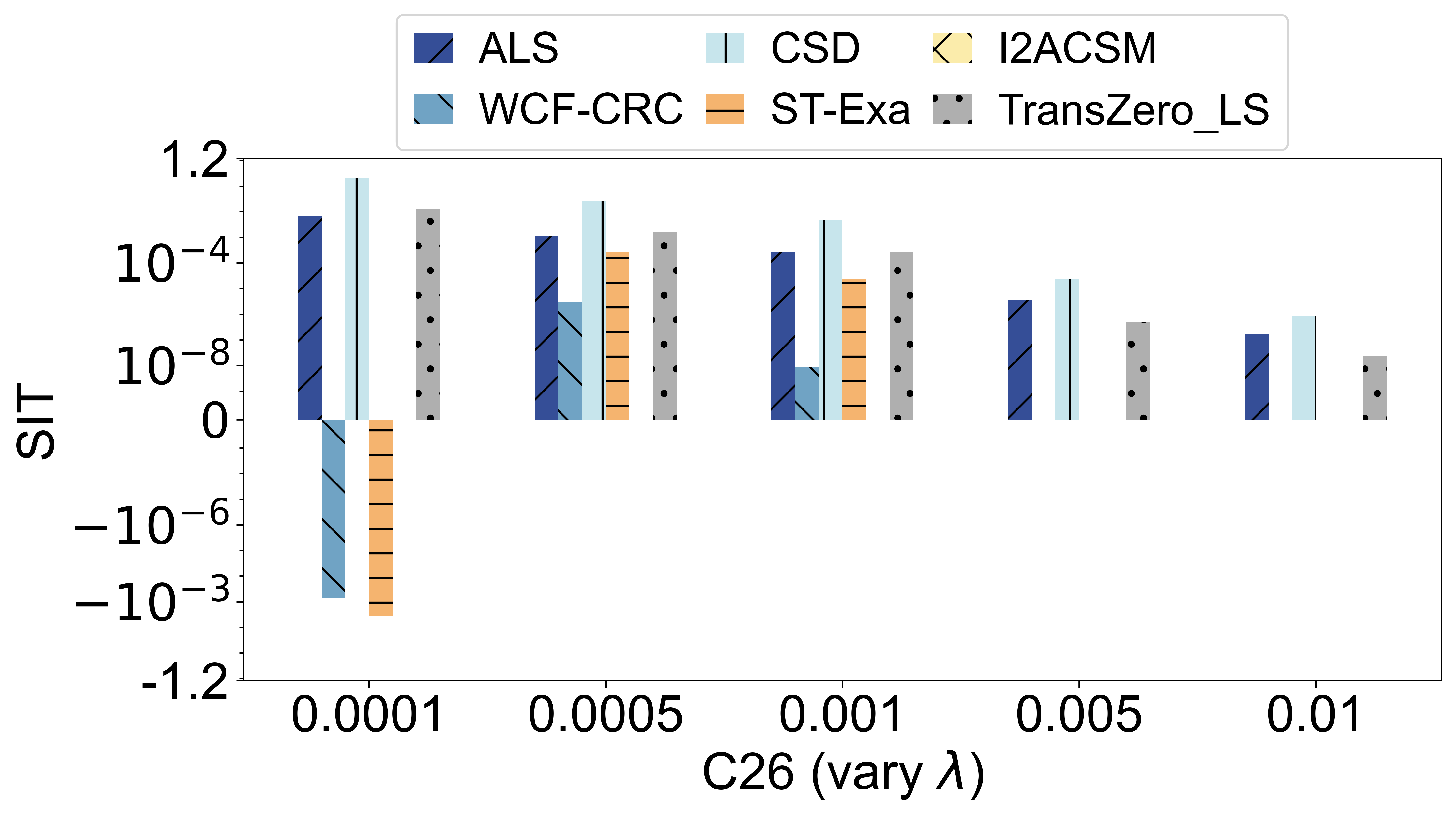}
        \label{fig: SIT_c26_lambda}
    \end{minipage}

    \begin{minipage}[t]{0.24\linewidth}
        \centering
        \vspace{-3ex}
        \includegraphics[width=\linewidth]{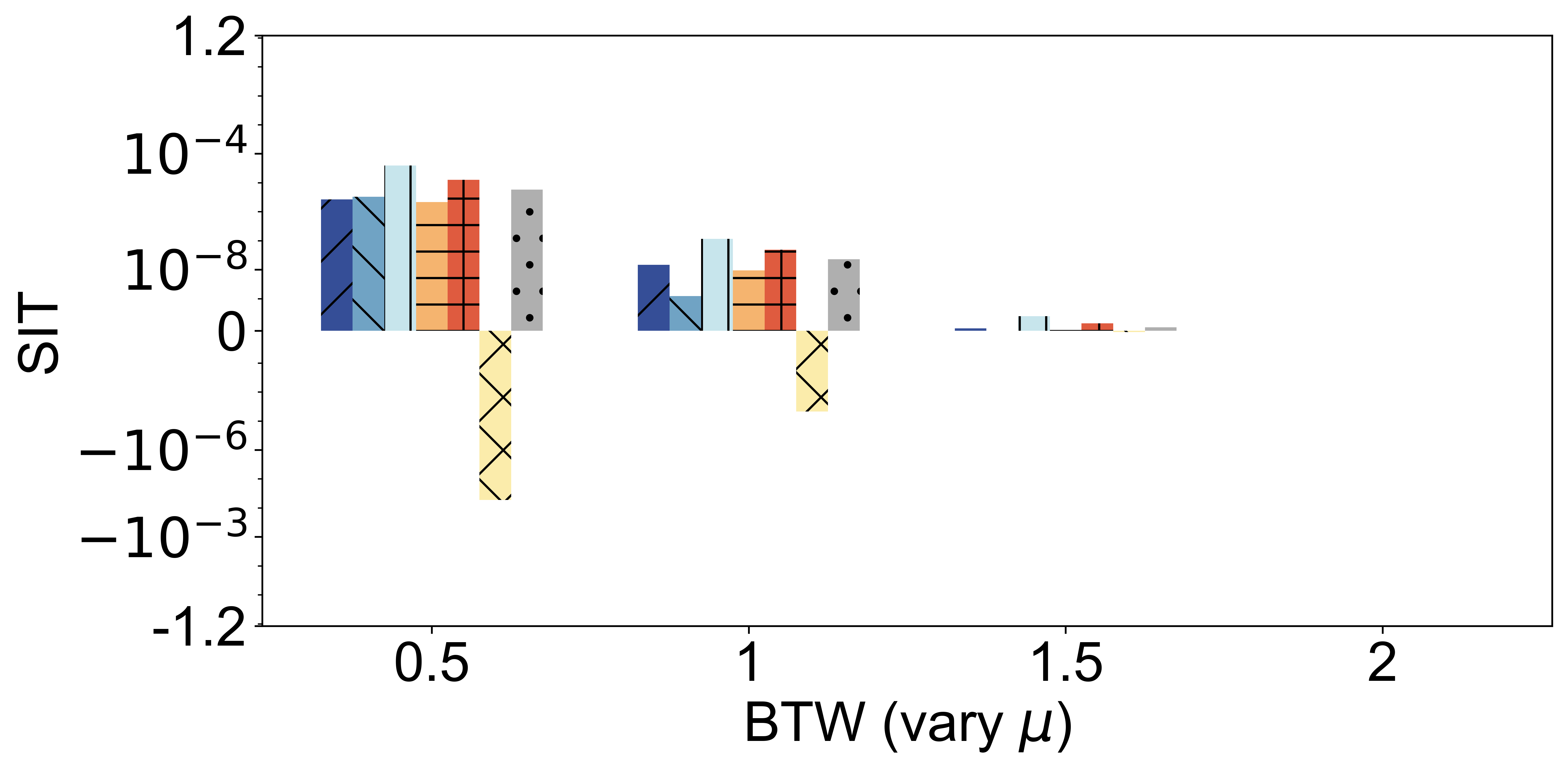}
        \label{fig: SIT_btw_mu}
    \end{minipage}
    \begin{minipage}[t]{0.24\linewidth}
        \centering
        \vspace{-3ex}
        \includegraphics[width=\linewidth]{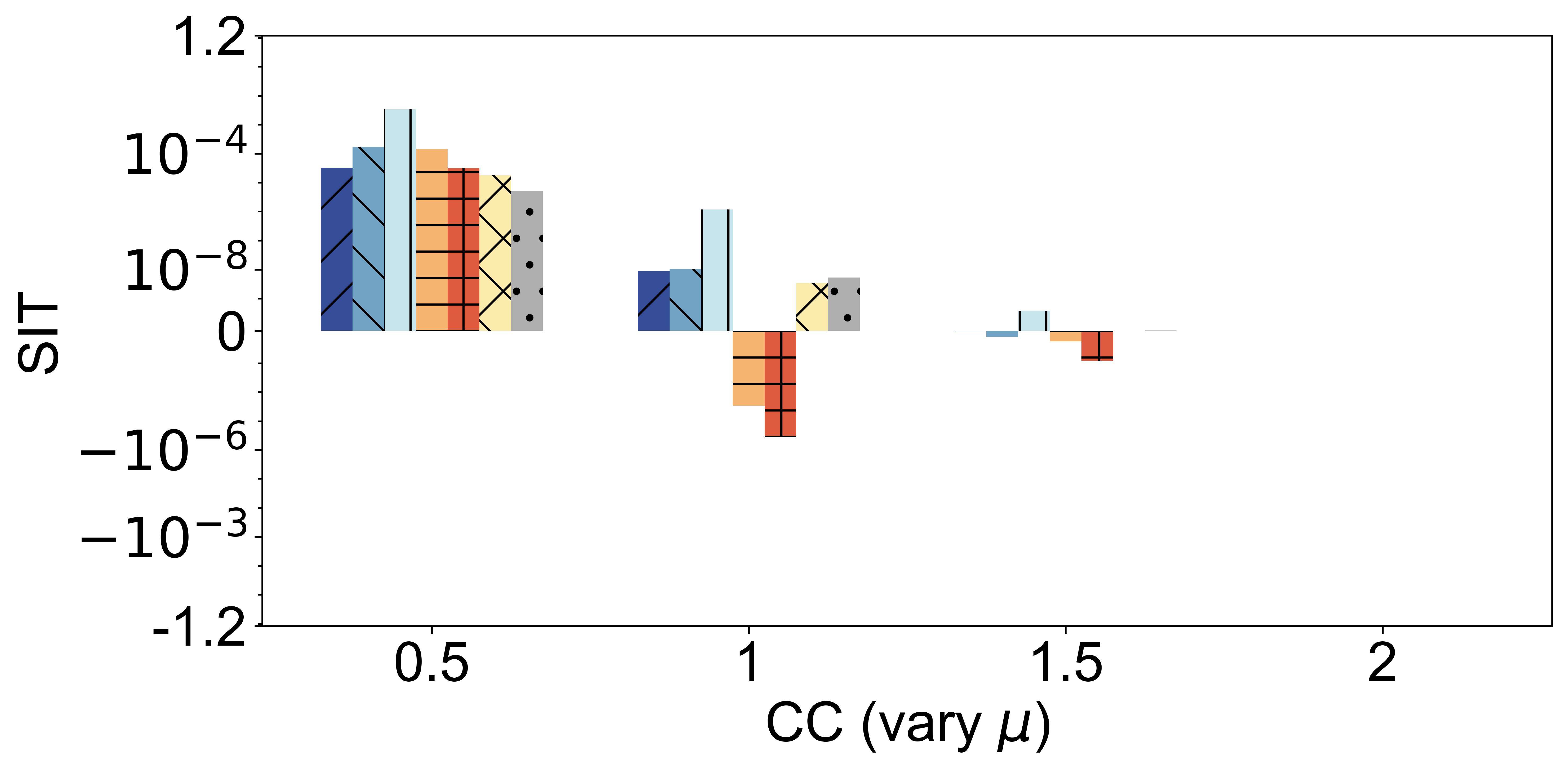}
        \label{fig: SIT_cc_mu}
    \end{minipage}
    \begin{minipage}[t]{0.24\linewidth}
        \centering
        \vspace{-3ex}
        \includegraphics[width=\linewidth]{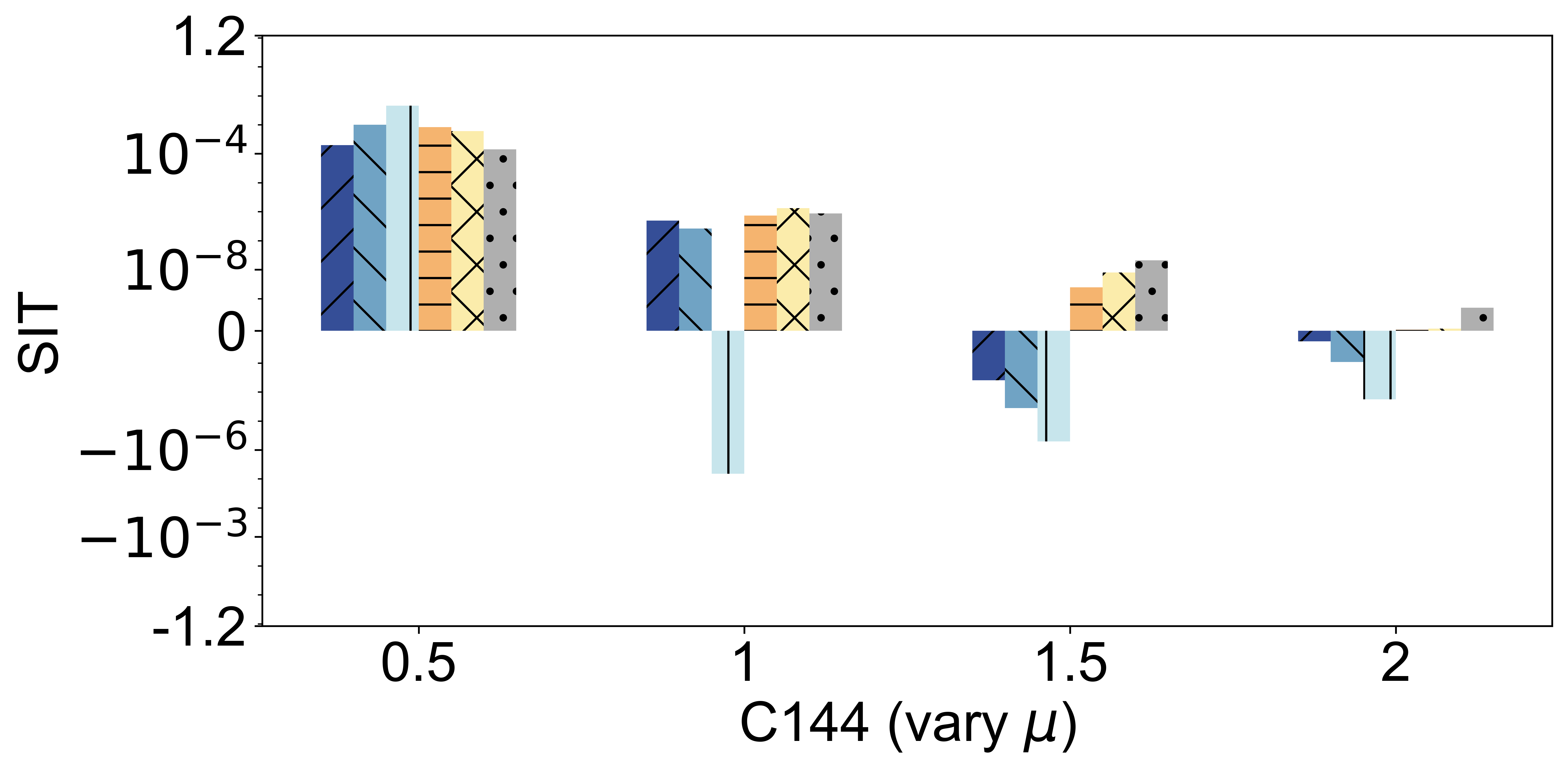}
        \label{fig: SIT_c144_mu}
    \end{minipage}
    \begin{minipage}[t]{0.24\linewidth}
        \centering
        \vspace{-3ex}
        \includegraphics[width=\linewidth]{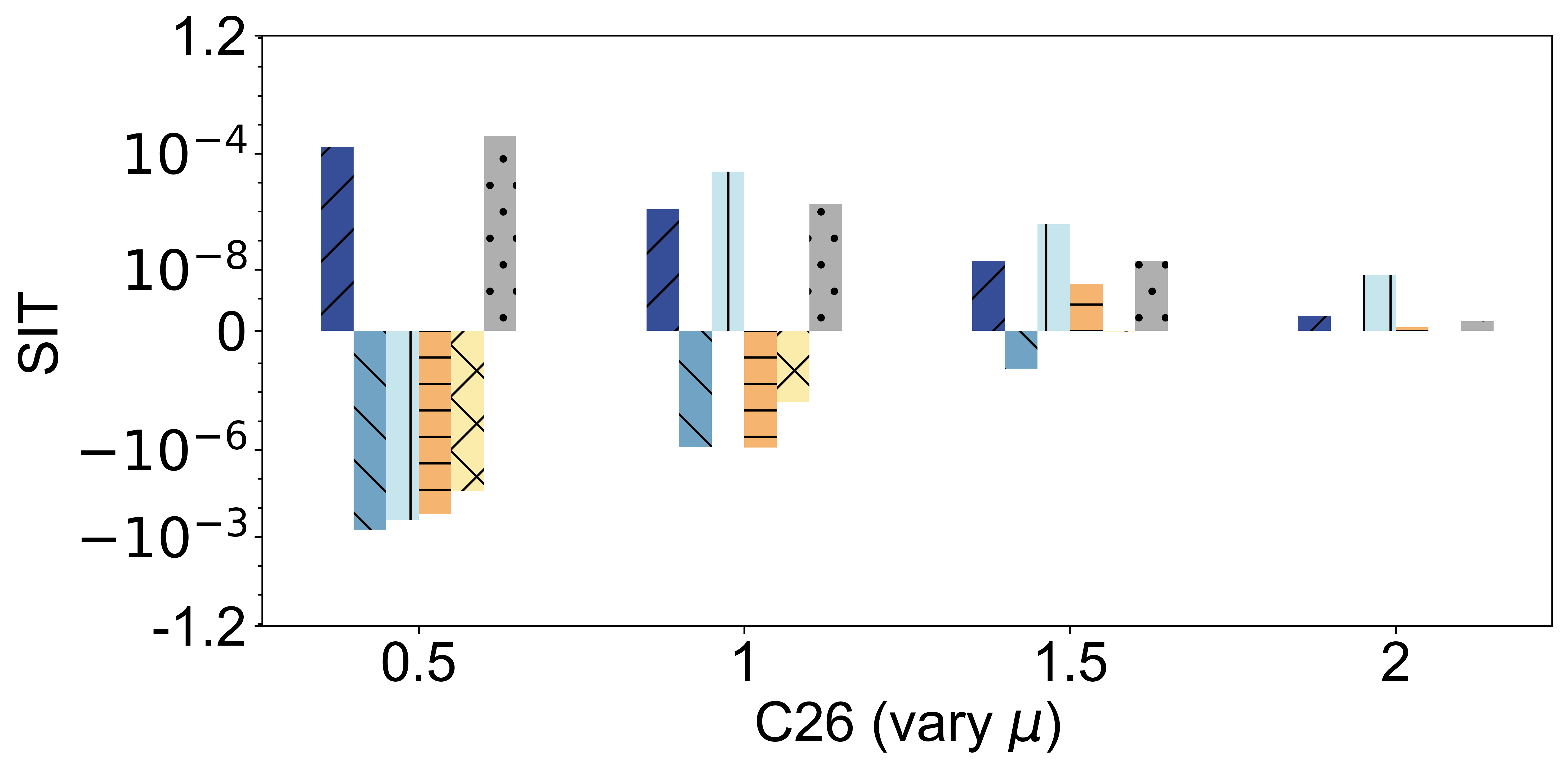}
        \label{fig: SIT_c26_mu}
    \end{minipage}
    
    \vspace{-4ex}
    \caption{SIT of communities (top row: community cohesiveness; bottom row: impact of time-decay functions).}
    \label{fig: SIT}
    \vspace{-3ex}
\end{figure*}

\vspace{-2ex}
\section{Experimental Evaluation} \label{sec: exp_results}
In this section, we conduct an in-depth evaluation of the representative CS algorithms and report key results. The reader may refer to~\cite{cohesiontech2025} for detailed results. We first assess the structural cohesiveness of their identified communities, then evaluate their quality using five psychology-informed cohesiveness measures. Learning-based algorithms were run on a Linux server with an Intel Xeon Platinum 8255C (2.50 GHz) and an RTX 3080 GPU (10 GB VRAM). Other experiments were conducted on a Linux server with an Intel Xeon Platinum 8352V (2.10 GHz) and 120 GB RAM. Recall that we do not focus on efficiency and scalability analysis here, as our goal is to understand the effectiveness of CS algorithms. The reader may refer to~\cite{fang2020survey} for studies related to the former.

\begin{figure*}[t]
    \setlength{\abovecaptionskip}{0.2cm}
    \begin{minipage}[t]{0.24\linewidth}
        \centering
        \includegraphics[width=\linewidth]{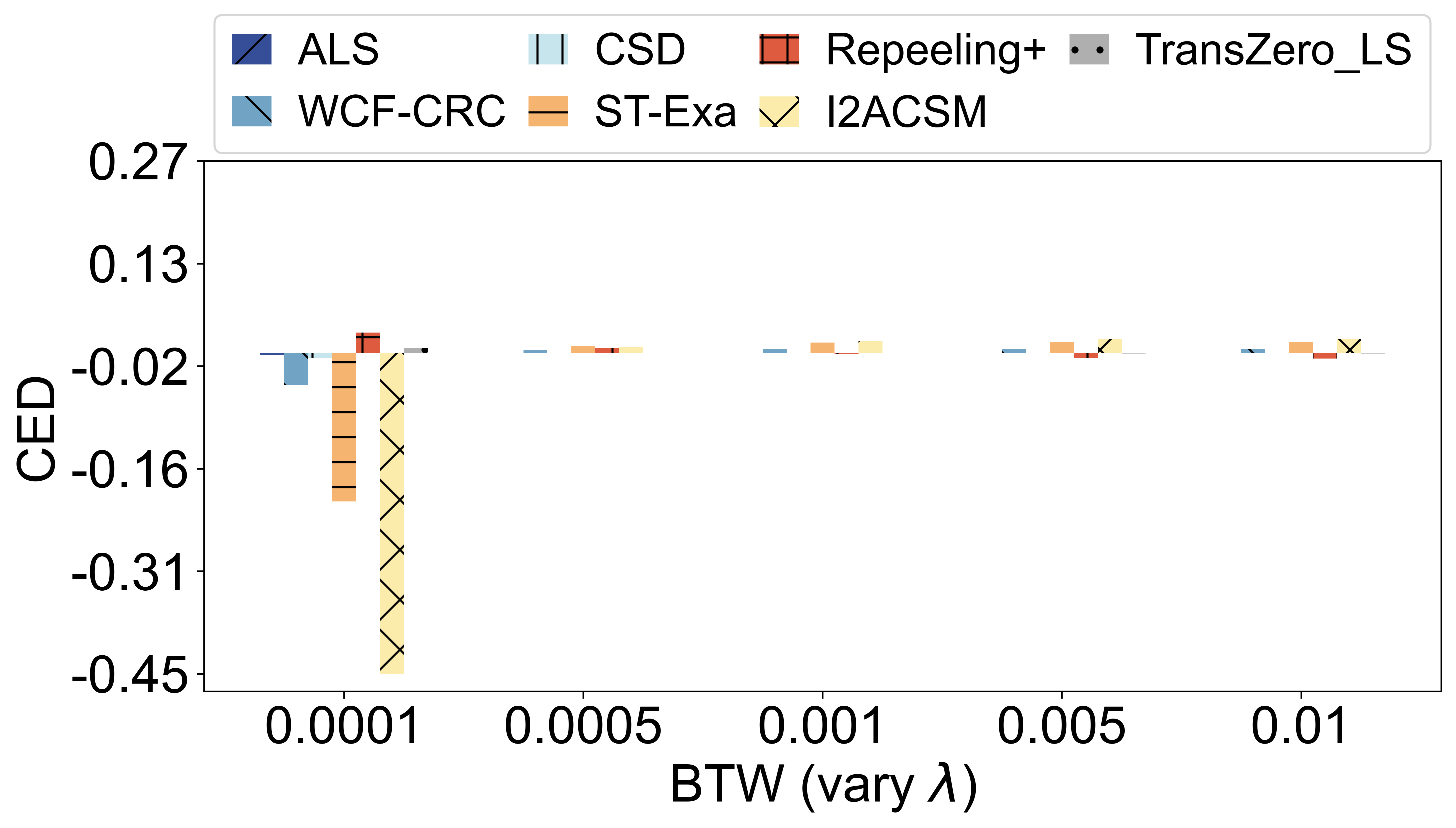}
        \label{fig: CED_btw_lambda}
    \end{minipage}
    \begin{minipage}[t]{0.24\linewidth}
        \centering
        \includegraphics[width=\linewidth]{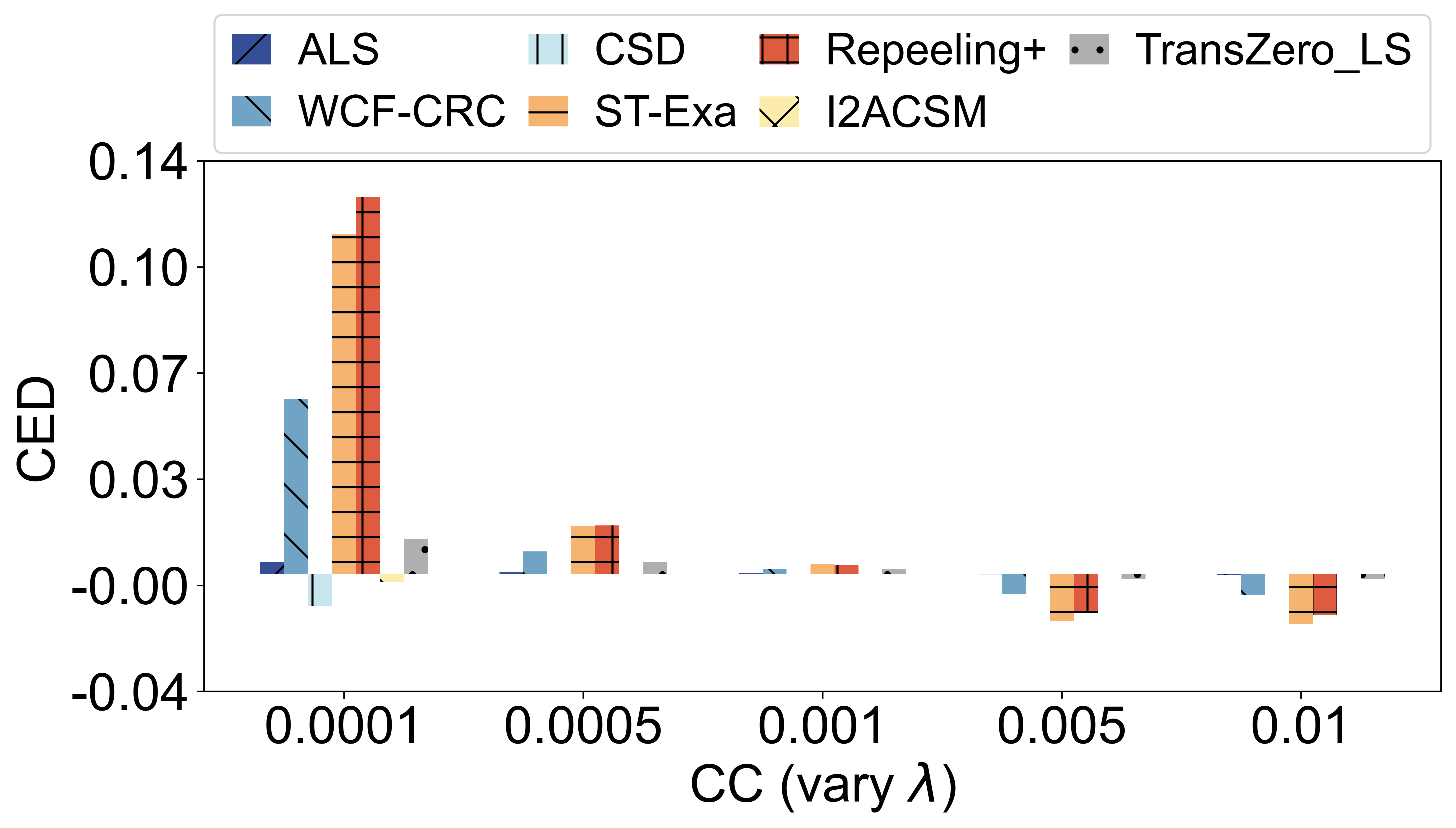}
        \label{fig: CED_cc_lambda}
    \end{minipage}
    \begin{minipage}[t]{0.24\linewidth}
        \centering
        \includegraphics[width=\linewidth]{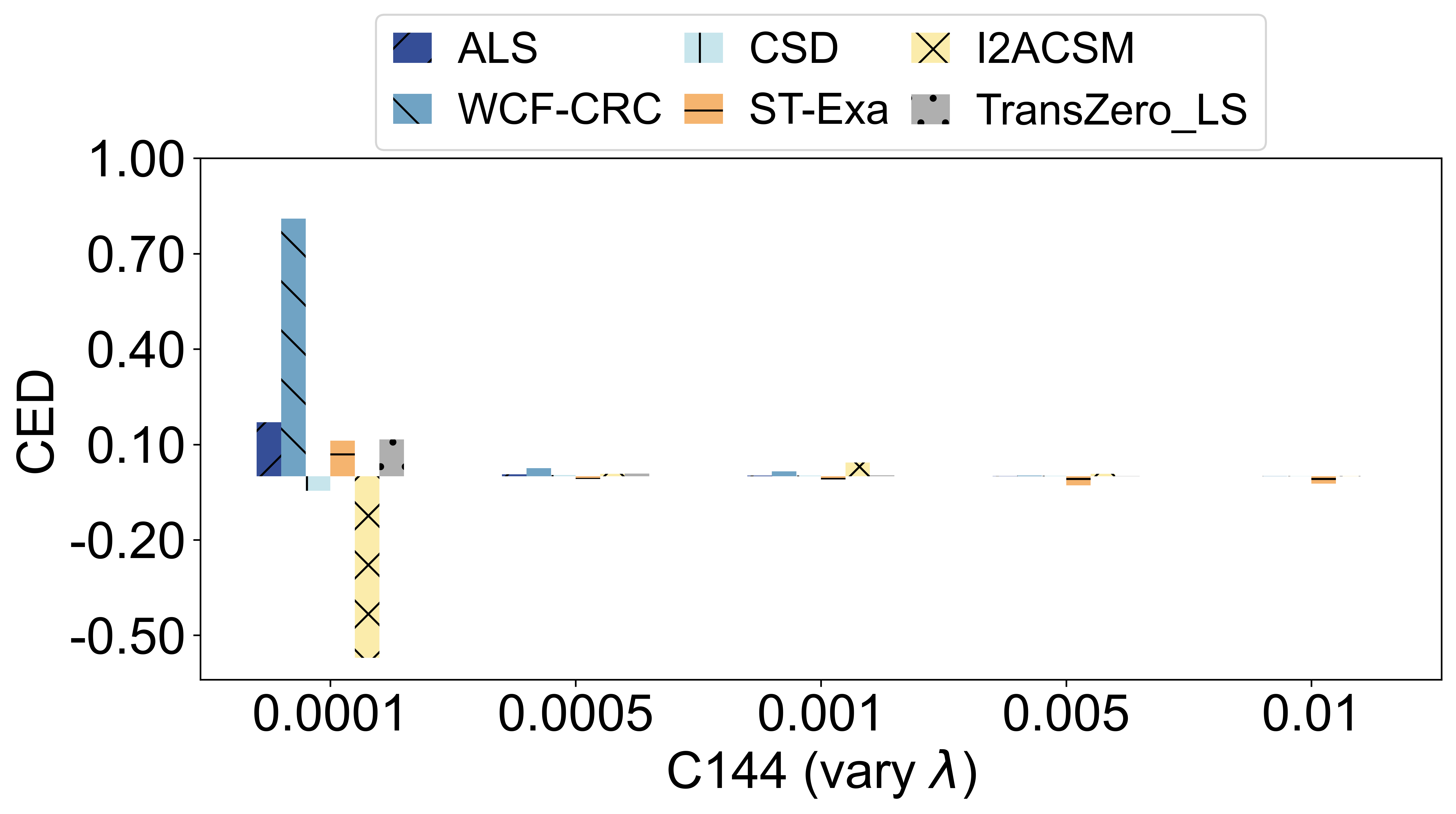}
        \label{fig: CED_c144_lambda}
    \end{minipage}
    \begin{minipage}[t]{0.24\linewidth}
        \centering
        \includegraphics[width=\linewidth]{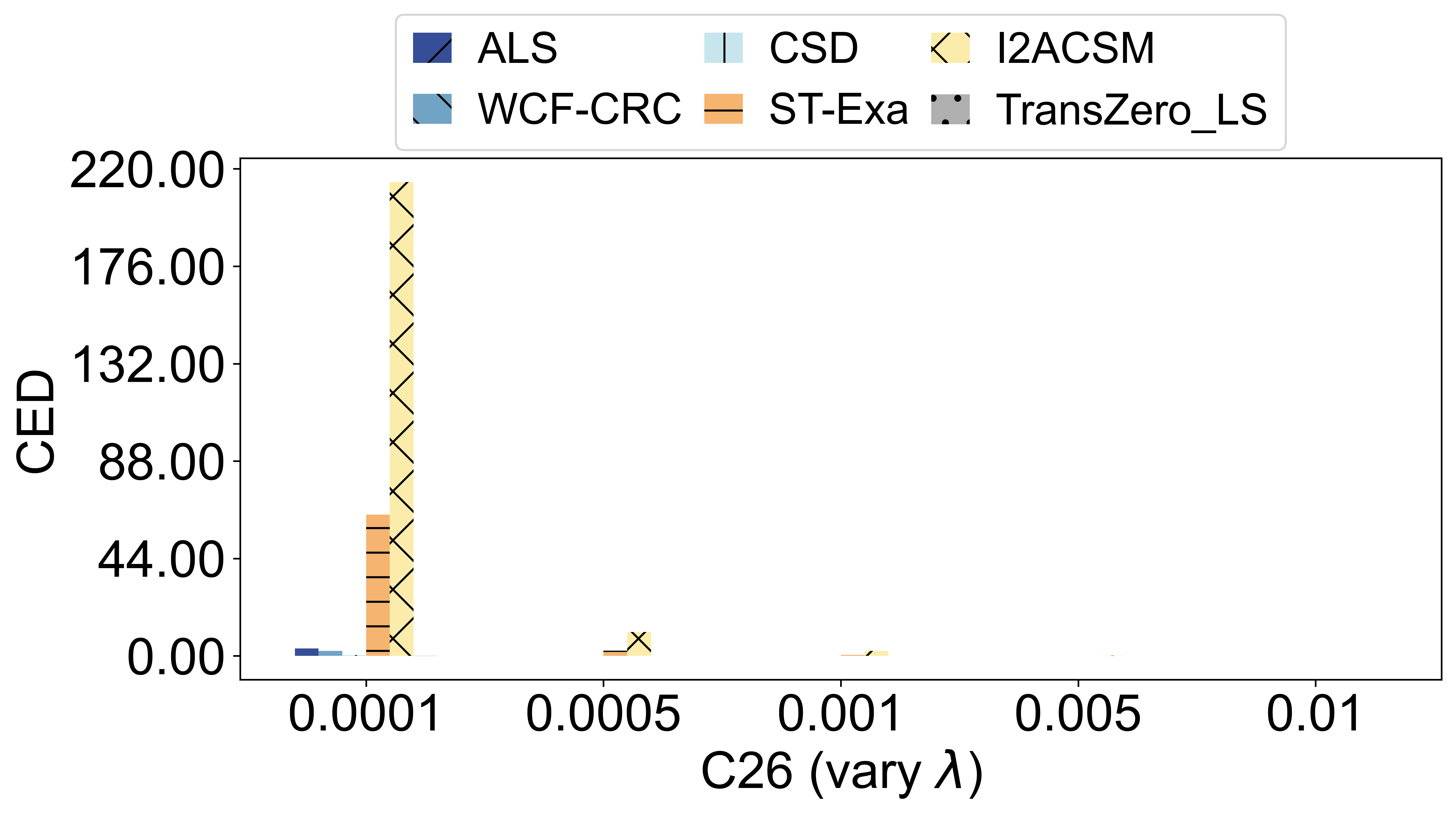}
        \label{fig: CED_c26_lambda}
    \end{minipage}
    
    \begin{minipage}[t]{0.24\linewidth}
        \centering
        \vspace{-3ex}
        \includegraphics[width=\linewidth]{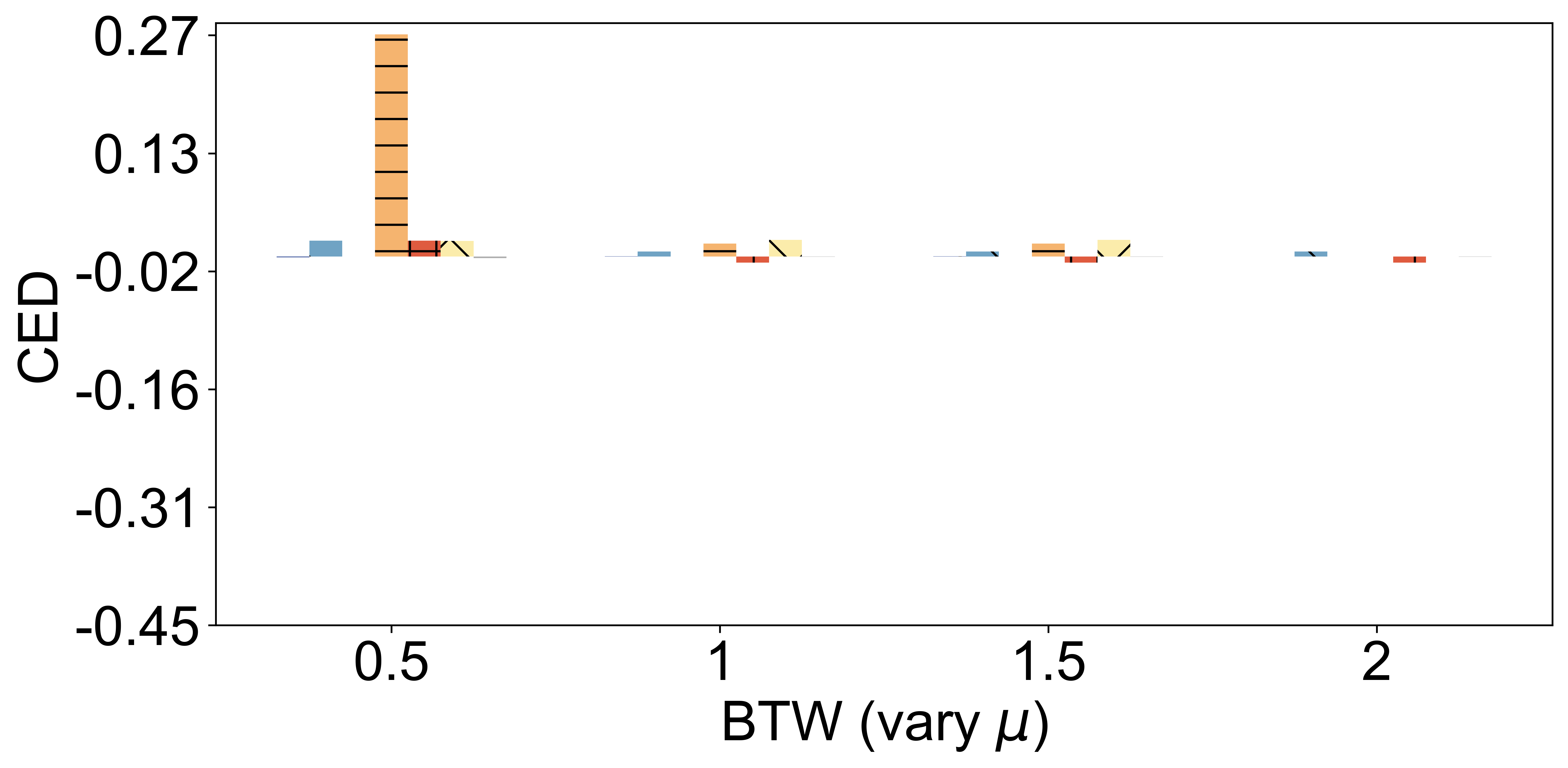}
        \label{fig: CED_btw_mu}
    \end{minipage}
    \begin{minipage}[t]{0.24\linewidth}
        \centering
        \vspace{-3ex}
        \includegraphics[width=\linewidth]{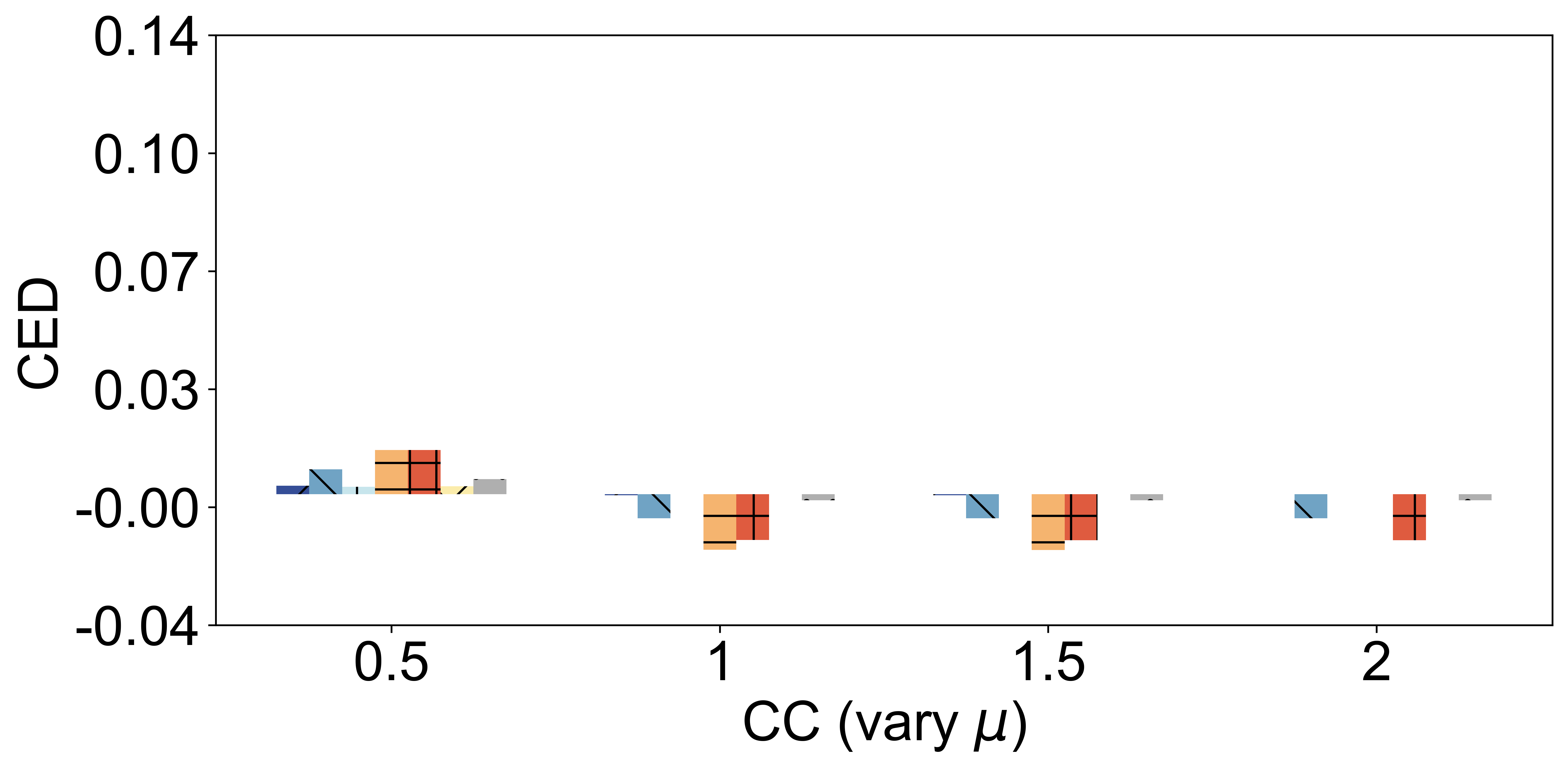}
        \label{fig: CED_cc_mu}
    \end{minipage}
    \begin{minipage}[t]{0.24\linewidth}
        \centering
        \vspace{-3ex}
        \includegraphics[width=\linewidth]{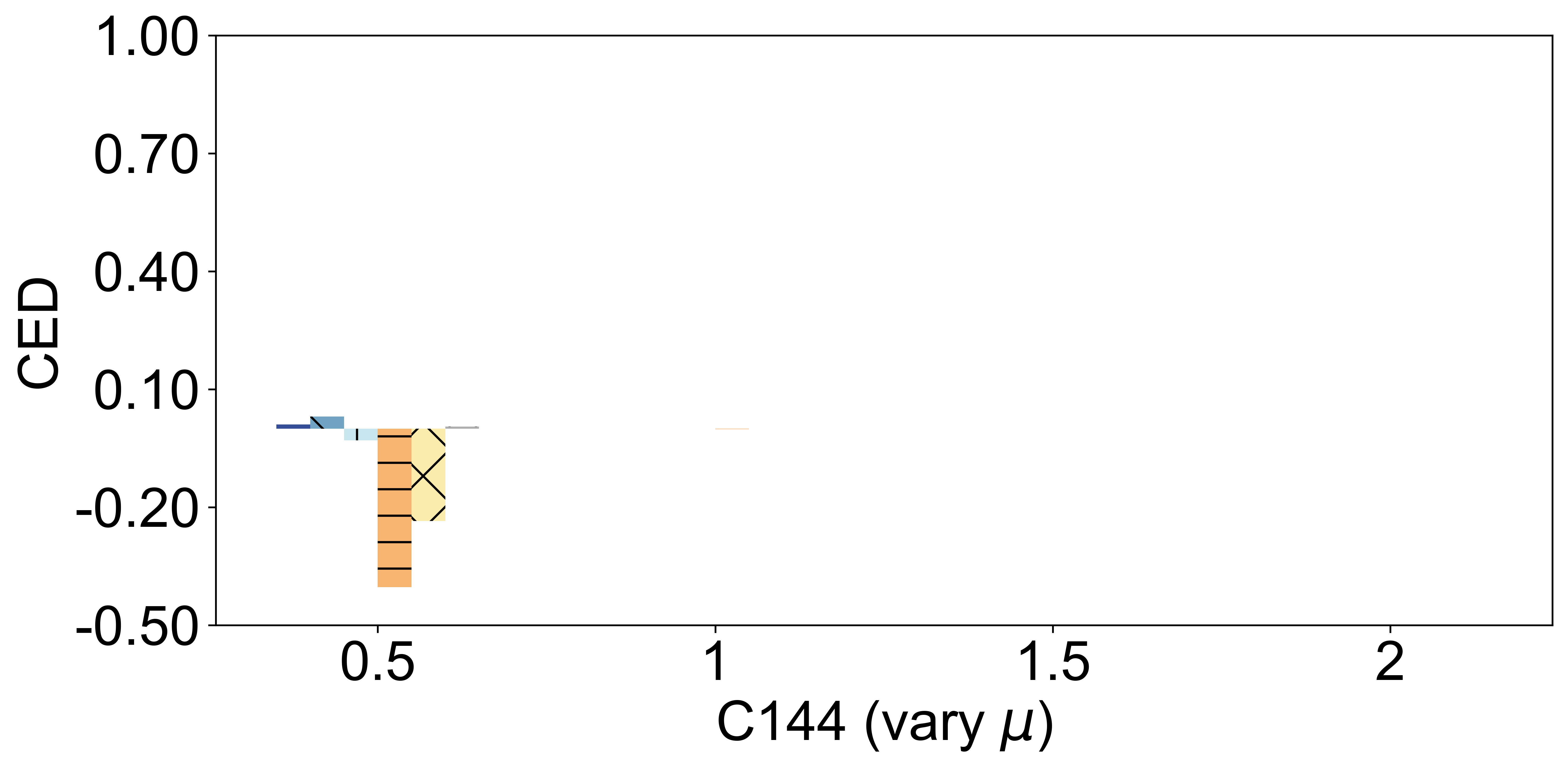}
        \label{fig: CED_c144_mu}
    \end{minipage}
    \begin{minipage}[t]{0.24\linewidth}
        \centering
        \vspace{-3ex}
        \includegraphics[width=\linewidth]{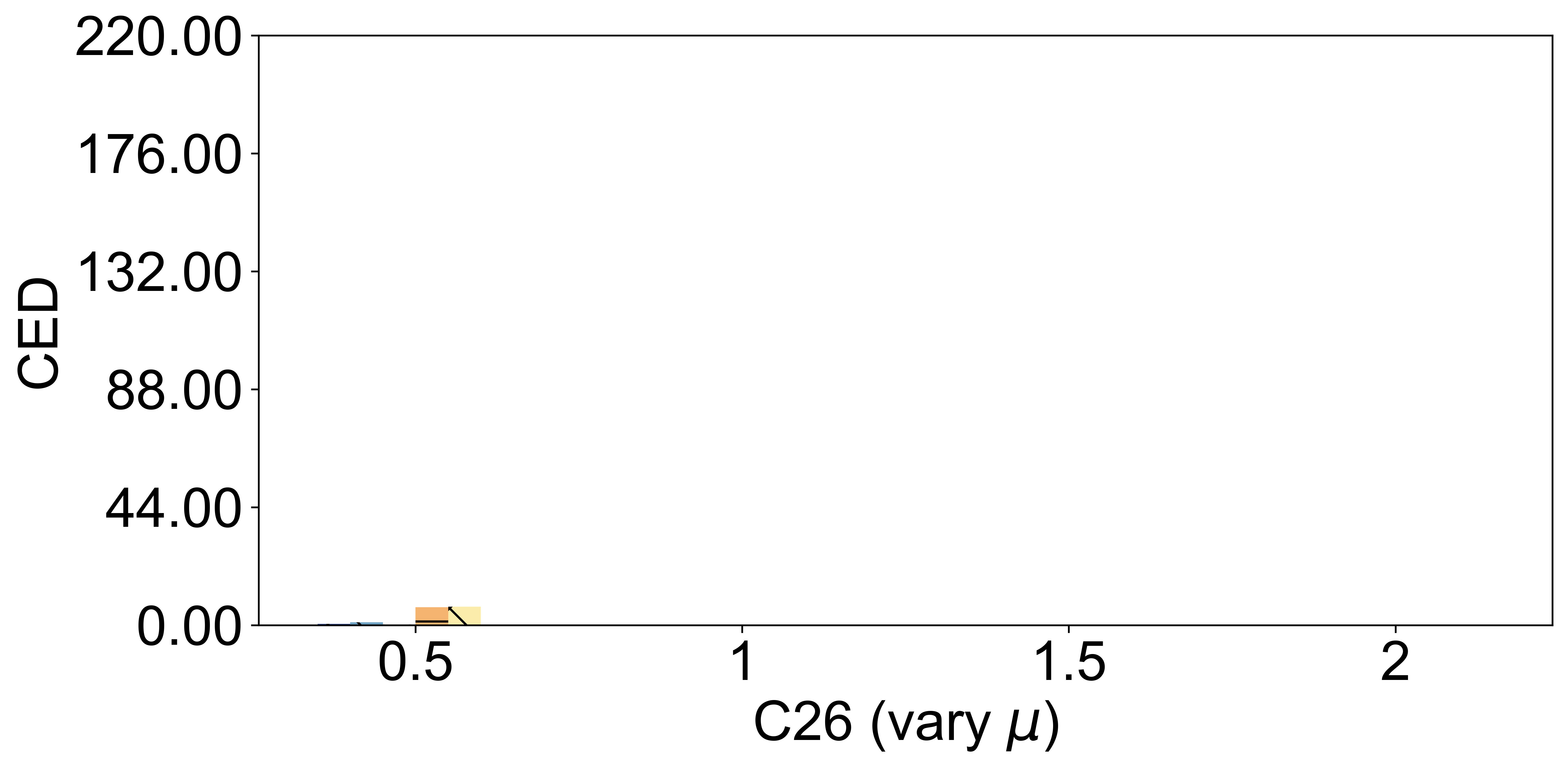}
        \label{fig: CED_c26_mu}
    \end{minipage}
    \vspace{-3ex}
    \caption{CED of communities (top row: community cohesiveness; bottom row: impact of time-decay functions).}
    \label{fig: CED}
    \vspace{-1ex}   
\end{figure*}

\vspace{-1ex}
\subsection{Structural Cohesiveness}
We use three commonly used evaluation metrics to assess the structural cohesiveness of CS algorithms \cite{fang2020survey, lu2022time, xu2022efficient}: diameter ($\mathbf{d}$), size, and minimum degree ($Deg_{min}$). Here, the $Deg_{min}$ calculation includes self-loops and parallel edges. Generally, higher community quality is indicated by lower diameter and size, and higher minimum degree~\cite{fang2020survey}. We also report the \textit{query hit rate} $Q_{hit}(\%)$, which is the ratio of queries that successfully return communities across all parameter combinations. Algorithms that fail to complete within 5 days are marked as \textit{INF} and excluded from the evaluation.

Table \ref{tab: structural} reports the results. We can make the following observations: (1) although \textsf{ALS}, \textsf{Repeeling+}, and \textsf{TransZero-LS} have high $Q_{hit}$, they tend to generate larger yet sparser communities characterized by greater degrees and sizes, and lower $Deg_{min}$ values; (2) \textsf{I2ACSM} identifies very small communities across all datasets; and (3) compared to aforementioned algorithms, \textsf{WCF-CRC}, \textsf{CSD}, and \textsf{ST-Exa} are better at identifying structurally cohesive communities. However, \textsf{WCF-CRC} and \textsf{CSD} generally achieve lower hit rates, indicating difficulty in identifying target communities. Among all algorithms, \textsf{ST-Exa} successfully returns communities for all queries and demonstrates superior performance across all datasets, partially due to its size-constrained nature. Note that the communities (or node lists) returned by \textsf{TransZero-GS} are disconnected, likely because its candidate selection ignores edge connections. Hence, its results are left blank.

The poor performance of \textsf{ALS} may stem from the smaller time scales used in our datasets, which hinder its ability to find nodes with greater temporal proximity. However, the original paper does not discuss the time scale and its impact on CS results. Additionally, identifying extremely large or small communities may arise from graph sparsity after multigraph transformations. The unpromising performance of the two learning-based methods may be attributable to insufficient node features in our datasets, limiting CSGphormer's ability to learn adequate information for calculating community scores. On the other hand, this highlights these methods' stringent data requirements and their limitations when applied to datasets not specifically designed for model training.

\begin{figure}[t]    
    \setlength{\abovecaptionskip}{0.2cm}
    \begin{minipage}[t]{0.49\linewidth}
        \centering
        \includegraphics[width=\linewidth]{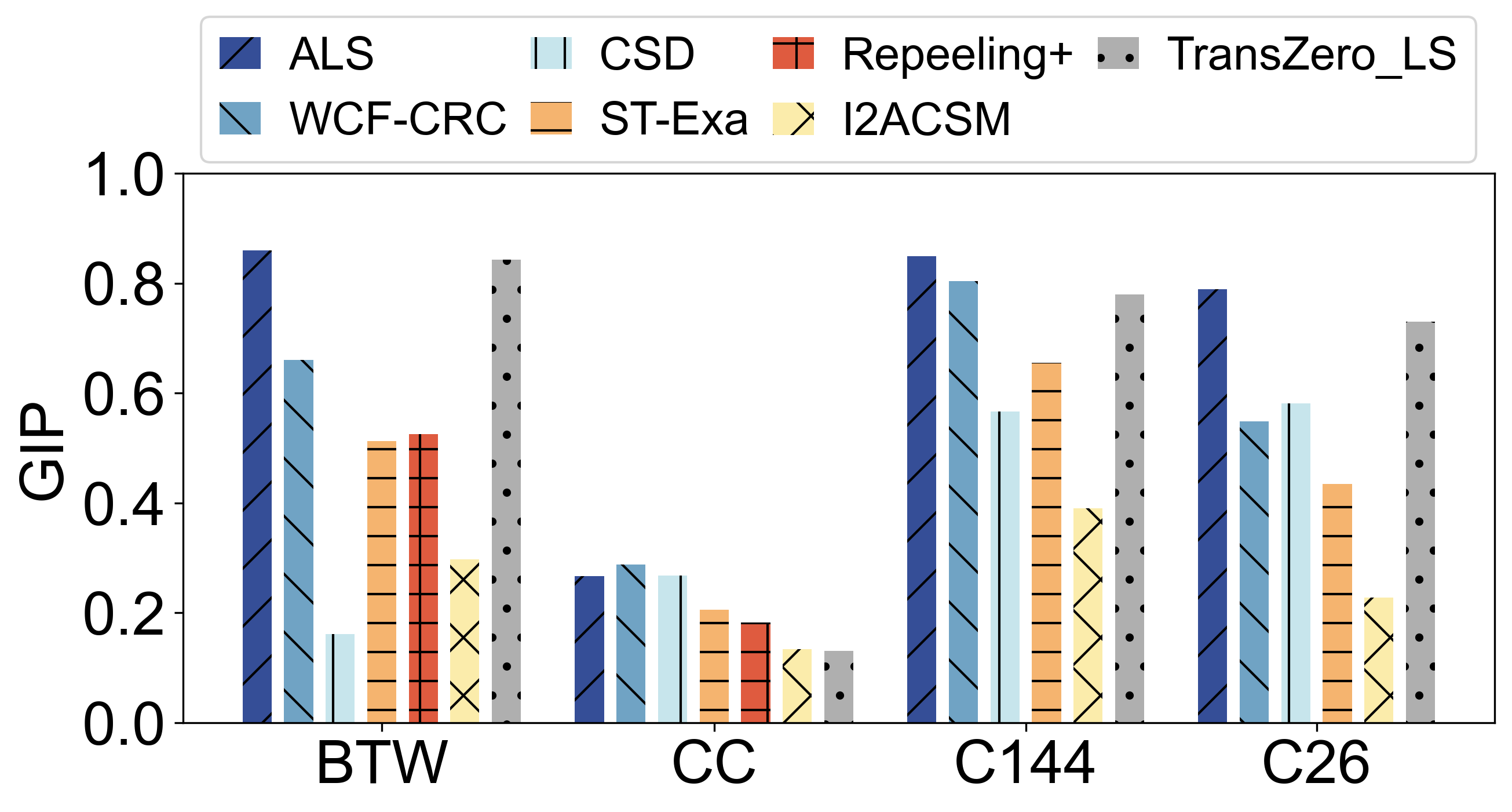}
        \label{fig: GIP}
    \end{minipage}
    \begin{minipage}[t]{0.49\linewidth}
        \centering
        \includegraphics[width=\linewidth]{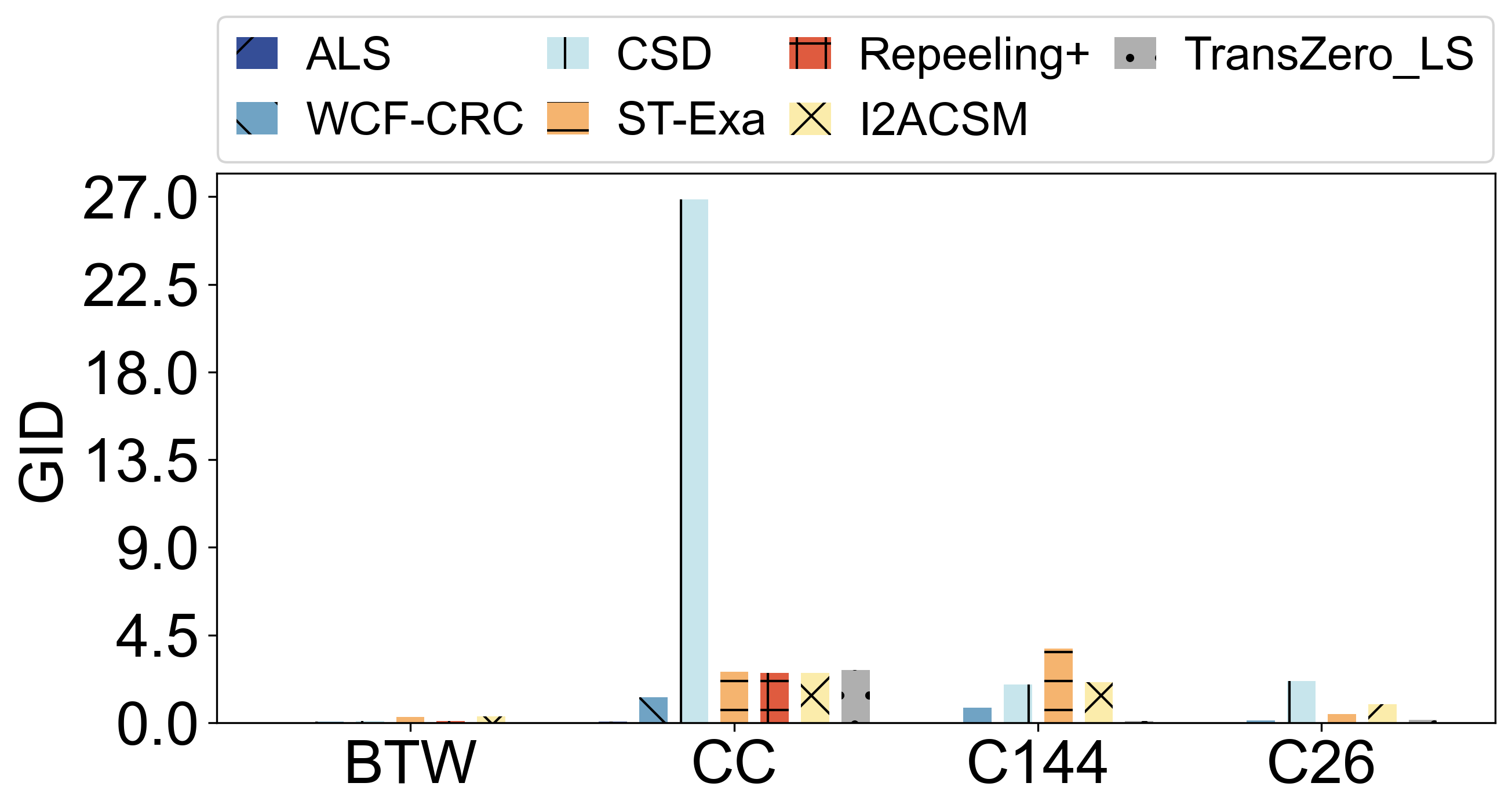}
        \label{fig: GID}
    \end{minipage}
    \vspace{-3ex}
    \caption{GIP and GID of CS communities.}
    \label{fig: GI-S}
    \vspace{-3ex}
\end{figure}

\vspace{-1ex}
\subsection{Psychology-informed Cohesiveness}
Next, we report evaluation results using our psychology-informed cohesiveness measures.

\textbf{Enjoyment Index (EI).} The top row of Figure \ref{fig: EI} illustrates that community cohesiveness, as indicated by EI values, varies across algorithms and is affected by the importance assigned to recent activities. For example, communities identified by \textsf{Repeeling+} on the \textsf{BTW} show higher enjoyment when $\lambda \in \{0.0001, 0.0005, 0.001\}$. However, increasing the weight of recent activities reduces enjoyment due to recent negative interactions. Notably, communities identified by \textsf{CSD} mostly exhibit positive EI values across all datasets and are less affected by the $\lambda$.

We also examine the impact of a different time-decay function on EI, as shown in the bottom row of Figure \ref{fig: EI}. Here, we use the polynomial decay function, $\Phi(t - t_{cur}) = (t_{cur} - t + 1)^{-\mu}$ ($\mu > 0$), which models a slower decline in the importance of past activities \cite{cormode2009forward}. We vary $\mu$ from 0.5 to 2 in steps of 0.5. With a stronger and more persistent impact of recent interactions, EI values become more consistent across decay rates, and communities identified by \textsf{CSD} still largely maintain positive EI scores. Overall, none of the algorithms retrieve communities with consistently high EI scores.

\textbf{Sentimental Interaction Tendency (SIT).} SIT scores are sensitive to decay rates and the choice of decay functions. Nevertheless, the SIT results are distinguishably different from the EI results. Figure \ref{fig: SIT} shows results across the four datasets. On \textsf{BTW}, as $\lambda$ increases, no algorithms effectively capture mutual relationships among users. Furthermore, results on \textsf{C144} reveal that communities with predominantly positive EI scores can still include negative mutual relationships (negative SIT values), and vice versa. Although \textsf{CSD} and \textsf{Repeeling+} consider directed edges, and \textsf{ALS} and \textsf{WCF-CRC} incorporate temporal information, their communities are still inadequate to achieve higher SIT scores.  

\textbf{Comparative Enjoyment Degree (CED).} Figure \ref{fig: CED} plots the CED results. When $\lambda = 0.0001$, some communities exhibit strong internal enjoyment, while others prefer external interactions, as indicated by the polarity of the CED scores. However, as the decay rate increases, the CED values for all communities converge toward zero, meaning the algorithms cannot find communities that increasingly prefer internal interactions over time. Similar to EI and SIT, CED is highly sensitive to the choice of decay function, reflecting the complexity of user interactions, as CED is calculated based on sentiment, timing, and where interactions occur. No algorithms consistently capture these complex interaction dynamics.

\textbf{Group Interaction Preference (GIP).} The left plot in Figure \ref{fig: GI-S} reports the GIP values. \textsf{ALS}, \textsf{WCF-CRC}, and \textsf{TransZero-LS} tend to identify highly interactive communities. However, as noted in Table \ref{tab: structural}, these high GIP values may result from the large community sizes returned by these algorithms, which include most interactions. Similar patterns are observed with higher GIP values for \textsf{CSD} communities on the \textsf{C144} and \textsf{C26} datasets and lower scores for \textsf{I2ACSM} communities across all datasets.

\textbf{Group Interaction Density (GID).} The right plot in Figure \ref{fig: GI-S} reports that users within communities from \textsf{CSD}, \textsf{ST-Exa}, and \textsf{I2ACSM} generally engage more with each other. Note that these high-GID communities also have smaller sizes and higher $Deg_{min}$.

\begin{figure*}[t]
    \setlength{\abovecaptionskip}{0.2cm}
    \begin{minipage}[t]{0.24\linewidth}
        \centering
        \includegraphics[width=\linewidth]{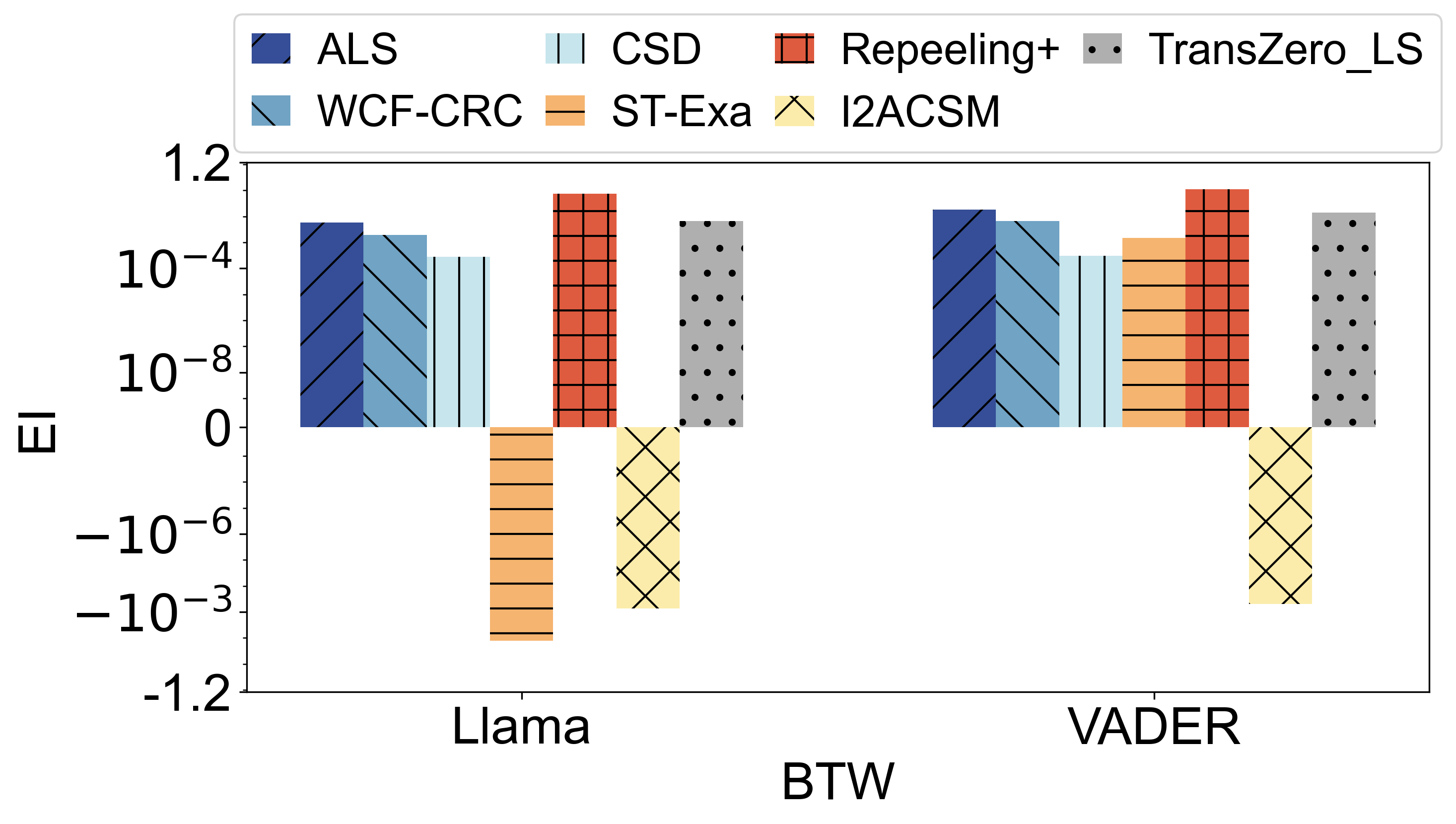}
        \label{fig: EI_btw_senti}
    \end{minipage}
    \begin{minipage}[t]{0.24\linewidth}
        \centering
        \includegraphics[width=\linewidth]{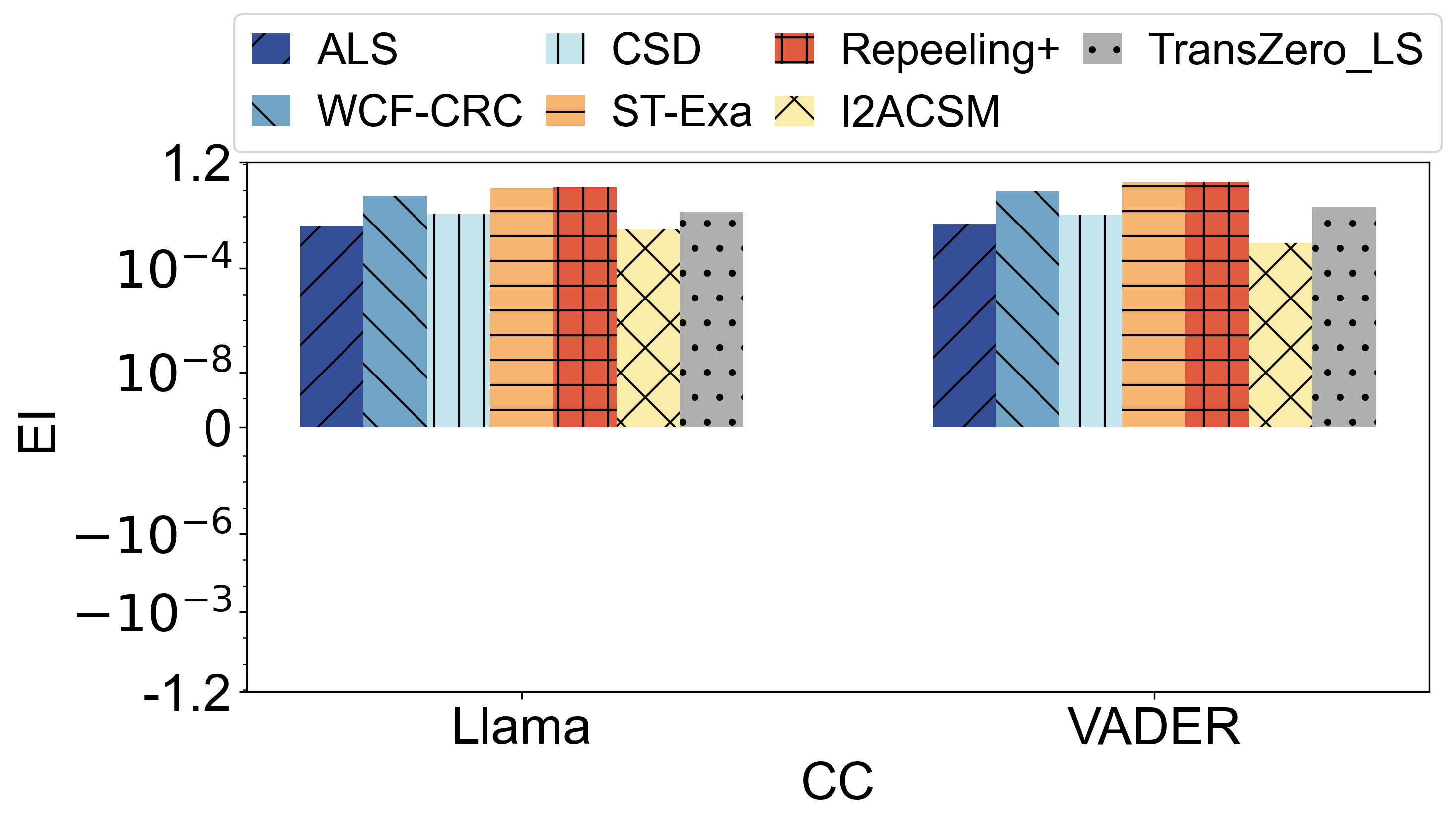}
        \label{fig: EI_cc_senti}
    \end{minipage}
    \begin{minipage}[t]{0.24\linewidth}
        \centering
        \includegraphics[width=\linewidth]{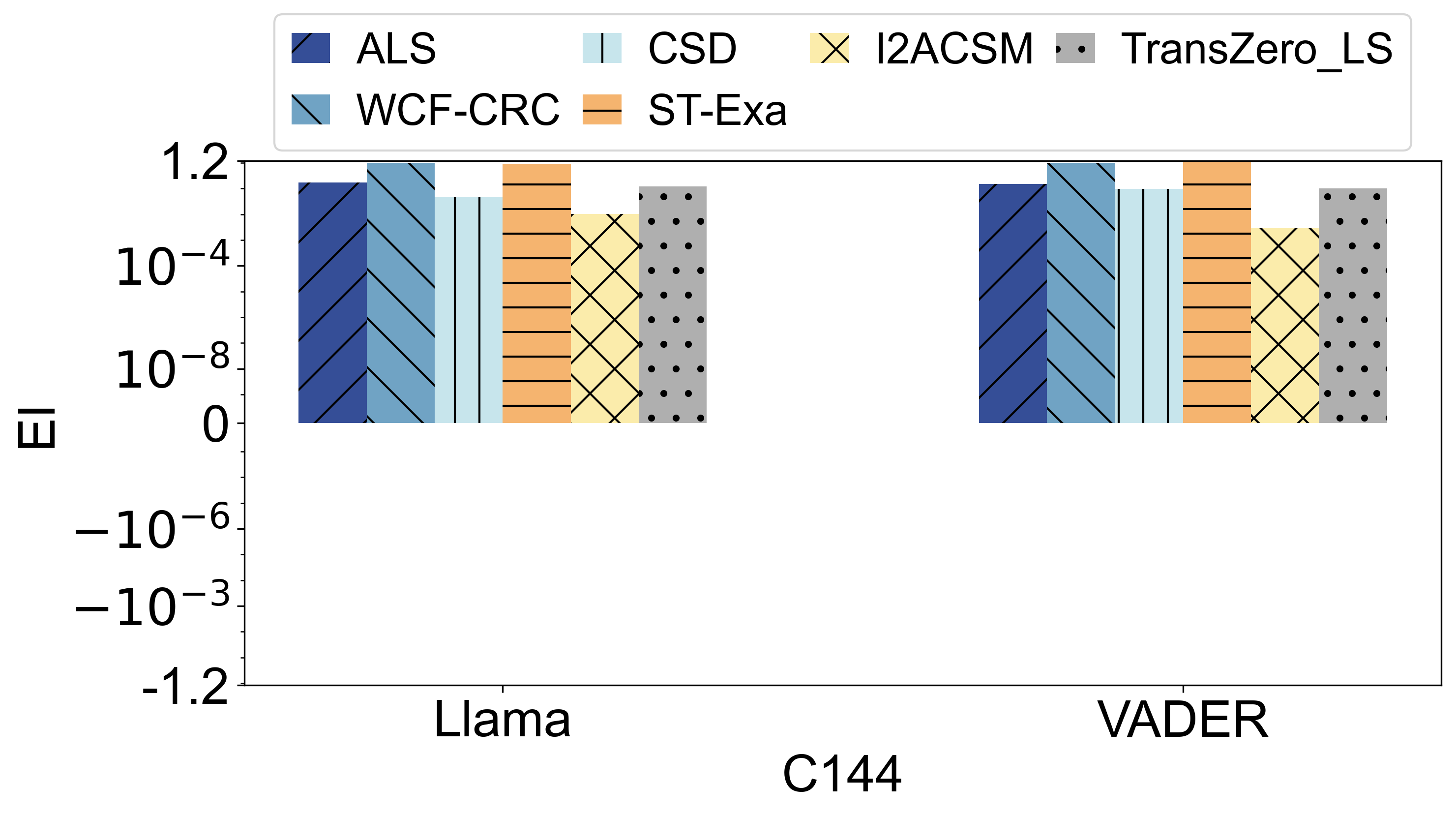}
        \label{fig: EI_c144_senti}
    \end{minipage}
    \begin{minipage}[t]{0.24\linewidth}
        \centering
        \includegraphics[width=\linewidth]{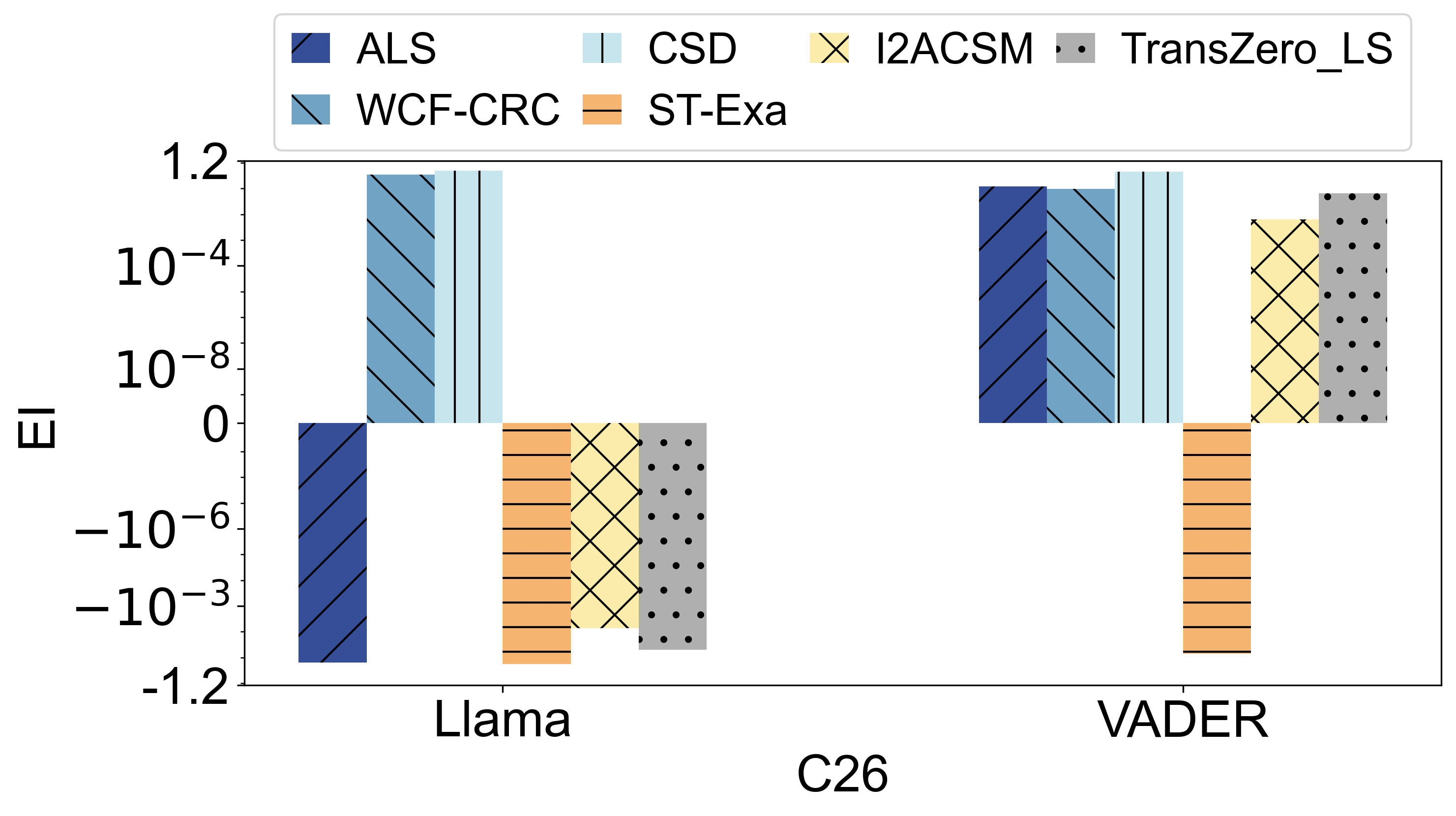}
        \label{fig: EI_c26_senti}
    \end{minipage}

    \begin{minipage}[t]{0.24\linewidth}
        \centering
        \vspace{-3ex}
        \includegraphics[width=\linewidth]{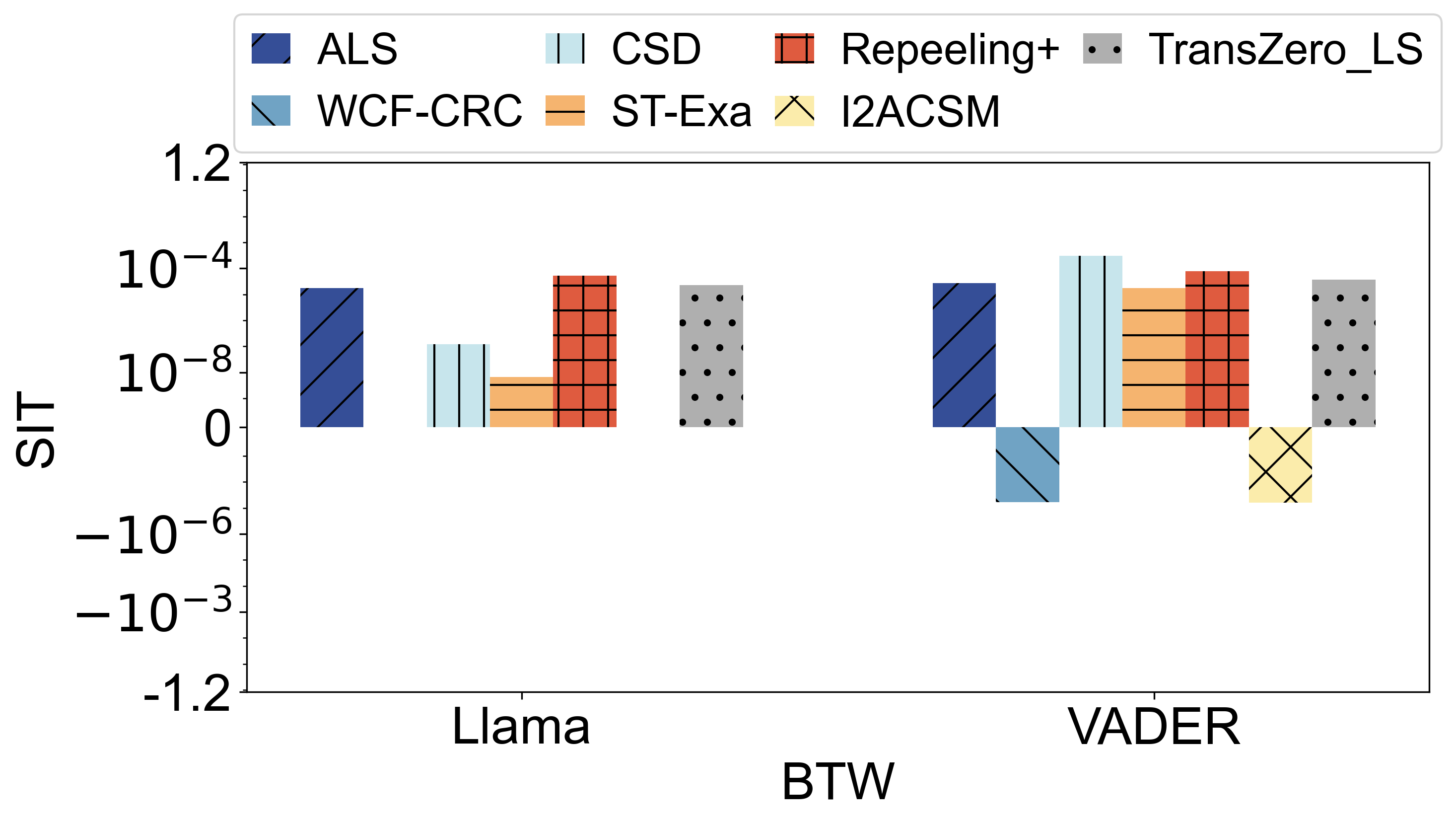}
        \label{fig: SIT_btw_senti}
    \end{minipage}
    \begin{minipage}[t]{0.24\linewidth}
        \centering
        \vspace{-3ex}
        \includegraphics[width=\linewidth]{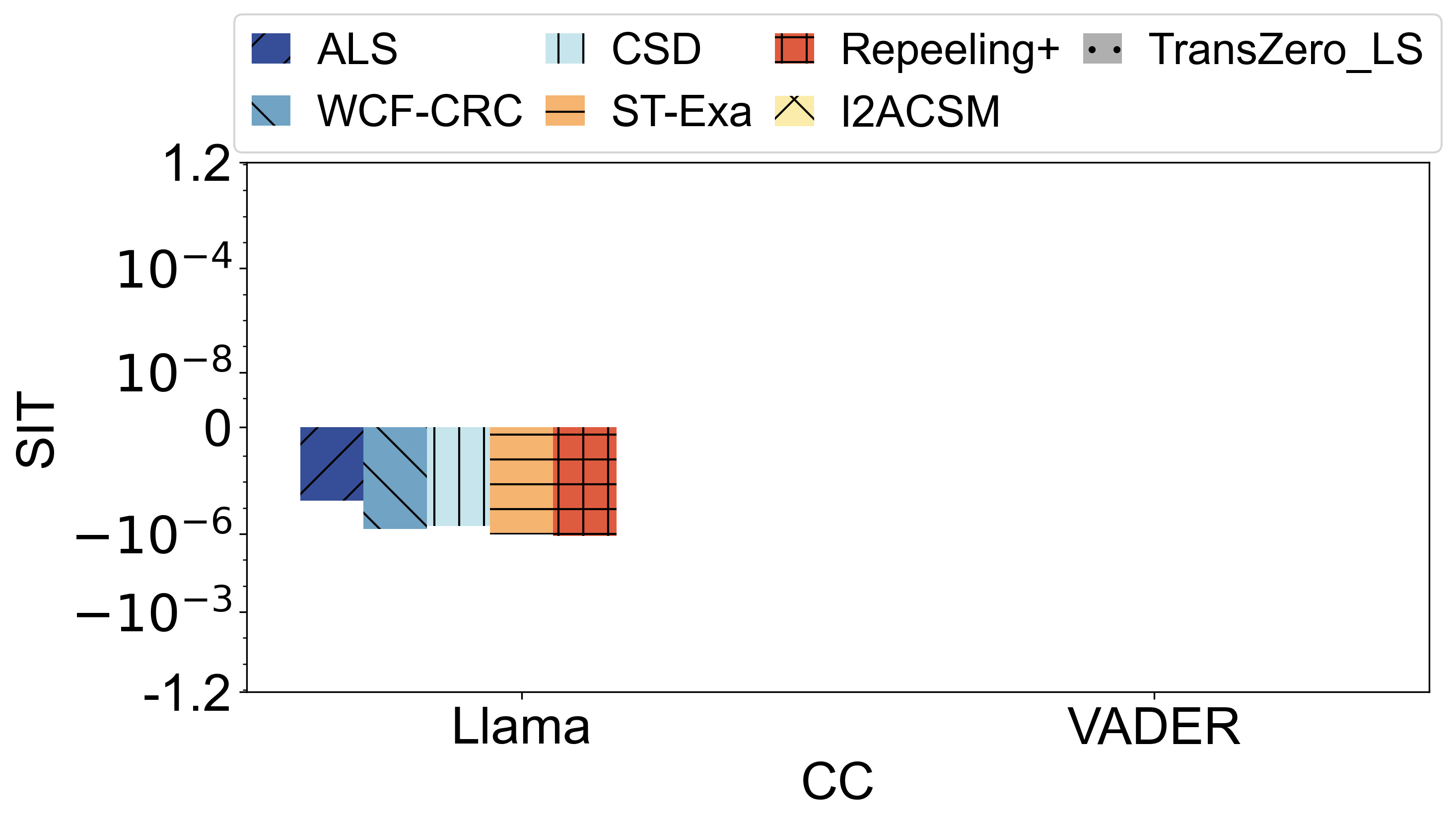}
        \label{fig: SIT_cc_senti}
    \end{minipage}
    \begin{minipage}[t]{0.24\linewidth}
        \centering
        \vspace{-3ex}
        \includegraphics[width=\linewidth]{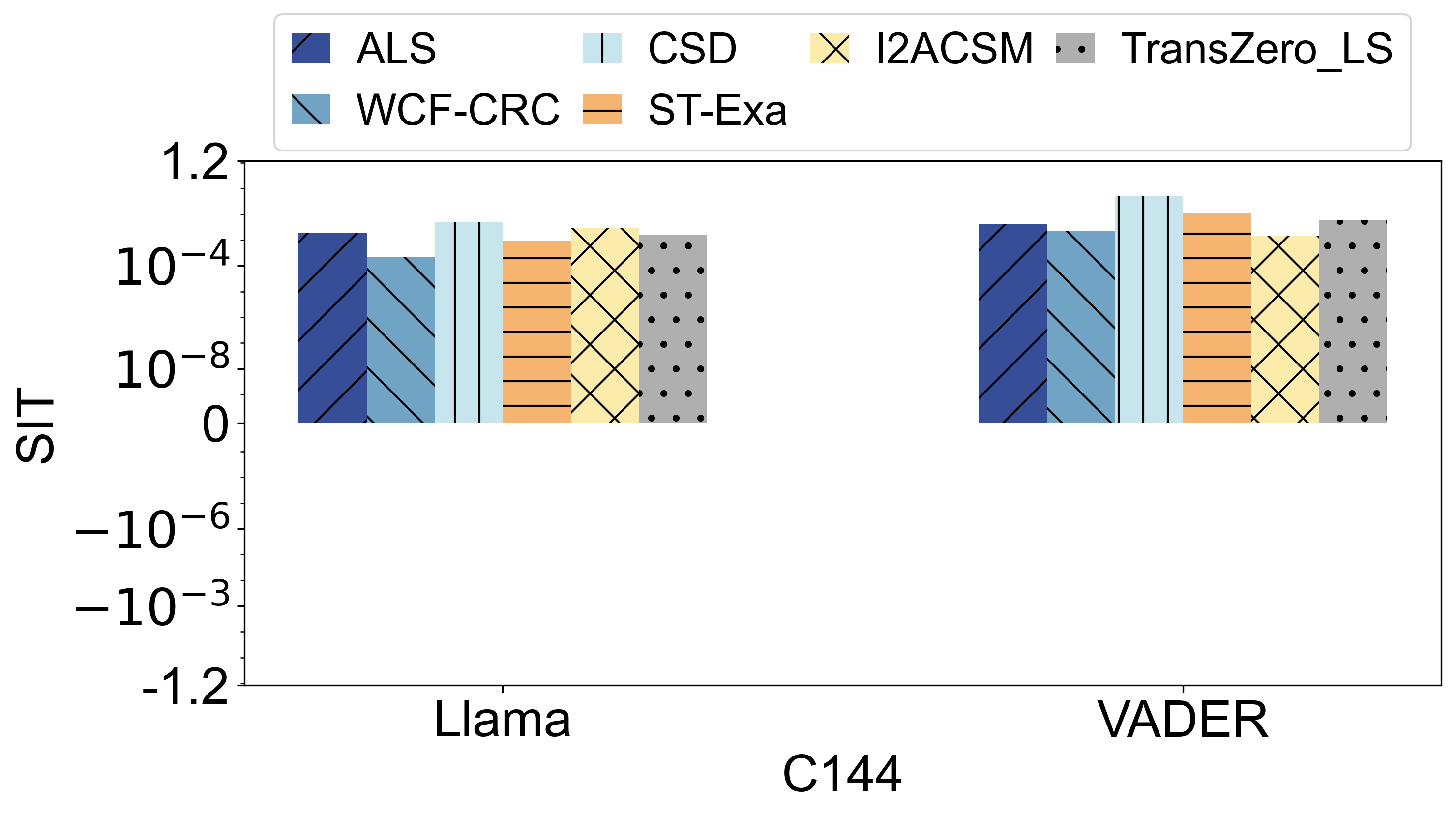}
        \label{fig: SIT_c144_senti}
    \end{minipage}
    \begin{minipage}[t]{0.24\linewidth}
        \centering
        \vspace{-3ex}
        \includegraphics[width=\linewidth]{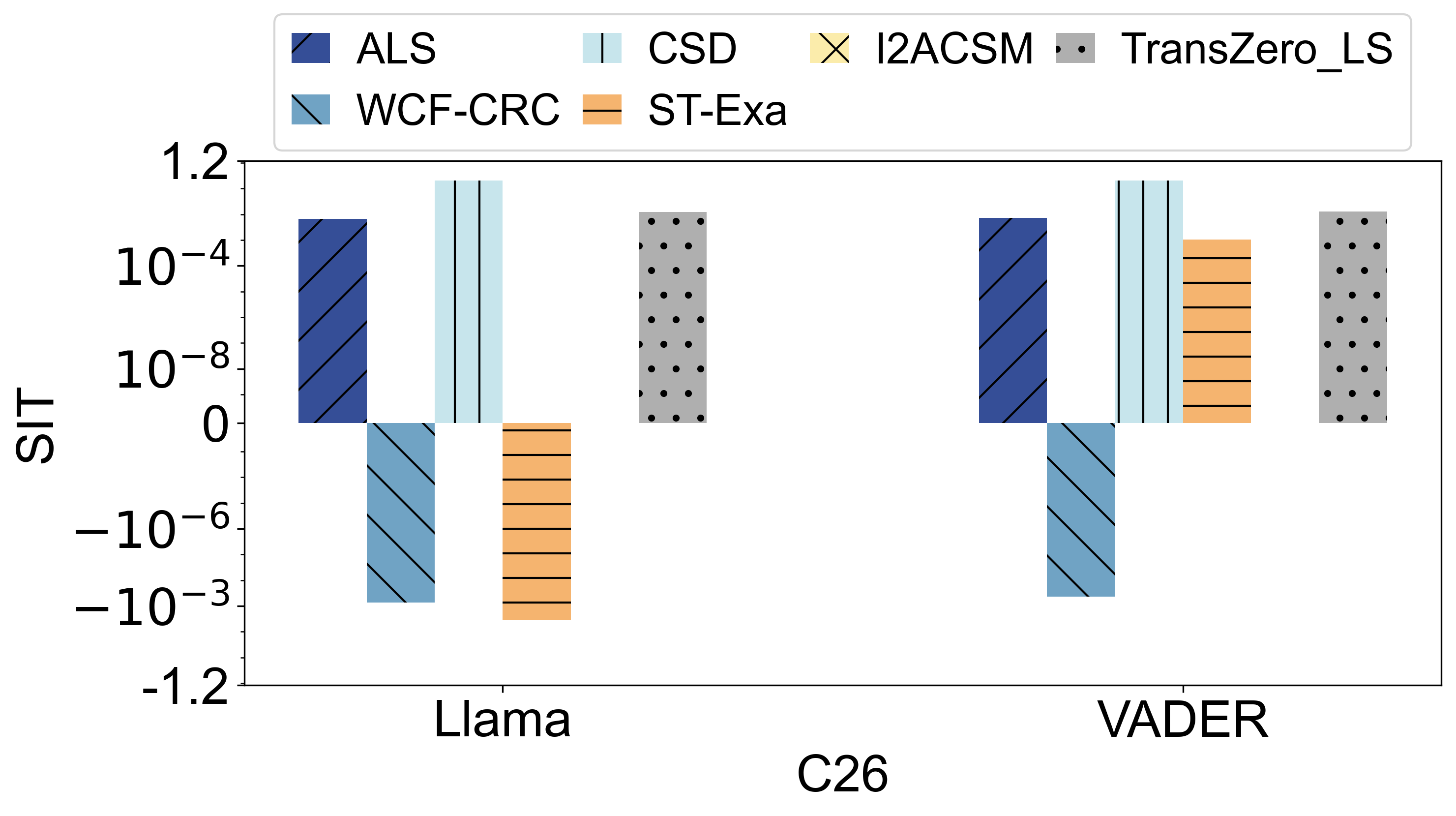}
        \label{fig: SIT_c26_senti}
    \end{minipage}

    \begin{minipage}[t]{0.24\linewidth}
        \centering
        \vspace{-3ex}
        \includegraphics[width=\linewidth]{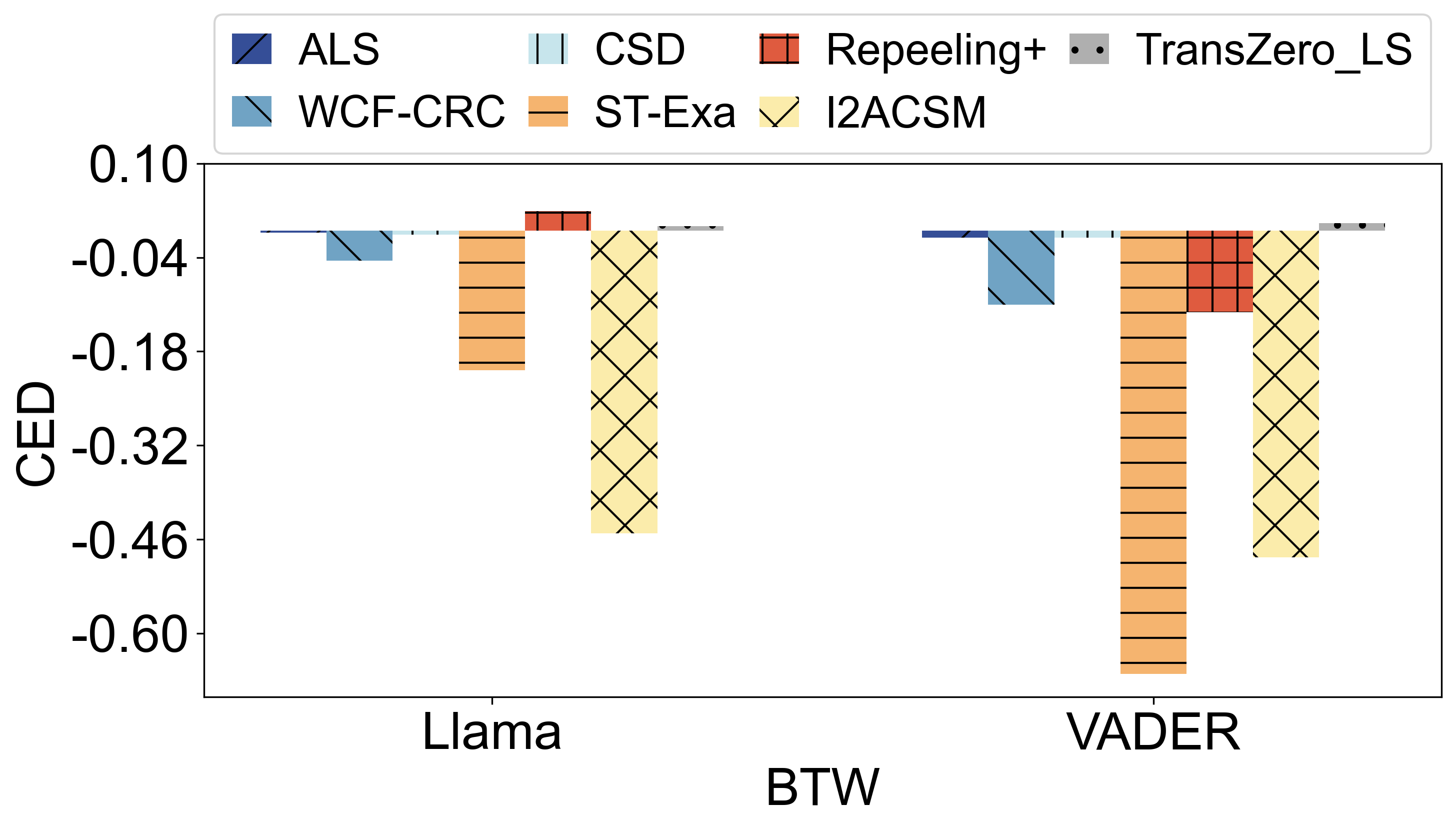}
        \label{fig: CED_btw_senti}
    \end{minipage}
    \begin{minipage}[t]{0.24\linewidth}
        \centering
        \vspace{-3ex}
        \includegraphics[width=\linewidth]{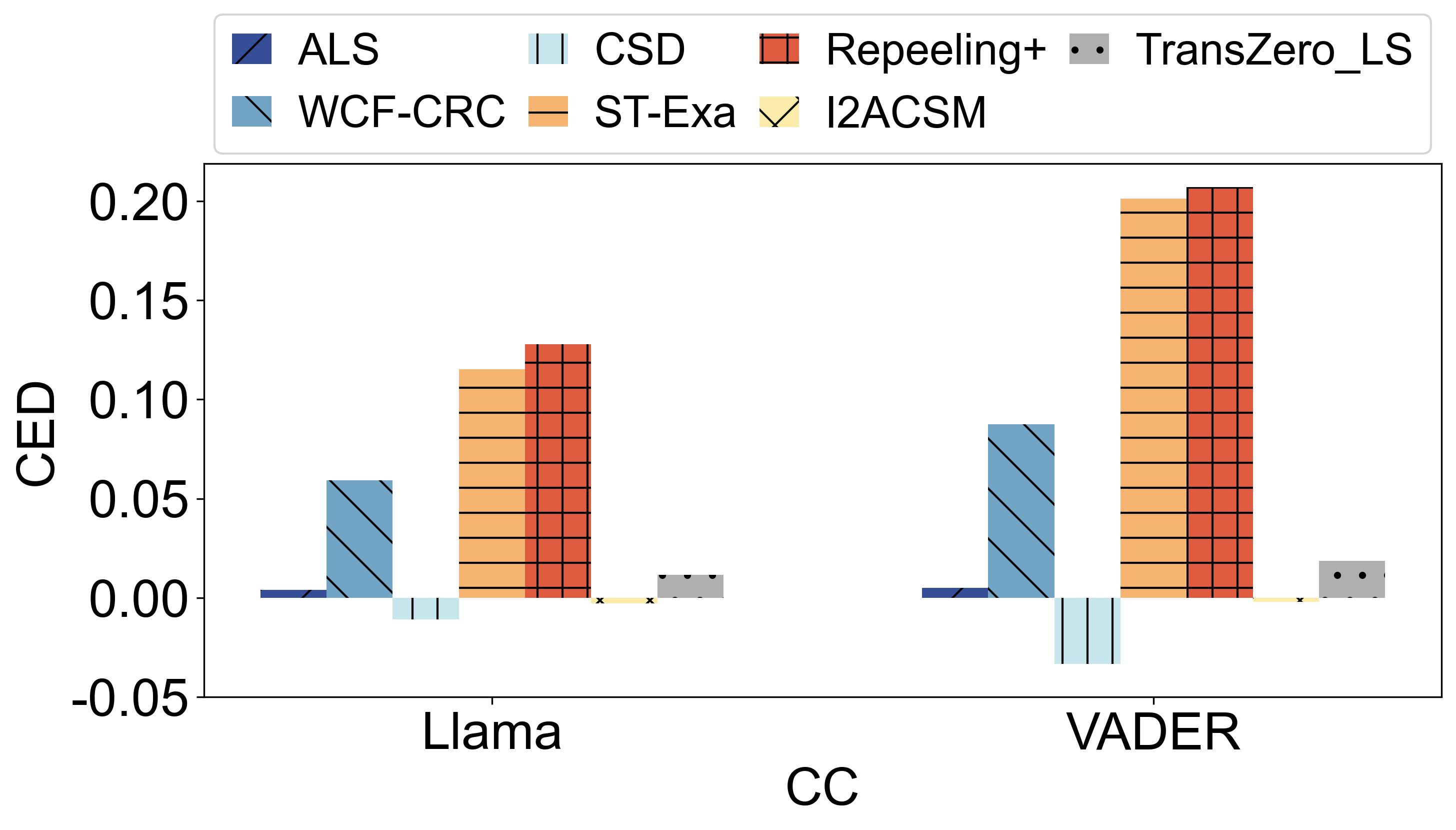}
        \label{fig: CED_cc_senti}
    \end{minipage}
    \begin{minipage}[t]{0.24\linewidth}
        \centering
        \vspace{-3ex}
        \includegraphics[width=\linewidth]{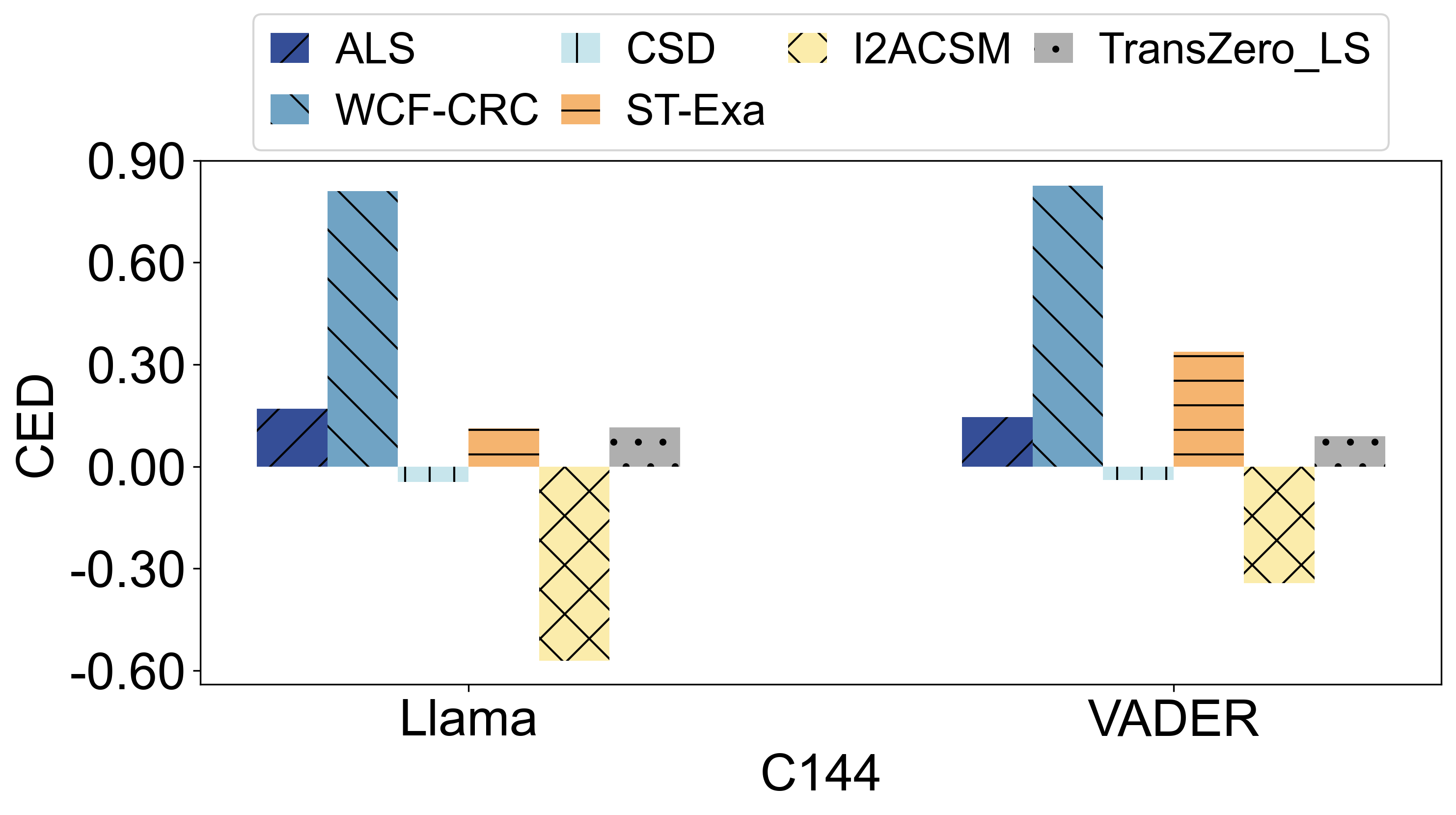}
        \label{fig: CED_c144_senti}
    \end{minipage}
    \begin{minipage}[t]{0.24\linewidth}
        \centering
        \vspace{-3ex}
        \includegraphics[width=\linewidth]{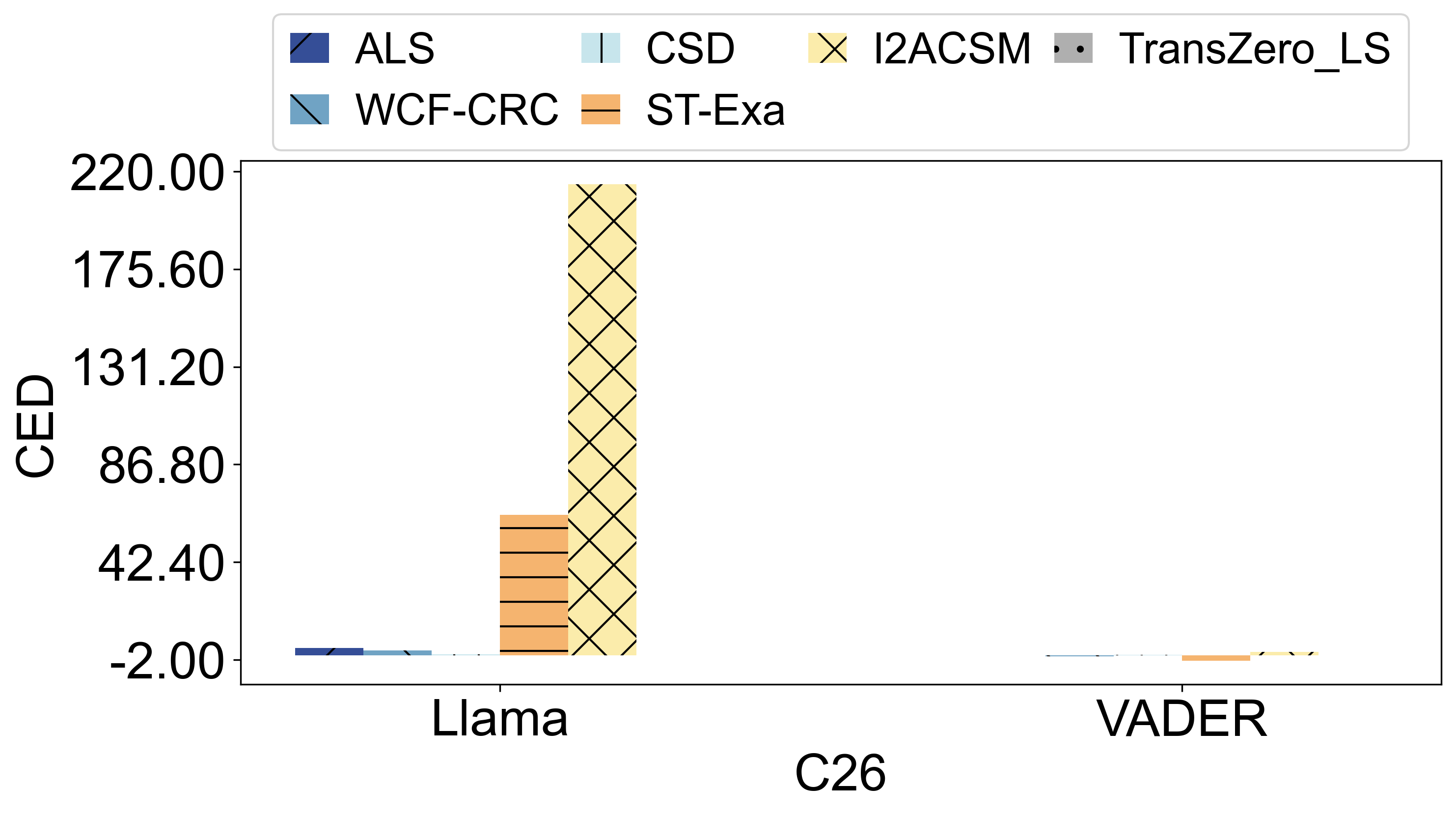}
        \label{fig: CED_c26_senti}
    \end{minipage}
   
    \vspace{-4ex}
    \caption{Impact of sentiment analysis techniques on ATG-S measures.}
    \label{fig: senti_tech}
    \vspace{-2ex}
\end{figure*}

\vspace{-3ex}
\subsection{Impact of Sentiment Analysis Techniques}

To examine the impact of sentiment analysis techniques on evaluation and to demonstrate the broad applicability of our cohesiveness metrics, we adopt a popular lexicon-based approach, \textsf{VADER}~\cite{hutto2014vader}, to relabel the datasets. We recompute sentiment-related cohesiveness measures (\ie EI, SIT, and CED), and compare these results with those from \textsf{Llama3}-labeled datasets. Figure \ref{fig: senti_tech} presents results across four datasets. In most cases, the cohesiveness scores are largely unaffected by sentiment analysis techniques. However, when the labeling technique changes to \textsf{VADER}, the \textsf{ST-Exa} and \textsf{I2ACSM} communities respond more strongly, with their values either switching polarity (\eg \textsf{ST-Exa}'s EI value on \textsf{BTW}) or showing obvious changes (\eg \textsf{ST-Exa}'s and \textsf{I2ACSM}'s CED values on \textsf{C144}).

\begin{figure}[t]
    \begin{minipage}[t]{0.485\linewidth}
        \centering
        \includegraphics[width=\linewidth, height=2cm]{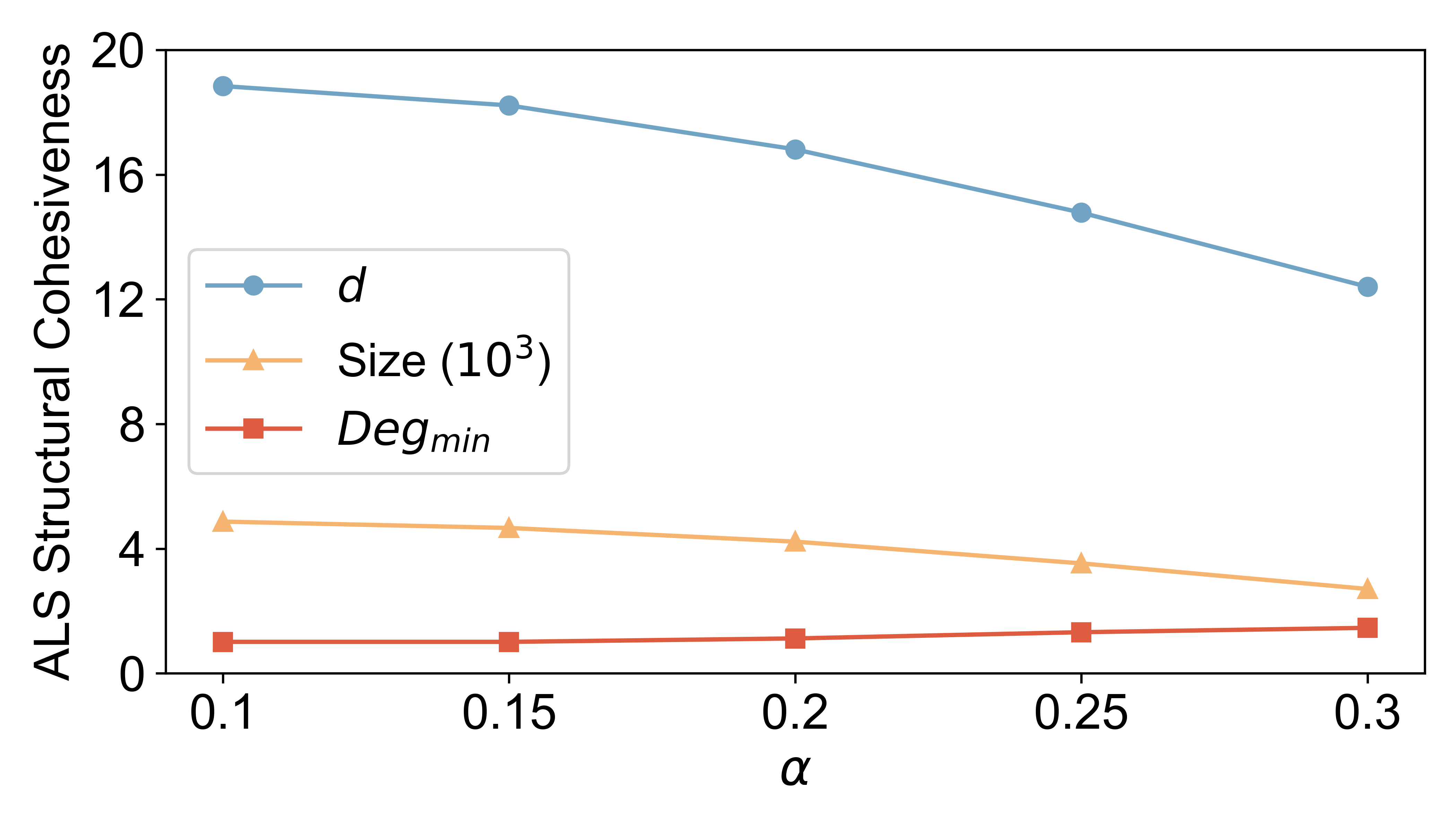}
        \vspace{-5ex}
    \end{minipage} \hspace{-2mm}
    \begin{minipage}[t]{0.492\linewidth}
        \centering
        \includegraphics[width=\linewidth, height=2cm]{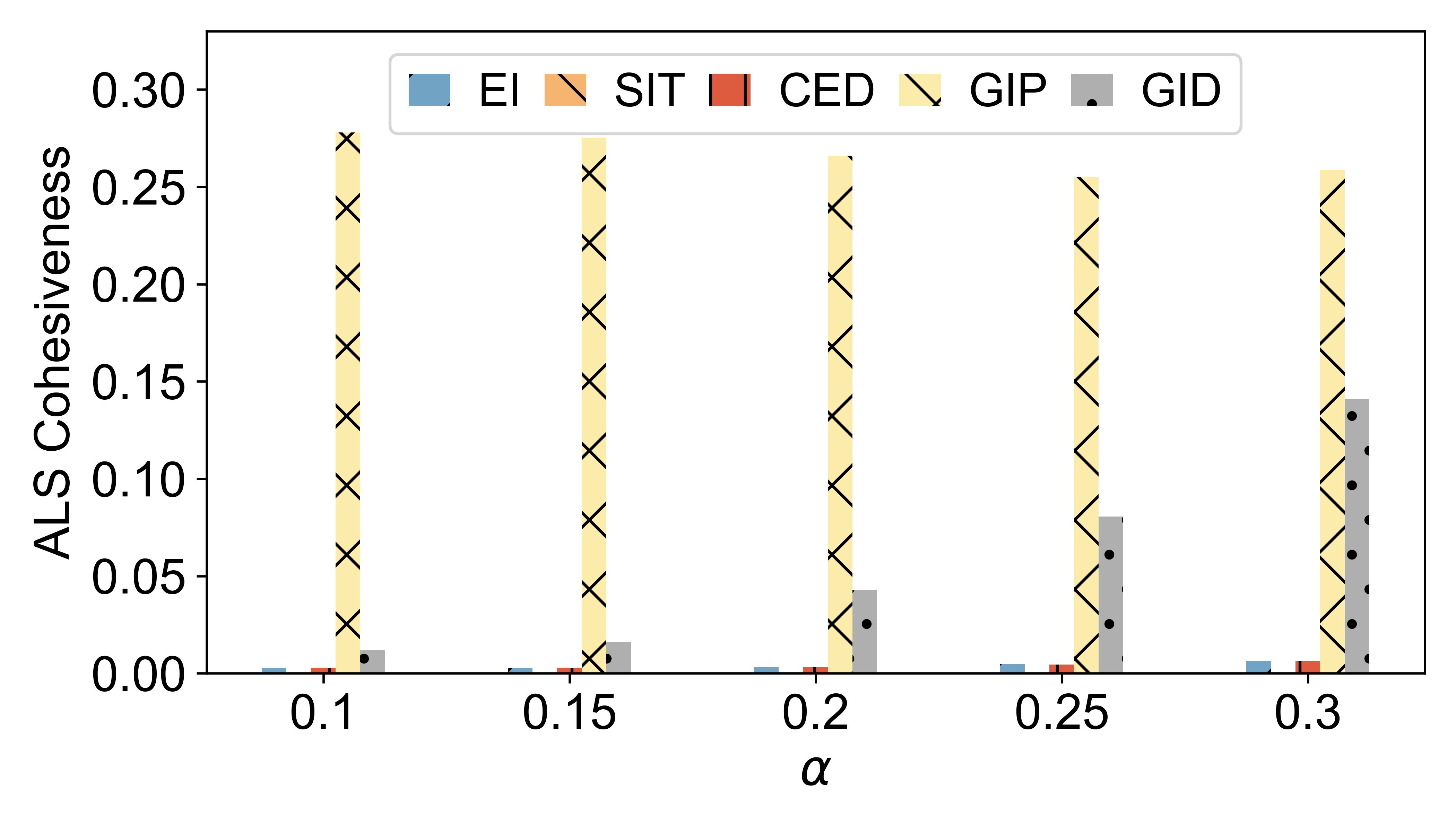}
        \vspace{-5ex}
    \end{minipage}\vspace{-1ex}
    
    \begin{minipage}[t]{0.495\linewidth}
        \centering
        \includegraphics[width=\linewidth, height=2cm]{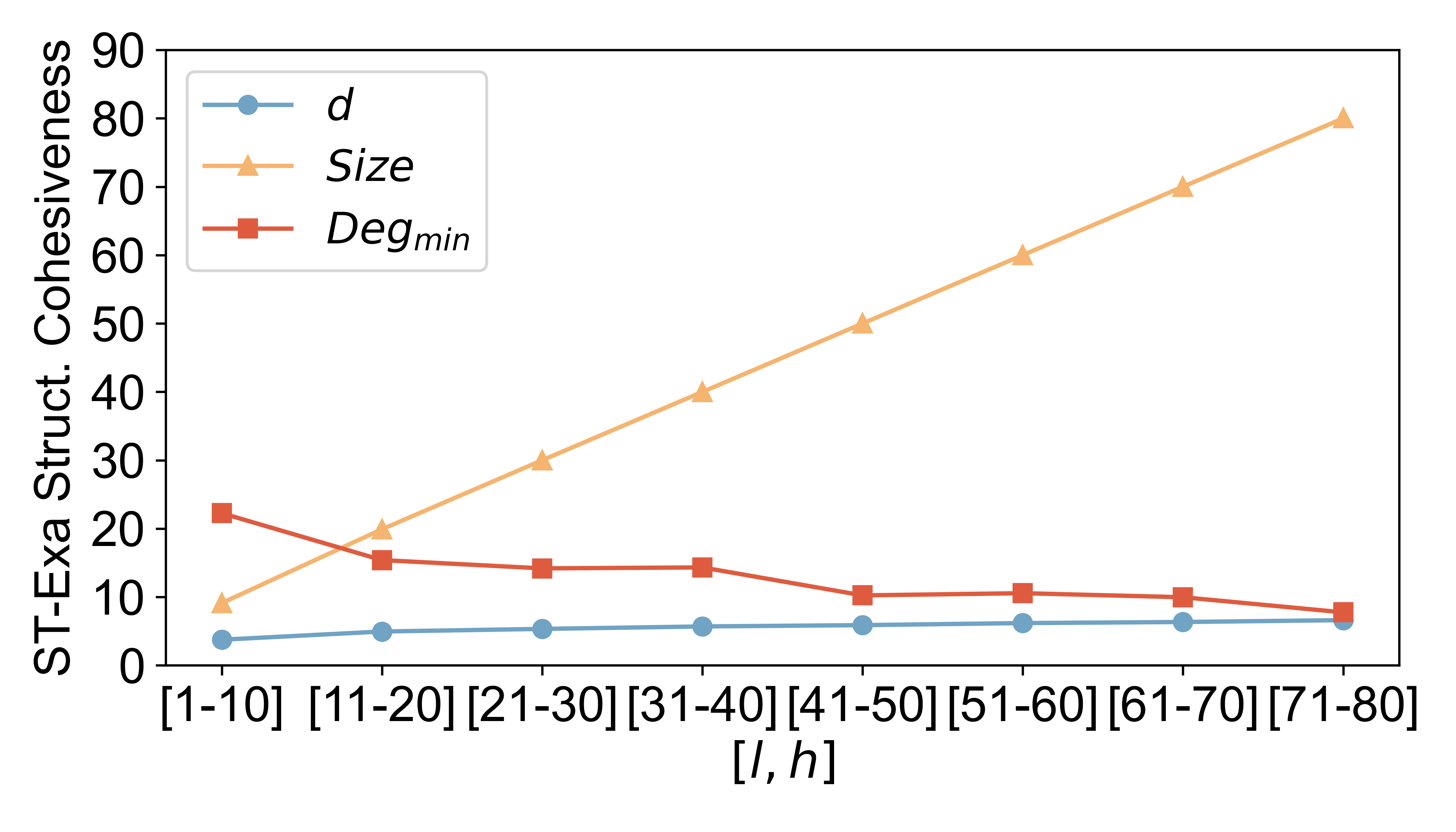}
        \vspace{-5ex}
    \end{minipage} \hspace{-2mm}
    \begin{minipage}[t]{0.492\linewidth}
        \centering
        \includegraphics[width=\linewidth, height=2cm]{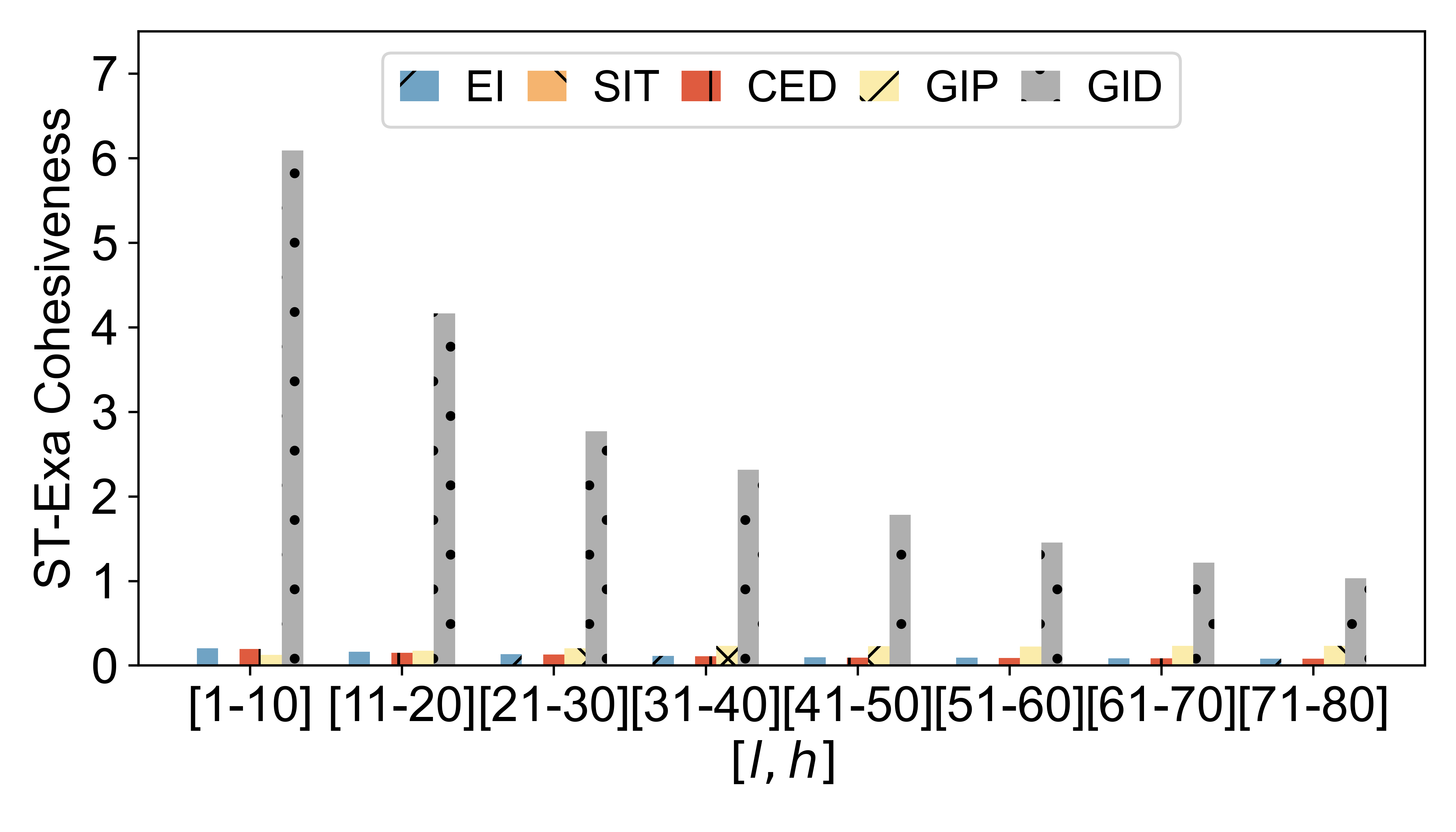}
        \vspace{-5ex}
    \end{minipage}
    \vspace{-3ex}
    \caption{Impact of parameter selection.}
    \label{fig: Params_select}
    \vspace{-3ex}
\end{figure}

\vspace{-1ex}
\subsection{Impact of Parameter Selection}
Next, we discuss how parameter selection affects structural and psychology-informed cohesiveness using two single-parameter algorithms, \textsf{ALS} and \textsf{ST-Exa}, on the \textsf{CC} dataset.

For \textsf{ALS}, increasing the \textit{teleportation probability} $\alpha$ results in smaller communities with higher structural cohesiveness and increasing psychological cohesiveness, except GIP. Notably, GID improves significantly with larger $\alpha$, while other scores show minor increases (Figure \ref{fig: Params_select}, top row). For \textsf{ST-Exa}, members of smaller, structurally dense communities consistently achieve higher EI, CED, and GID. In contrast, members of larger communities tend to engage more, as observed in slightly higher GIP (Figure \ref{fig: Params_select}, bottom row).

\begin{figure*}[t]
    \begin{minipage}[t]{0.28\linewidth}
        \centering
        \includegraphics[width=\linewidth]{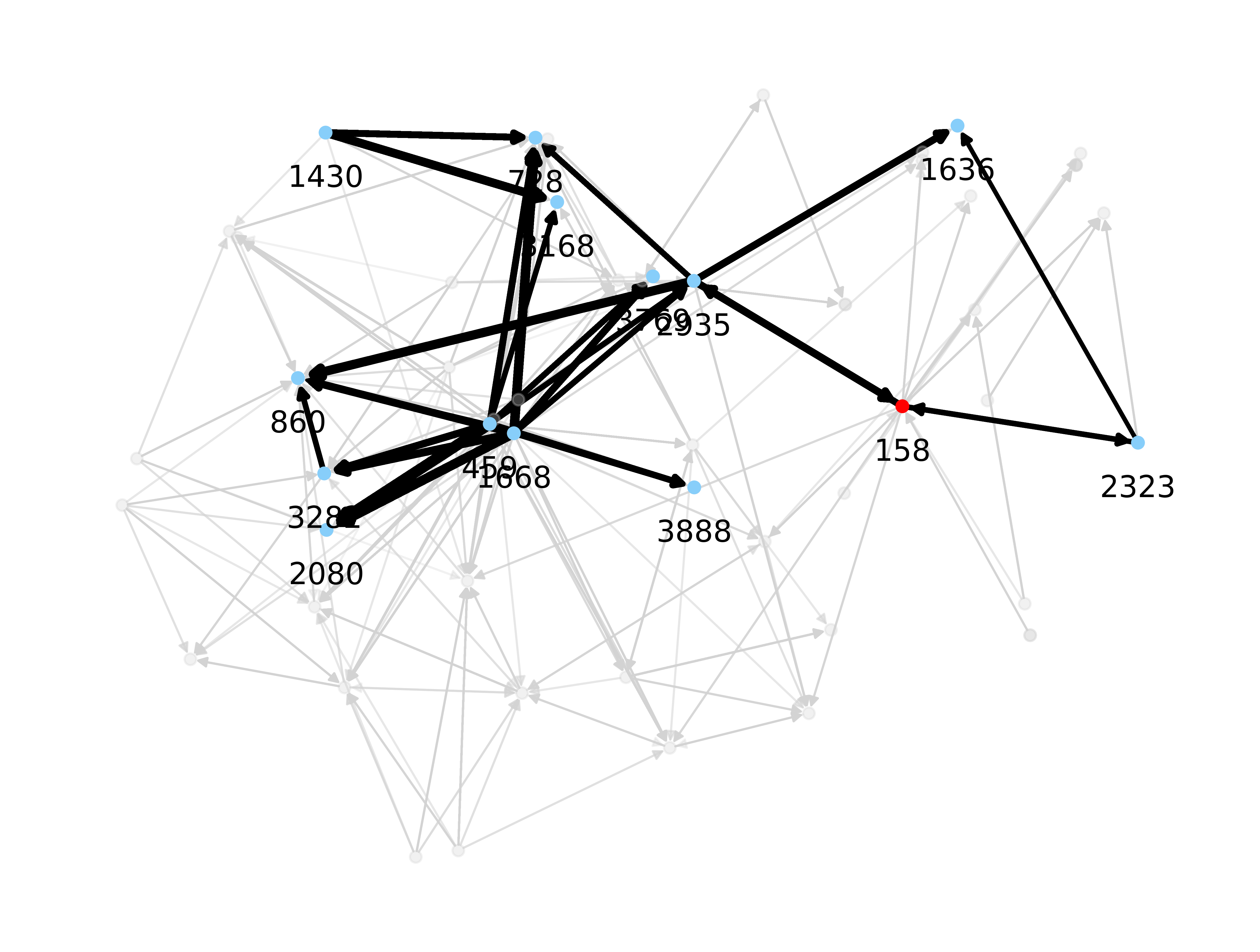}
        \vspace{-5ex}
        \subcaption{WCF-CRC}
        \label{fig: case_CRC}
    \end{minipage} \hspace{-2mm} 
    \begin{minipage}[t]{0.28\linewidth}
        \centering
        \includegraphics[width=\linewidth]{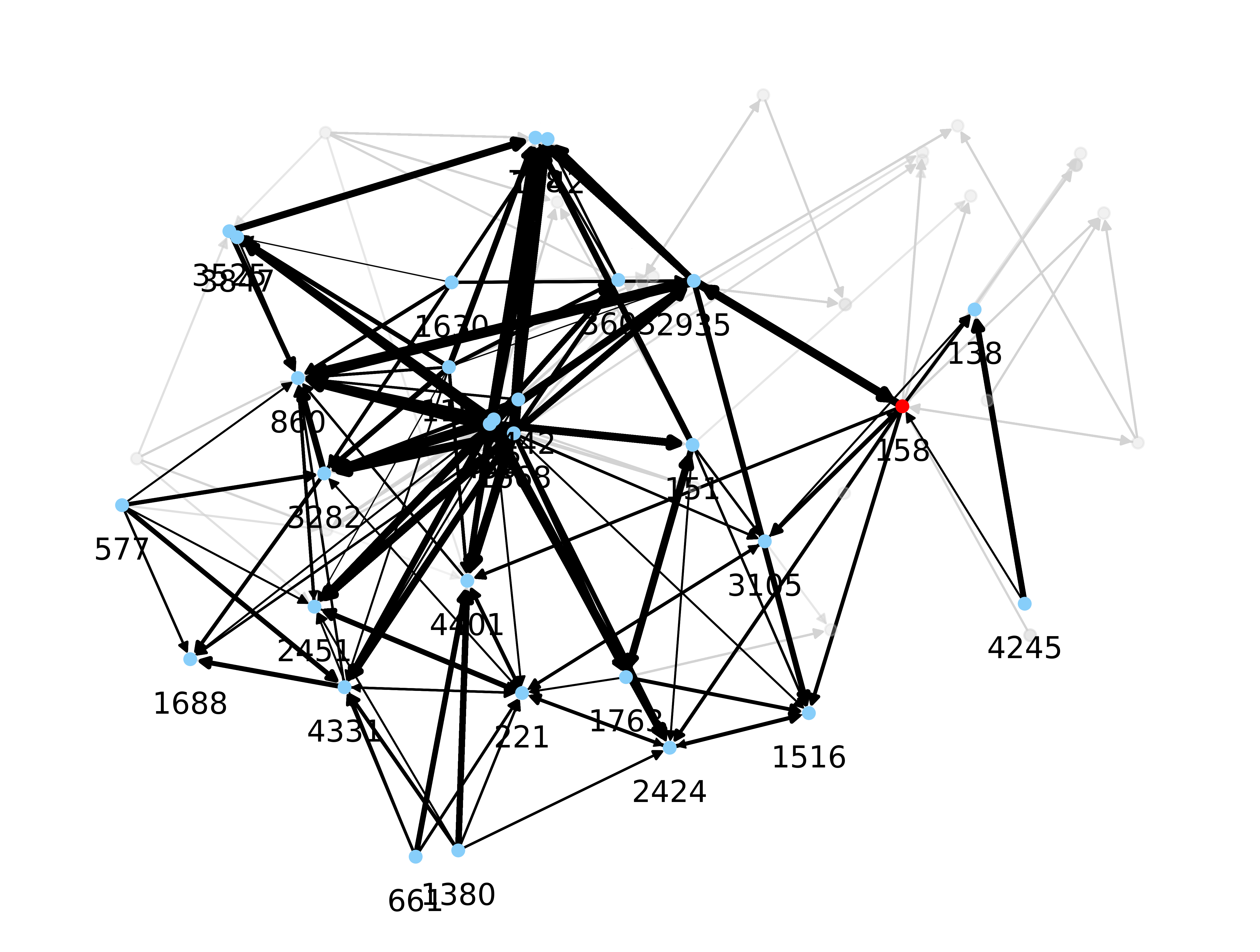}
        \vspace{-5ex}
        \subcaption{ST-Exa}
        \label{fig: case_SETxa}
    \end{minipage} \hspace{-2mm} 
    \begin{minipage}[t]{0.28\linewidth}
        \centering
        \includegraphics[width=\linewidth]{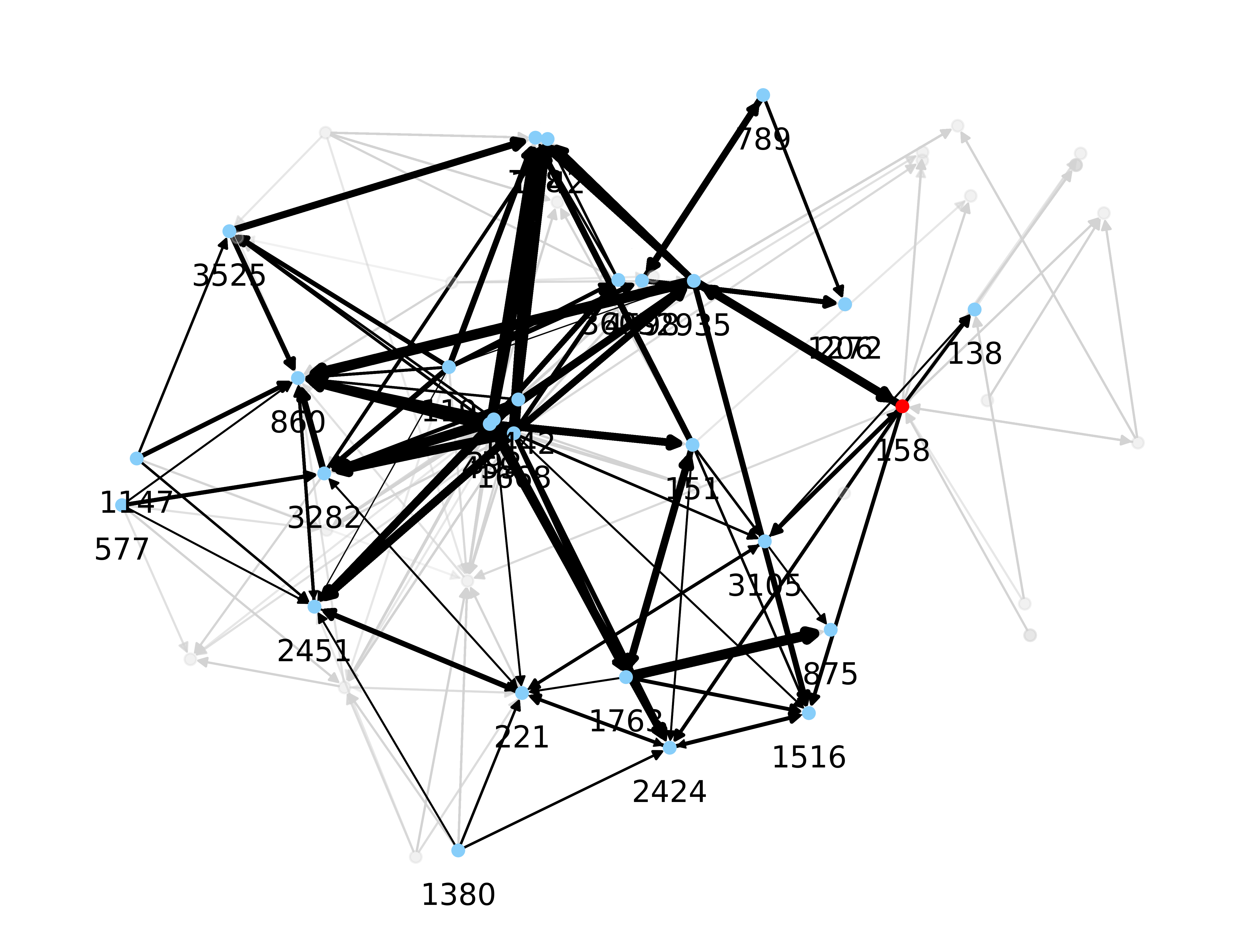}
        \vspace{-5ex}
        \subcaption{Repeeling+}
        \label{fig: case_Repeeling}
    \end{minipage} \hspace{-2mm}

    \begin{minipage}[t]{0.26\linewidth}
        \centering
        \includegraphics[width=\linewidth]{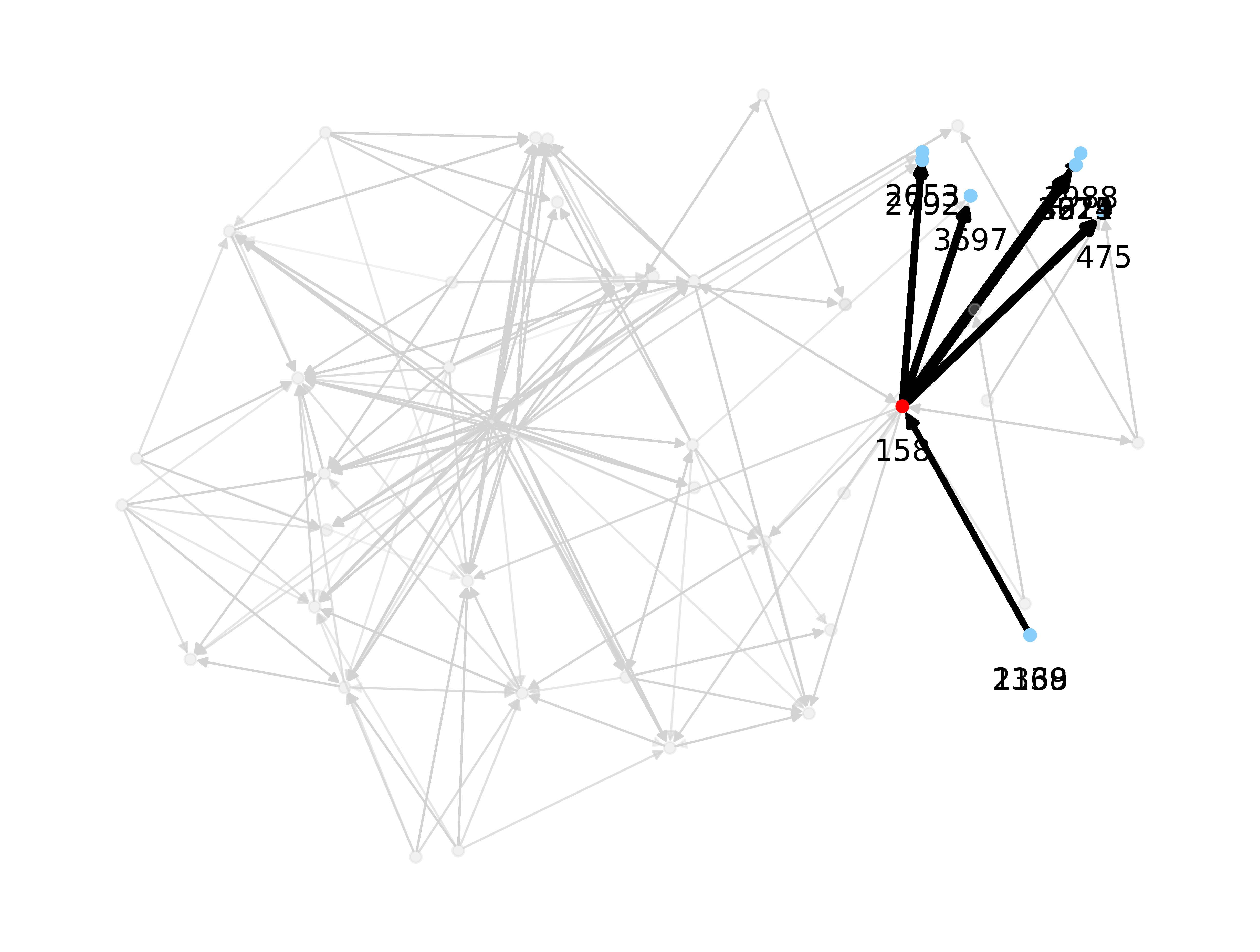}
        \vspace{-5ex}
        \subcaption{I2ACSM}
        \label{fig: case_I2ACSM}
    \end{minipage} \hspace{-2mm} 
    \begin{minipage}[t]{0.26\linewidth}
        \centering
        \includegraphics[width=\linewidth]{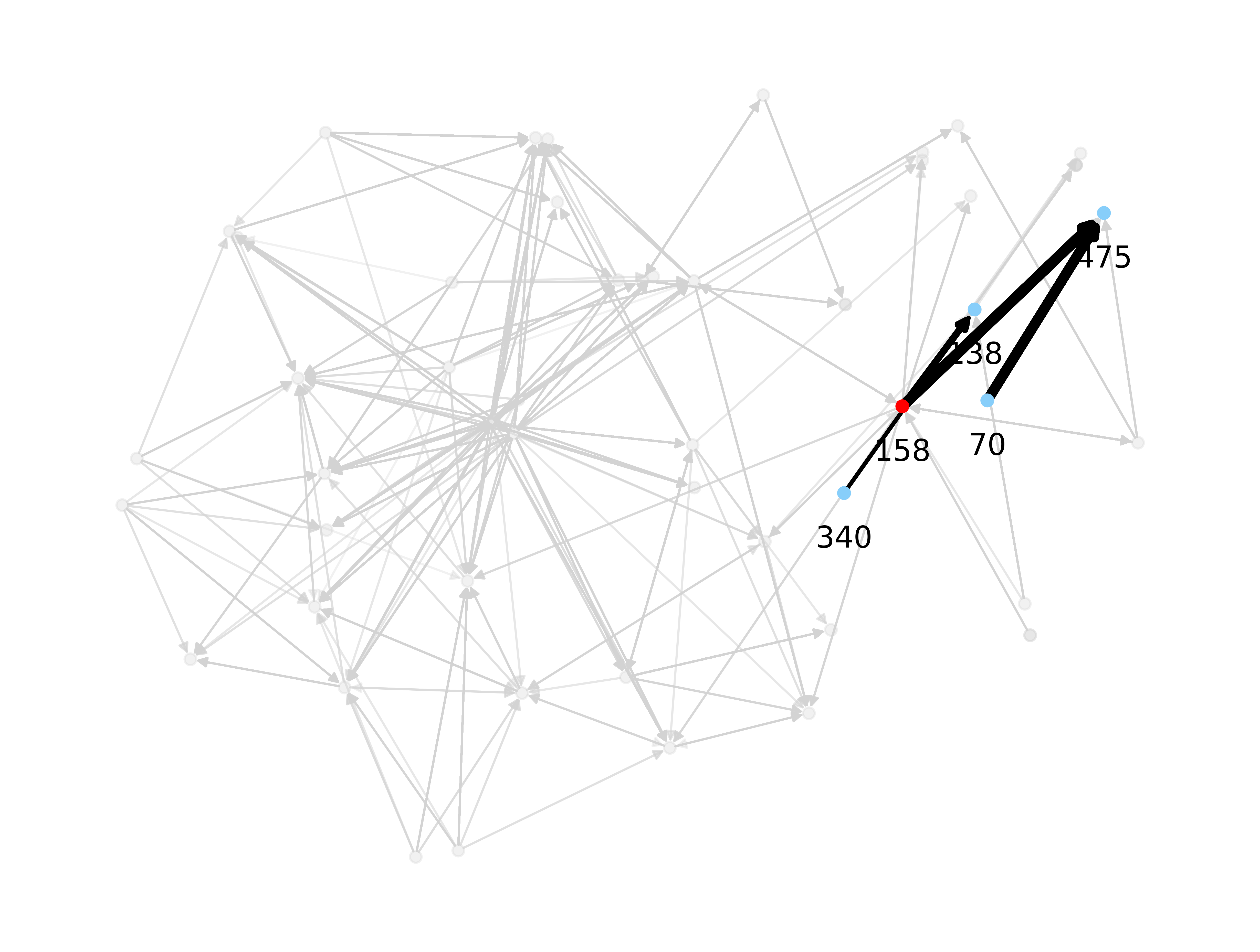}
        \vspace{-5ex}
        \subcaption{TransZero\_LS}
        \label{fig: case_LS}
    \end{minipage}
    \vspace{-2ex}
    \caption{Case study on the \textsf{CC} dataset with the query node marked in red.}
    \label{fig: case_study}
    \vspace{-2ex}
\end{figure*}

\begin{figure}[t]
    \centering
    \vspace{-1ex}
    \includegraphics[width=0.80\linewidth]{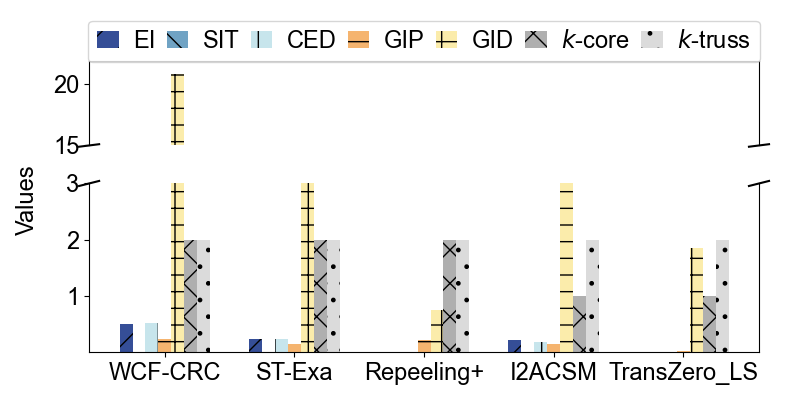}
    \vspace{-3ex}
    \caption{Comparison of two types of cohesiveness.}
    \label{fig: case_coheisveness}
    \vspace{-2ex}
\end{figure}

\vspace{-1ex}
\subsection{Case Study}

Finally, we present a case study using the \textsf{CC} dataset to compare CS results from various algorithms, evaluating cohesiveness with our proposed measures and two additional structural metrics: $k$-core and $k$-truss. The structural metrics are computed after converting the \textsf{CC} into an undirected graph. Here, we present the results for the query node \textsf{``158''}, as most algorithms return its communities. To simplify visualization, we only display algorithms that return communities of size 5-30. If algorithms produce multiple communities under different parameters, the largest one is selected. Besides, self-loops within the networks are not displayed, and edge widths reflect the number of interactions between users (See Figure \ref{fig: case_study}).

We observe that communities returned by different algorithms vary significantly. The first three algorithms identify larger communities with more internal connections. While the communities from \textsf{I2ACSM} and \textsf{TransZero-LS} exhibit frequent interactions between users, their overall connectivity is sparse. For structural cohesiveness, the first three communities are 2-core and 2-truss. However, despite being truss-based, \textsf{Repeeling+} and \textsf{I2ACSM} fail to find more prominent $k$-truss communities. \textsf{I2ACSM} and \textsf{TransZero-LS} communities only meet the 1-core requirement. \textit{In other words, existing CS techniques fail to return consistent results for the same query node.}

Interestingly, our comparison of two types of cohesiveness across five communities reveals no explicit correlation between them (Figure \ref{fig: case_coheisveness}). The 2-core and 2-truss communities identified by the first three algorithms lack consistent higher psychological cohesiveness. Furthermore, all methods struggle to capture psychological cohesion. \textit{That is, existing techniques return communities whose features are misaligned with the notion of cohesion in social psychology}.

\vspace{-2ex}
\section{Reflections \& Future Work}
In this paper, we conduct the first study to extensively evaluate the cohesiveness of communities in online social networks produced by state-of-the-art CS algorithms. Our study reveals that communities identified by existing methods for the same query node lack consensus, and their structural and psychological cohesiveness vary greatly, with no clear correlation between them. Despite the notion of cohesion being strongly rooted in social psychology theories, none of these generic techniques performs well on online social networks \wrt the five psychological cohesiveness measures that are inspired by research in group cohesion. This is primarily because the existing one-size-fits-all cohesiveness measures of CS techniques do not systematically incorporate theories from social psychology into their design. This issue is further compounded by the fact that current CS techniques are either not assessed using ``ground-truth'' communities or rely on communities that are built without considering theories of group cohesion.

Despite these significant results, our evaluation has some limitations. Firstly, the design of our measures is constrained by sentiment analysis tools, and the simple three-class classification in our sentiment analysis may overlook nuanced sentiments inherent in specific comments. Secondly, our study is based on a single social network platform (\ie \textit{X}). However, given the fundamental limitations in the design of existing cohesiveness measures for human-interaction datasets (Section~\ref{subsec: compre}), we believe that the key insights reported in this paper will also hold in other large online social networks. Thirdly, our study cannot use ground-truth communities for benchmarking because no public ground-truth datasets are created based on cohesion theories.

We believe that the findings of our study can inspire a reevaluation of current community search algorithms by incorporating insights from social psychology, leading to the development of more effective solutions. Our proposed measures may be incorporated into search techniques to better identify cohesive communities in online social networks. In particular, it is paramount to create ground-truth community datasets for community search that are grounded in social psychology theories. This will not only enable the validation of our proposed psychology-informed cohesiveness measures but also assist existing CS techniques in developing more effective cohesiveness measures and algorithms for practical usage. Lastly, while this work concentrates on the community search problem, it is clear that our findings could also inspire a rethinking of various \textit{community-centric} social search and analytics problems involving human interactions, such as community detection~\cite{bedi2016community, javed2018community}.

\balance
\bibliographystyle{ACM-Reference-Format}
\bibliography{ref}

\end{document}